\documentstyle[epsf]{mn}
\def\mch{M$\rm^{c}$Hardy\,}
\def\etal{\it et al.~\rm}
\def\del{$\Delta_{X-O}$}
\def\deg{$^{\circ}$}
\def\ecs{ergs cm$^{-2}$ s$^{-1}$~}
\def\lesssim{\mathrel{\hbox{\rlap{\hbox{\lower4pt\hbox{$\sim$}}}\hbox{$<$}}}}
\def\gtrsim{\mathrel{\hbox{\rlap{\hbox{\lower4pt\hbox{$\sim$}}}\hbox{$>$}}}}

\title[The Cosmic Soft X-ray Background]
{\bf THE ORIGIN OF THE COSMIC SOFT X-RAY BACKGROUND: 
OPTICAL IDENTIFICATION OF AN EXTREMELY DEEP ROSAT SURVEY}

\author[M$\rm^{c}$Hardy, I.M., \etal]
{ I M M$\rm^{c}$Hardy$^{1}$, L R Jones$^{1,2}$, M R Merrifield$^{1}$, 
K O Mason$^{3}$, A M Newsam$^{1}$, \and  R G Abraham$^{4,5}$,
G B Dalton$^{7}$, F Carrera$^{3,6}$, P J Smith$^{3}$, 
M Rowan-Robinson$^{8}$, \and G A Wegner$^{9}$, T J Ponman$^{2}$,
H J Lehto$^{10}$, G Branduardi-Raymont$^{3}$, \and
G A Luppino$^{11}$, G Efstathiou$^{7}$, D J Allan$^{2}$, J J Quenby$^{8}$. \\
$^{1}$ Department of Physics and Astronomy, The University, Southampton SO17 1BJ\\
$^{2}$ School of Physics and Space Research, University of Birmingham,
Edgbaston, Birmingham B15 2TT.\\
$^{3}$ Mullard Space Science Laboratory, University College London, 
Holmbury St Mary, Dorking RH5 6NT\\
$^{4}$ Dominion Astrophysical Observatory, 5071 West Saanich Road, Victoria
V8X 4M6, British Columbia, Canada.\\
$^{5}$ Institute of Astronomy, Madingley Road, Cambridge CB3 0HA.\\
$^{6}$ Instituto de Fisica de Cantabria, 39005 Santander, Spain.\\
$^{7}$ Department of Astrophysics, Keble Road, Oxford OX1 3RH.\\
$^{8}$ Department of Physics, Imperial College, Prince Consort Road,
London SW7 2BZ.\\
$^{9}$ Department of Physics and Astronomy, Dartmouth College, 
Hanover, NH 03755, USA.\\
$^{10}$ Tuorla Observatory, Turku University, Piikki\"{o}, FIN-21500, Finland.\\
$^{11}$ University of Hawaii, Institute for Astronomy, 2680 Woodlawn Drive,
Honolulu, Hawaii 96822, USA.
}
\date{{\bf Note:} As accepted by Monthly Notices, 
September 1997.}

\begin{document}
\maketitle

\begin{abstract}
We present the results of the deepest optically identified X-ray
survey yet made. The X-ray survey was made with the ROSAT PSPC and
reaches a flux limit of $1.6 \times 10^{-15}$ erg cm$^{-2}$ s$^{-1}$
(0.5 -2.0 keV). Above a flux limit of $2 \times 10^{-15}$ erg
cm$^{-2}$ s$^{-1}$ we define a complete sample of 70 sources of which
59 are identified. For a further 5 sources we have tentative
identifications and for a further 4 the X-ray error-boxes are blank to
R=23 mag.  At brighter fluxes ($\geq 10^{-14}$ erg cm$^{-2}$ s$^{-1}$)
we confirm the results of previous less deep X-ray surveys with 84\%
of our sources being QSOs. However at the faint flux limit the survey
is dominated by a population of galaxies with narrow emission lines
(NELGs).  In addition at intermediate fluxes we find a small number of
groups and clusters of galaxies at redshifts generally $>0.3$.  Most
of these groups are poor systems of low X-ray luminosity and the
number which we find is consistent with a zero evolutionary scenario,
unlike the situation for high luminosity clusters at the same
redshift.  To a flux limit of $2 \times 10^{-15}$ erg cm$^{-2}$
s$^{-1}$ QSOs contribute $>31\%$ of the cosmic soft X-ray
background (XRB), groups/clusters contribute $\sim 10\%$ and NELGs
contribute $\sim8\%$. However the QSO differential source count slope
below $ 10^{-14}$ erg cm$^{-2}$ s$^{-1}$ is $\sim$-1.4, severely
sub-Euclidean, as is the (poorly defined) group/cluster slope, whereas
the differential NELG slope is close to Euclidean ($\sim-2.4$).  If
the NELG source counts continue to rise at that slope, all of the
remaining cosmic soft XRB will be explained by a flux limit of $\sim1-2
\times 10^{-16}$ erg cm$^{-2}$ s$^{-1}$ with NELGs
contributing about one quarter of the XRB. The average NELG X-ray
spectrum is harder than that of the QSOs, and similar to that of the
remaining unresolved cosmic (XRB) suggesting that NELGs will also be
substantial contributors to the XRB at higher energies.  The observed
NELGs lie in the redshift range 0.1-0.6 and have $M_{R}=-20$ to $-23$,
approximately 3 magnitudes more luminous than typical field
galaxies. They have predominantly blue colours, and some are
definitely spirals, but the presence of some ellipticals cannot yet be
ruled out. Many are in interacting or disturbed systems.  The NELGs
have optical spectra similar to those of the majority of the field
galaxy population at a similar redshift and may simply be the more
luminous members of the emission line field galaxy population.  Based
on optical line ratios and X-ray/optical ratios, the NELGs, both as a
sample and within individual galaxies, appear to be a mixture of
starburst galaxies and true AGN.

\end{abstract}
 
 
\section{INTRODUCTION}
 
The origin of the extragalactic X-ray background (XRB) is one of the
key questions in astrophysics. The excellent fit of the microwave
background, as observed by COBE, to a pure blackbody undistorted by
Compton scattering from hot electrons, has ruled out a major
contribution to the XRB ($\geq$few \%) from diffuse hot gas (Mather
\etal 1990), so a large collection of discrete sources must be
responsible. Previous X-ray surveys with ROSAT have been dominated by
QSOs (Shanks \etal 1991; Boyle \etal 1993) which, to a flux limit of
$2 \times 10^{-14}$ erg cm$^{-2}$ s$^{-1}$ (the band 0.5-2.0 keV is
used throughout this paper), can account directly for $>$30\% of the
cosmic soft XRB. Here, following preliminary reports by Jones \etal
(1995) and \mch (1995) which listed most of the major conclusions, we
present full details of the deepest optically identified X-ray survey
yet made which clarify the origin of the remainder of the cosmic soft
XRB.

The X-ray data consist of a 115 ksec ROSAT PSPC observation reaching a
flux limit of $1.6 \times 10^{-15}$ erg cm$^{-2}$ s$^{-1}$ (Section
2). This observation was made in two parts and source counts from the
first 73 ksec of the observation have already been given by
Branduardi-Raymont \etal (1994 - hereafter GBR).  Deep optical CCD
images (Section 3.1) have been taken in V and R of the inner 15 arcmin
radius of the ROSAT field to provide optical identifications and
colours. Optical spectra (Section 3.2) have subsequently been taken of
sources in a statistically complete sub-section comprising $\sim80\%$
of the 15 arcmin radius survey area. Deep radio surveys at 20 and 6 cm
have also been made of the survey area and some preliminary results
are presented here.

The main purpose of this paper is not to recalculate the integral
source counts of GBR using an increased dataset, although we do
present such counts, but to discuss the identification content of the
survey (Section 4) and to consider the implications for the origin of
the soft X-ray background. Within a region comprising $\sim80\%$ of
the full 15 arcmin radius X-ray survey region, we have optical spectroscopic
observations of almost all likely optical identifications for almost
all X-ray sources. We refer to this regions as our (spectroscopically)
`complete area' and all discussion of the identification content of
our survey refers to this area.  We show that although QSOs dominate at
the brighter flux levels, as found in previous ROSAT surveys, galaxies
mainly with narrow emission lines (NELGs) dominate at the faintest
flux levels.  We also note that a small number of sources are
definitely identified with groups or clusters of galaxies.  The X-ray
spectra of the various classes of identifications are discussed in
Section 5 where they are compared with the spectrum of the remaining
unresolved XRB. In Section 6 we calculate the contribution of the
various classes of identified sources to the soft XRB and show that,
if the NELG source counts extrapolate to
$1-2 \times 10^{-16}$ erg cm$^{-2}$ s$^{-1}$ they will account for
almost all of the remaining soft XRB. We also discuss their possible
contribution to the XRB at higher energies. The optical spectra of the
NELGs are compared with those of other narrow emission line galaxies
such as starburst or Seyfert 2 galaxies in Section 7 and in Section 8
we compare the NELGs with field galaxies in a similar redshift
range. In Section 9 we summarise our results.

\section{ROSAT OBSERVATIONS}

Observations with the ROSAT position sensitive proportional counter
(PSPC) were made at position RA 13 34 37.0 Dec +37 54 44 (J2000), in a
region of sky of extremely low obscuration ($N_{H} \sim 6.5 \times
10^{19}$ cm$^{-2}$). The low obscuration was initially found in
the neutral hydrogen survey of Stark \etal (1992) and confirmed by
analysis of the IRAS 100$\mu$ cirrus maps which showed very low
emission in the chosen direction. The observations were made in two
parts.  The source counts resulting from the first 73 ksec observation
in June 1991 have already been reported by GBR who provide details of
the absorption measurements. Fluctuation analysis of these same data
were presented by Barcons \etal (1994). These two initial papers show
that the source density flattens off below a Euclidean distribution at
fluxes less than $1.6 \times 10^{-14}$ erg cm$^{-2}$ s$^{-1}$.

A further observation of 42ksec was made with the PSPC in June and July 1993.
Together these two observations comprise the second deepest
ROSAT X-ray survey made. The deepest PSPC survey, of $\sim$150ksec duration,
is in the direction of the first Lockman hole (10h 48m +57$^{\circ}$) by
Hasinger \etal (1993) but optical results have not yet been published
from that survey.

\subsection{Source Searching} 

Initial analysis of the combined AO1 and AO4 data were carried out using the
UK ASTERIX software system to create an image from the combined
datasets. Following GBR we have rejected data
with high anticoincidence rate (Master Veto Rate $>$ 170) or bad
aspect ratio. This has left 115ksec of `clean' data. To optimise the
search for point sources we have restricted the energy band to 0.5-2.0
keV. Below 0.5 keV there is a substantial contribution to the
background from diffuse emission in our own galaxy and above 2.0 keV the
particle background dominates.  Source searching was carried out
using our own software, based on the Cash (1979) statistic,
including the radial variation of the psf.
Our algorithm is essentially the same as the standard Starlink PSS source
detection algorithm but improves the fitting of the local background.
We set our detection criterion such that $\sim$one of the sources in our
survey is expected to be false. This is approximately the same as
using a 3.5$\sigma$ significance detection threshold.  

Flux variability between the AO1 and AO4 observations is small with
only $\sim$20\% of the sources showing variability, with a maximum
detected variation of $\sim$20\% amplitude. We leave discussion of
source variability to a future paper and here consider time-averaged
fluxes.

To convert between count rate and flux we follow GBR and assume that
the X-ray spectra of most sources are described by power laws of
energy index $\alpha$=+1.1 (where the flux, $S$, at an energy, $E$, is
given by $S(E)\propto E^{-\alpha}$) and that the absorbing column is
$6.5 \times 10^{19}$ cm$^{-2}$.  The conversion then is that 1 pspc
(0.5-2.0 keV) count s$^{-1}$ = $1.14 \times 10^{-11}$ erg cm$^{-2}$
s$^{-1}$. Varying the assumed column by 20\% (eg to $8 \times 10^{19}$
cm$^{-2}$) has no measureable effect on the conversion but varying
$\alpha$ has a small effect. For example, assuming $\alpha$=0.5 gives
1 pspc (0.5-2.0 keV) count s$^{-1}$ = $1.18 \times 10^{-11}$ erg
cm$^{-2}$ s$^{-1}$. In section 5 we will see that, although an energy
index of $\sim1$ is correct for the QSOs which comprise the majority
of the bright sources, there is a population of faint galaxies with
flatter spectra, $\alpha \sim 0.5$. However as the uncertainty in the
count/flux conversion is a very small effect, particularly for faint
sources where photon counting errors are noticeably larger, we
continue to use the conversion factor appropriate to $\alpha$= 1.1 for
all sources.

\subsection{Simulations}

\subsubsection{Completeness of the Survey}

We have conducted simulations to determine the level of completeness
of our survey and the accuracy of our positions.  The simulations were
carried out in a manner similar to that described in GBR.
Each simulated observation consists of a single square frame 34 arcmin along
each side with 4 arcsec square pixels. Sources are positioned randomly
about the frame and assigned a flux between $2.0 \times 10^{-16}$ and
$5.0 \times 10^{-14}$ erg cm$^{-2}$ sec$^{-1}$ drawn from the observed
source counts of the survey with a linear extrapolation below $2.0
\times 10^{-15}$ erg cm$^{-2}$ sec$^{-1}$. We simulated 150 frames
giving a total area of $\sim$ 250 times that of the present deep survey.
 
The source detection algorithm is then applied to the simulated
observation frame.  In the simulations we use only the on-axis PSPC
PSF throughout. The (small) effect of the radial variation of the
psf and the effective area are considered below.  The source fitting
procedure produces a list of positions of `found' sources $(x_f \pm
\sigma_x, y_f \pm \sigma_y)$ with measured fluxes ($F_f$). The
observed positions and fluxes are then compared with the list of known
`real' (ie input)  positions $(x_r, y_r)$ and fluxes ($F_r$) in
order to find likely matches. The comparison was performed
as follows. Each `real' source brighter than $1.0 \times 10^{-15}$
erg cm$^{-2}$ sec$^{-1}$ is read in and compared to each of the
`found' sources within a 1 arcmin radius using a $\chi^2$-like statistic:
\begin{equation}
  S = {\left ( {{x_f - x_r} \over {\sigma_x}} \right ) }^2 +
      {\left ( {{y_f - y_r} \over {\sigma_y}} \right ) }^2 +
      {\left ( {{F_f - F_r} \over {\sqrt{F_r}}} \right ) }^2
\end{equation}
Each `real' source is then matched with the best-fit (minimum-$S$)
`found' source and the resulting real/found pairs are removed from
their respective source lists. The matching process continues until
only unmatched `real' and `found' sources remain. Unmatched `real'
sources are referred to as missing sources and unmatched `found'
sources are referred to as spurious sources.  In order to avoid
edge-effects, all `real' and `found' sources within 1 arcmin of the
edge of the simulated frames are removed from the analysis.

The statistic `S' is distributed very much like the $\chi^2$
statistic, ie the large majority of matches have a low value of S
corresponding to `real' and `found' sources of very similar flux and
position.  A very small number of matches have large values of S, but
that is just what we would expect in any real observation. We
therefore do not impose any upper limit on the accepted value of
S. Our only cut-offs are that the matches should occur within 1 arcmin
and that we do not match `real' sources of flux less than $1.0 \times
10^{-15}$ erg cm$^{-2}$ sec$^{-1}$. We could easily impose an
upper cut-off on S, such as the equivalent of a $5\sigma$ confidence
limit, but it would have a negligible effect on the number of
matched sources.

\begin{table*}
\centering
\caption {\bf SURVEY COMPLETENESS: Missed and Spurious Sources in the
`Complete Area'}
\begin{tabular}{lllllll}
Measured Flux  & 2.00-2.49&2.50-3.14 & 3.15-4.09&4.10-5.99&6.0-9.99&10.0-17.4\\
($\times 10^{-15}$ erg cm$^{-2}$ s$^{-1}$) &&&&&&\\
&&&&&&\\ 
Missed sources& 1.6 &  1.8     &   2.0    &   0.8   & 0.6    &   0.3 \\
Spurious sources&     1.4  & 0.4 &   0.2  &  0  &   0  &   0 \\
\end{tabular}
\label{tab:completeness}
\end{table*}

\begin{figure}
\begin{center}
  \leavevmode
  \epsfxsize 1.0\hsize
  \epsffile{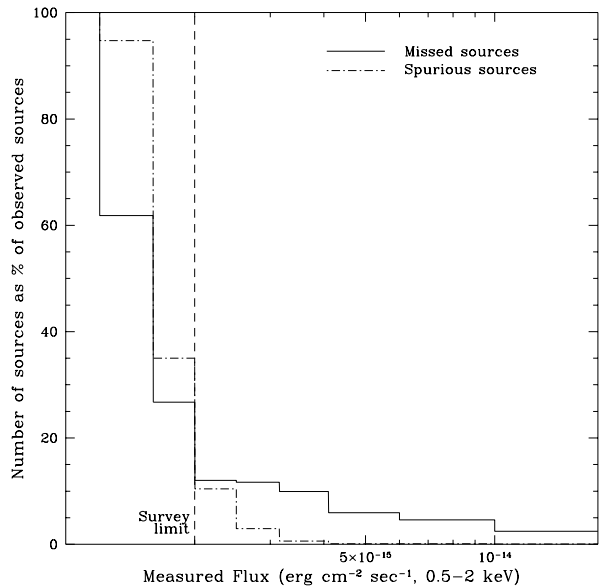}
\end{center}
\caption{The percentage of sources expected to be missed in our survey
as a result of source confusion and statistical noise, as a function
of detected source flux.  The flux bin size is varied so that a
similar number of observed sources lie in each bin. The number of
spurious sources expected is also displayed. These results are
produced from simulations covering an area $\sim250 \times$ the area
of the present deep survey.  See text for further details of the
simulations.  }
\label{fig:completeness}
\end{figure}

The completeness of our survey is shown in
figure~\ref{fig:completeness}.  Here we plot the percentage of missed
and spurious sources as a function of measured (ie observed) source
flux. The bin sizes are varied in order to have approximately equal
numbers of sources in each bin. Below $2 \times 10^{-15}$ erg
cm$^{-2}$ s$^{-1}$ the number of spurious sources rises rapidly. The
number of missed sources also rises rapidly below that flux, which we
therefore choose as our survey limit.

\begin{figure*}
\begin{center}
  \leavevmode
  \epsfxsize 1.0\hsize
  \epsffile{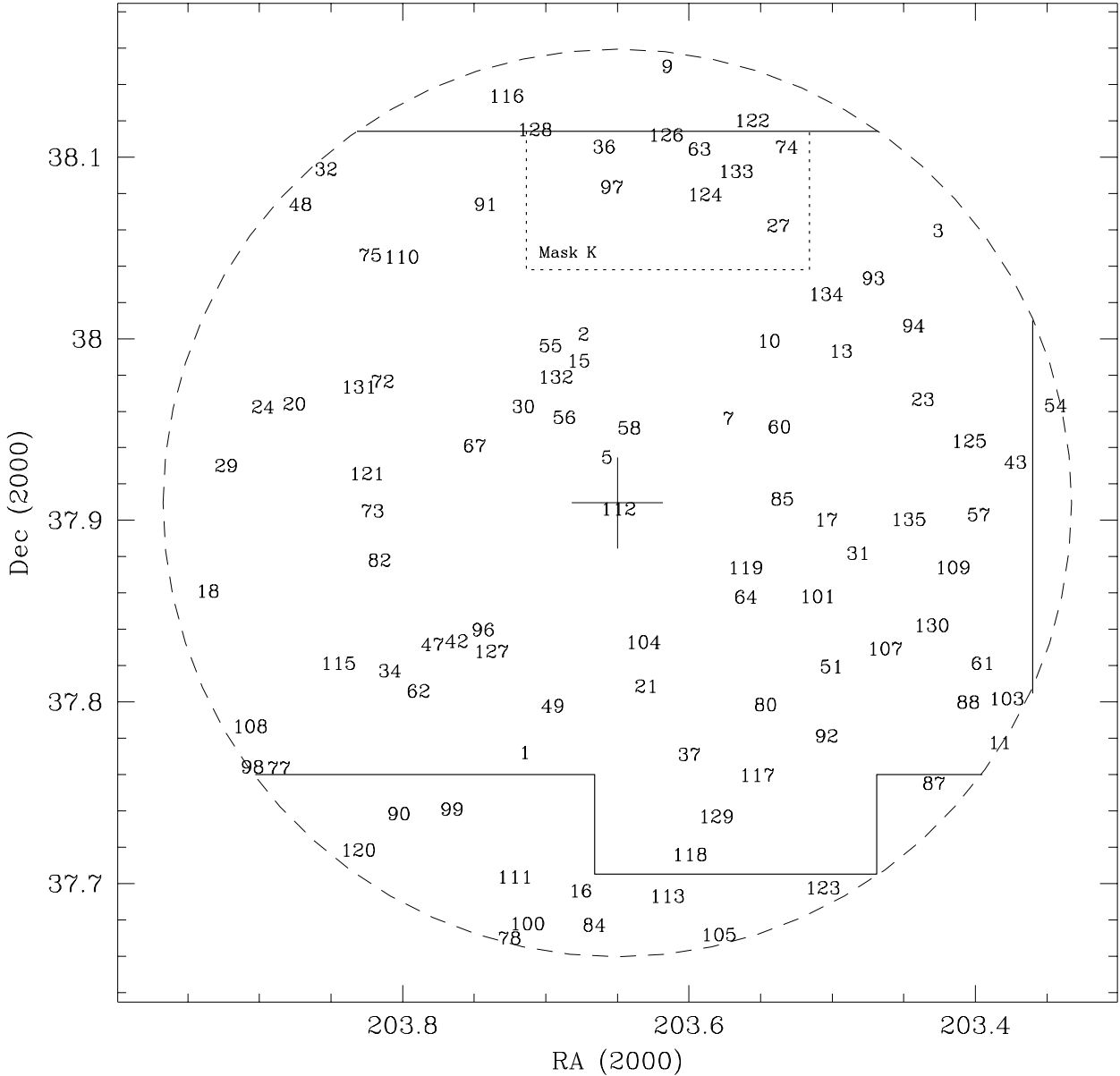}
\end{center}
\caption{Map of the ROSAT source distribution with numbers identifying
individual sources.  Right ascension and declination are both given in
degrees.  Source 47 has been shifted west by 30 arcsec to avoid
overlapping source 42. The solid lines delineate the area covered by
our spectroscopically complete survey which includes 84\% of the total
15 arcmin radius area (dashed line).  To avoid a contrived area, only lines
of constant right ascension or declination have been used in
delineating the area. CFHT MOS Mask K is noted with dotted lines. If
it is left out our complete area comprises 78\% of the 15 arcmin radius
area.}
\label{fig:map}
\end{figure*}

Above $2 \times 10^{-15}$ erg cm$^{-2}$ s$^{-1}$ there are 96 sources
in the full 15 arcmin diameter survey area (and there are 105 sources
above a flux limit of $1.6 \times 10^{-15}$ erg cm$^{-2}$ s$^{-1}$
although, of course, we are very incomplete at that flux limit).  The
distribution of these sources is shown in figure~\ref{fig:map}, in
which figure we also show the shape of the `complete area', which is
described in section 4.  There are 70 sources above a flux limit of $2
\times 10^{-15}$ erg cm$^{-2}$ s$^{-1}$ in the complete area. The
actual number of missed and spurious sources expected in the complete
area is given in Table~\ref{tab:completeness}. Reassuringly, the total
number of spurious sources down to a flux limit of $2 \times 10^{-15}$
erg cm$^{-2}$ s$^{-1}$, by which flux we expected to have detected
$\sim$1 spurious source, is 2.

Above a flux of $2 \times 10^{-15}$ erg cm$^{-2}$ s$^{-1}$ we miss a
total of $\sim7$ real sources. We have examined many of the simulated
source maps by eye and, in the large majority of cases, sources are
missed because they happen to lie close to a brighter source, ie they
are confused. The number of confused sources is approximately what one
expects from `back of the envelope' calculations.  We can slightly
decrease the number of missed sources by adjusting the
parameters of the source searching algorithm, but only at the expense
of detecting more spurious sources.

In addition to the loss of sources due to confusion, and statistical
noise at low flux levels, we may lose a small number of the faintest
sources at large off-axis angles where the psf is larger and the
effective collecting area is less.  The psf and effective area do not
change by more than a few percent until one reaches 12 arcmin, the
radius chosen by GBR to produce the source counts from the AO1
observation.  However by 15 arcmin off-axis the effective area has
dropped to 90\% of its on-axis value for a typical 1 keV source. By
taking into account how the psf increases, and how the collecting area
of ROSAT decreases with radius, and the exact shape of our survey
area, we have estimated how the effective area of our survey varies
with flux and hence derived an approximate area correction factor.
The correction is small, applies only at the very faintest fluxes and
has little effect on derived source count slopes.  Our best estimate
is that the effective area of our spectroscopically complete survey
area (figure~\ref{fig:map}) is about 30\% lower than the geometric
area at $2 \times 10^{-15}$ erg cm$^{-2}$ s$^{-1}$ but that the
effective area is the same as the geometric area by about $2.4 \times
10^{-15}$ erg cm$^{-2}$ s$^{-1}$.  The fact that the area correction factor
(ie not taking account of any confusion losses)
cannot be much greater than 30\% at the faintest fluxes is supported
by the observation that the third faintest source in our list, source
133 which has a flux of $2.1 \times 10^{-15}$ erg cm$^{-2}$ s$^{-1}$,
lies 11.6 arcmin off-axis and approximately 70\% of our
spectroscopically complete area lies within 11.6 arcmin of the axis.

\subsubsection{Accuracy of Fluxes}

\begin{figure}
\begin{center}
  \leavevmode
  \epsfxsize 1.0\hsize
  \epsffile{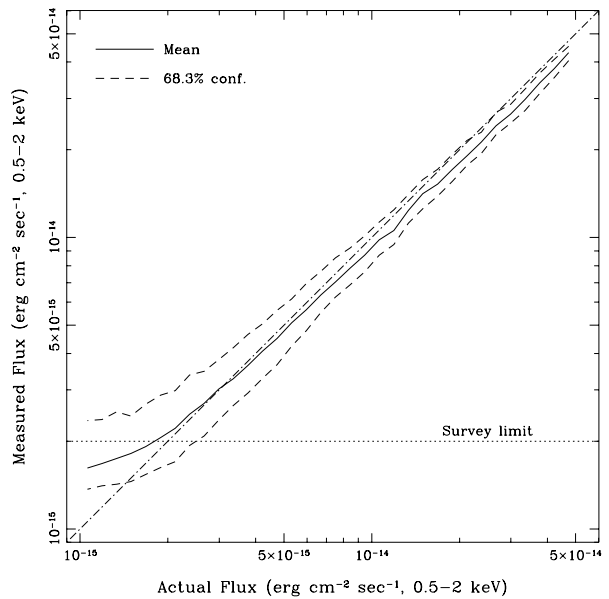}
\end{center}
\caption{
Fluxes of the `real' (ie actual) and corresponding `found' (ie measured) sources.
The mean value and 68\% confidence limits of the distribution are shown.
}
\label{fig:fluxes}
\end{figure}

In figure~\ref{fig:fluxes} we present the relationship between the
fluxes of the `real' and corresponding `found' sources.  The
relationship is quite close to 1:1 but there are some minor
deviations. In particular we notice that at high fluxes the measured
fluxes are systematically slightly lower ($\sim 10\%$) than the input
actual fluxes.  The reason is
probably associated with the marginally different psfs used in the
creation of the simulated images and in the source finding
algorithm. In order to create the simulated images we used an average
`observed' numerical psf. However, in order to greatly speed up the
computational process, we used an analytic psf in the source searching
procedure. Exactly the same analytic psf was used in the analysis of
the real data and in the simulations.  At lower fluxes the difference
between the measured and actual fluxes becomes less and at the survey
limit the mean measured flux just exceeds the actual flux and the
spread in measured flux for any input actual flux increases.  At
fluxes below the survey limit the measured flux substantially exceeds
the actual flux. The rise in measured flux relative to actual flux at
the lowest flux limits is well known (eg Hasinger \etal 1993) and
provides an additional criterion for choice of the survey flux limit.
As these systematic errors are small compared to the statistical
errors in the fluxes, we do not correct any of our observed fluxes.

\subsubsection{Positional Accuracy}

\begin{figure}
\begin{center}
  \leavevmode
  \epsfxsize 1.0\hsize
  \epsffile{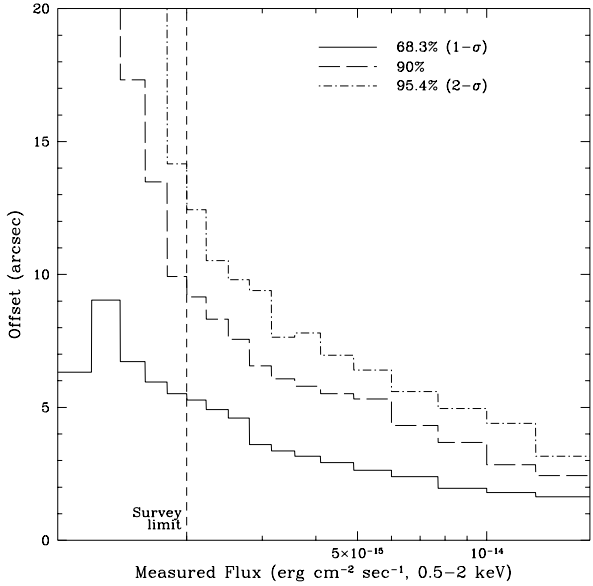}
\end{center}
\caption{Offsets, derived from simulations, between `real' (ie input) 
and `found' (ie detected) source positions as a function of
detected source flux. These offsets take proper account of the
displacement of detected source positions caused by source confusion.
}
\label{fig:positions}
\end{figure}

In figure~\ref{fig:positions} we present the difference in position
between the `real' and `found' sources, as a function of measured
flux.  We include all `found' sources, independent of whether they are
confused by the presence of a very close companion, and so the
positional errors are a true representation of the positional errors
in the actual survey sources. As the distributions are not entirely
gaussian, we present a number of different confidence regions for the
offsets.

Our standard source searching algorithm does not vary the width of the psf,
apart from as a function of off-axis angle, to take account of
extended or confused sources.  Examination of the simulated confused sources
shows that, unless the two constituent sources are of similar
flux, the source searching algorithm quite strongly favours the
position of the brighter source. Thus the main effect of confusion
is for sources to be missed rather than for the positions of
detected sources to be grossly displaced. 

When two sources of similar flux occur less than a beamwidth ($\sim25$
arcsec) apart, an extended source will be produced and, depending on
the separation and flux of the individual components, the extension
may be measurable and it may be possible to model the individual
components.  Although our standard source searching algorithm would
only detect one source, we are able to perform a more sophisticated
search for possible extension or multiple components in individual
cases. We take account of source spectra as the psf of soft sources is
larger than that of hard sources. We have not carried out a systematic
search of all sources for extension but have examined sources for
which there was some reason to consider extension.  Examples include
sources identified with clusters of galaxies or sources for which
there was more than one good candidate identification. Initial source
selection was performed by visual examination of the photon map.
Visual examination is only effective when the photon counting
statistics are good and so we only considered the brighter $\sim$half
of our sources, down to source number 60.  Of the 9 sources selected
visually as being possibly extended (see notes on sources in section 4.1 for
details), source 23 definitely consists of two separate sources of
approximate flux ratio 2:1, source 34 is definitely extended and is
almost certainly a distant cluster and 3 others (sources 5,9 and 51) have
weak-to-reasonable evidence for a possible companion of flux
considerably less than the primary source. The remaining sources
(10,11,17,43) are all perfectly consistent with a single point source,
although sources 10 and 11 are very soft and so the psf is large. In
all cases of extended/multiple X-ray sources, the possible optical
identifications are consistent with the X-ray results.

\subsection{The Source Counts}

\begin{figure}
\begin{center}
  \leavevmode
  \epsfxsize 1.0\hsize
  \epsffile{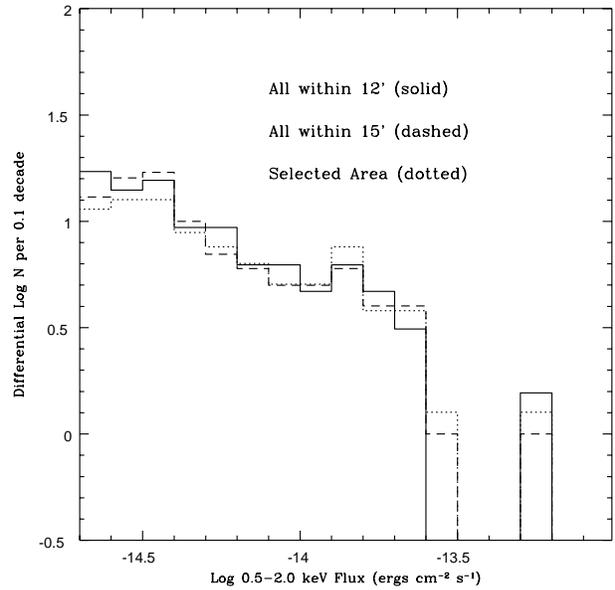}
\end{center}
\caption{Differential source counts for sources within the
full 15 arcmin radius X-ray survey area (dashed line), for sources
within 12 arcmin (solid line) and for sources within the
spectroscopically complete area (dotted line). All source counts have
been normalised to the full geometric area of the 15 arcmin radius
X-ray survey area. No corrections have been applied for confusion or
for slight incompleteness in the 15 arcmin and complete area counts in the
lowest flux bin.}
\label{fig:diffcount}
\end{figure}

In figure~\ref{fig:diffcount} we present the observed differential
source counts. We compare the counts within 12 arcmin radius (solid
line), where the area correction factor has no effect, with those
within the selected area (dotted line) and those out to the full 15
arcmin radius (dashed line). No corrections have been applied for confusion or
reduced effective area at low fluxes. If the area correction factor
were applied, it would only affect the lowest flux bin, raising its
value in the 15 arcmin radius and selected area cases to almost exactly the
same as that of the 12 arcmin counts. To avoid cluttering the diagram
errorbars have not been applied but, for the differential plot, they
are simply `root N' errors based on the numbers of objects in each
differential bin and easily cover the minor differences between the
three source counts. Thus we confirm that the area correction factor
has little effect on the overall source counts.

\begin{table*}
\caption {\bf DIFFERENTIAL SOURCE COUNT SLOPES AT FAINT FLUXES:
$2-16 \times 10^{-15}$ ergs cm$^{-2}$ s$^{-1}$}
\begin{tabular}{lllll}
              & Selected Area    & Full 15' radius & GBR & Hasinger93$^{*}$\\
              &                  &                 &            &           \\
Uncorrected   &  $-1.42^{+0.19}_{-0.26}$& $-1.65^{+0.16}_{-0.28}$& & \\
Corrected for Missing Area             &$-1.46^{+0.19}_{-0.28}$ 
&$-1.73^{+0.17}_{-0.29}$ & & \\
Corrected for Missing Area and 
Confusion&$-1.53^{+0.19}_{-0.28}$&$-1.80^{+0.17}_{-0.25}$
                                        & $-1.78\pm0.38$ & $-1.94\pm0.19$ 
\vspace{2mm}\\
\multicolumn{5}{@{}p{\hsize}@{}}{
$^{*}$ Note that Hasinger \etal (1993) determine a higher break flux
($2.66 \times 10^{-14}$ ergs sm$^{-2}$ s$^{-1}$) than the present work
and so are expected to measure a steeper source count slope.}
\end{tabular}
\label{tab:scounts}
\end{table*}

\begin{figure*}
\begin{center}
  \leavevmode
  \epsfxsize 1.0\hsize
  \epsffile{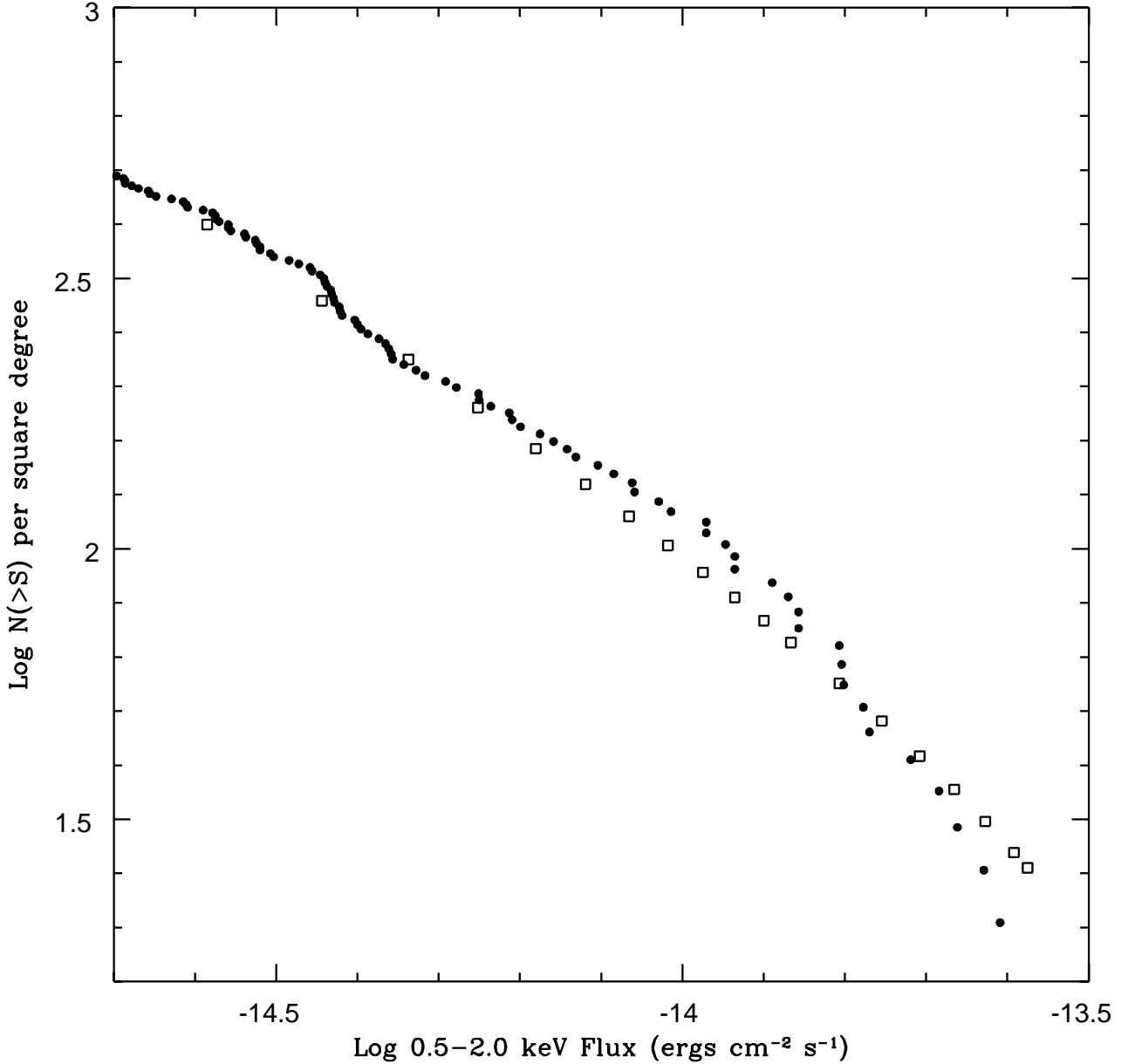}
\end{center}
\caption{Integral source counts from the full 15 arcmin radius area
(filled circles) compared with the best fit to the source counts
derived by Hasinger \etal 1993 (open squares) from a similar ROSAT PSPC survey.
Note that Hasinger \etal correct for incompletness at the high flux end
caused by the small area of their (and our) survey whereas we do not.
Also we plot our raw data but a (smoother) best fit to Hasinger
\etal's data. 
}
\label{fig:intcount}
\end{figure*}

In figure~\ref{fig:intcount} we present the uncorrected integral
source counts from the full 15 arcmin radius area.
For comparison with previously published counts we follow GBR and also
show the best fit to the source counts
of Hasinger \etal (1993).  Given that our sources are presented as raw
data whereas, following GBR, Hasinger {\it et al.}'s counts are presented as a
smooth best fit, we see very good agreement between the datasets. Our
source counts drop below those of Hasinger \etal at the highest fluxes
shown because, unlike them, we have not corrected for the fractional
lack of very bright sources of very low surface density which 
are not found in the small area of our survey.
We can see that both the integral and, more importantly, the
differential source counts, continue down to $2 \times 10^{-15}$ erg
cm$^{-2}$ s$^{-1}$ as a smooth extrapolation of the source counts
derived by Hasinger \etal, and GBR, at slighly higher fluxes $\sim 3 \times
10^{-15}$ erg cm$^{-2}$ s$^{-1}$, in agreement with the source counts
derived from fluctuation analysis by Barcons \etal (1994).  

Both GBR and Hasinger \etal fit the integral source counts with
separate power laws at faint and bright fluxes. GBR measure the break
flux between the two power laws to be $1.6 \times 10^{-14}$ erg
cm$^{-2}$ s$^{-1}$ and the slightly deeper data presented here confirm
that break flux.  Because of our incompleteness at bright fluxes we do
not attempt to measure the source count slope above the break flux.
However below the break flux we determine the source count slope by
maximum likelihood fitting to the differential source counts in the
flux range $2$ to $16 \times 10^{-15}$ erg cm$^{-2}$ s$^{-1}$.  The
results are presented in Table~\ref{tab:scounts}.  We can see that
none of the correction factors have much effect and that our fully
corrected results are in excellent agreement with the earlier
measurement of GBR. They are also in good agreement with the result of
Hasinger \etal given that Hasinger \etal determine a higher break flux
($2.66 \times 10^{-14}$ erg cm$^{-2}$ s$^{-1}$) and, as we move the
break flux to higher fluxes, we move onto the steeper part of the
integral source counts. We note that the differential slope for the
complete area is flatter than that for the full 15 arcmin counts
at about the $1 \sigma$ level. The difference just reflects the
random distribution of sources within the 15 arcmin radius area.
If we rotate the complete area by 90\deg with respect to the sky
we will include the faint sources which we presently miss to the
south and the source counts will be steeper than those of the
full 15 arcmin area.

\section{OPTICAL OBSERVATIONS}

\subsection{CCD Photometry}

The initial optical identification of the X-ray sources was made using
V and R band CCD observations of the survey area from the 88 inch
University of Hawaii Telescope and the 2.4m Michigan-Dartmouth-MIT (MDM)
Telescope in May 1992 and June 1993. The observations were typically of 10
minutes duration and reached R=23 mag. Seeing was typically 1
arcsecond. Subsequently the whole field was observed in the I band
with the 2.4m MDM Telescope and using the Hitchiker Camera on the 4.2m
William Herschel Telescope (WHT) on La Palma; the resultant observations
were of similar depth and resolution to the V and R band observations
and will be discussed elsewhere.
Somewhat deeper observations (R=24) were made on the WHT at
the auxiliary focus of some fields which were blank on the initial CCD
images. Optical coordinates were determined to better than 0.5 arcsec
by cross-referencing the CCD coordinate frame to that of FK4 stars
determined from scans of the Palomar Sky Survey plates by the
Cambridge automated plate measuring machine (APM).
 
In March 1995, we observed the whole ROSAT survey region
again in the R band on the Canada-France-Hawaii telescope using the
University of Hawaii 8K x 8K CCD array (Metzger \etal 1995). This device
enabled us to observe the whole 15 arcmin radius area at the same
time, with 0.22 arcsec pixels. A total of approximately one hour of
integration time was obtained in sub-arcsecond seeing conditions
resulting in a limiting magnitude of R $>$ 24.

\subsubsection{Registration of X-ray and Optical Coordinate Frames}

A correction for the ROSAT roll angle
error of 0.185 degrees (Briel \etal 1995) was applied initially. Subsequently,
in order to align accurately the X-ray and optical images, three
independent methods were used to measure the small ROSAT PSPC
systematic position error.  X-ray positions were compared to (a)
positions determined from APM measurements of Palomar plates of the
first few definite spectroscopic identifications which we obtained in
our initial spectral observations at the NOT and UH 88'' telescopes in
1992 (see section 3.2), (b) APM positions of the few bright optical
stars coincident with X-ray sources, (c) VLA positions of the 11
X-ray/radio coincidences. All three methods gave a consistent offset
of $13\pm1$ arcsec which was removed from all X-ray positions,
leaving only the statistical uncertainty which has been discussed
above (figure~\ref{fig:positions}).
Most ($\sim90\%$) of the X-ray sources were then identified with objects
brighter than R=23 mag.  R magnitudes were determined for all optical objects
within 30 arcseconds of the X-ray positions and the magnitudes of likely
identifications are listed in Table~\ref{tab:main}.

\subsection{Spectroscopy}

Low resolution ($10-15 \rm \AA$) spectra of 10 of the brightest optical
candidates were obtained on the NOT and on the UH 88" telescope in May
1992 and June 1992 respectively. The UH spectra covered 4000-9000
$\rm \AA$ and the NOT spectra covered only 6000-9000 $\rm \AA$ but were
adequate to confirm and classify the identifications and hence enable
us to tie our X-ray coordinate frame to the optical frame.
Subsequently spectra of the majority of the remaining fainter optical
candidates were obtained with the Multiple Object Spectrograph (MOS)
on CFHT and the ISIS spectrograph on the WHT. With MOS we used grism
O300 in first order with Loral 3 as the detector, covering $4000 -
9000 \rm \AA$ with $\sim15 \rm \AA$ resolution.
Approximately 15 spectra were taken at a time. Data were analysed
using IRAF. Observations with ISIS were made in single long-slit mode
to cover objects in gaps between MOS masks. ISIS covered a similar
wavelength range to MOS with slightly better resolution.  In the red
and blue arms of ISIS we used gratings R158R and R300B respectively,
with the 5400 dichroic, and Tektronics CCD detectors in both cases.
With both MOS and ISIS multiple exposures were made to enable cosmic
ray events to be identified and removed. The typically slit width was
1.5 arcsec.  Flux standards were observed and we observed at the
parallactic angle whenever possible (the large majority of cases) to
avoid gross distortions of the spectral shape.

\section{THE OPTICAL IDENTIFICATIONS}

As a result of the above observations we are able to define a sub-area
comprising 85\% of our full 15 arcmin radius survey area in which we are
nearly spectroscopically complete. The outline of this area is shown
in figure~\ref{fig:map}, together with the distribution of sources.
The shape of the sub-area is based on the distribution of CFHT MOS
masks. We have drawn the shape in as unbiased a manner as we can,
using only NS or EW lines to avoid producing a contrived shape. Thus
we include a small number of sources for which we have not yet
obtained optical spectra.  We note particularly a mask which we label
as `mask K'.  Conditions were somewhat worse than normal during the
observation of mask K and so we have only obtained identifications for
4 of the 8 sources covered by mask K. For the present purposes we do
not include the mask K sources in our statistical analysis, thereby
reducing our complete survey area to 79\% of the complete 15 arcmin radius,
ie 0.155 square degrees.  However all mask K sources are noted in
Table~\ref{tab:main} so that readers may reproduce the source counts with those
sources included, if they wish.

The process of optical identification is based primarily on the
positional coincidence between the X-ray source and the optical candidates.
For sources in the brighter half of our list we generally require
an X-ray/optical offset of less than 10 arcsec 
but for sources in the fainter half of the list we relax the
positional criterion slightly to $\sim15$ arcsec.  The second
criterion is optical magnitude - objects brighter than R=21 are
sufficiently rare that they are likely candidates but objects of
R$\geq23$ are not unusual. For the brighter sources there was usually
only one reasonable candidate satisfying the above criterian but for
the fainter sources there was sometimes more than one.  A tertiary
criterion was morphology. We had initially expected that almost all of
the X-ray sources would be identified with QSOs and so even if there
was a galaxy of $R\leq$21 located close to the X-ray centroid, we also
took optical spectra of possible stellar candidates at fainter
magnitudes and at X-ray/optical offsets greater than we considered
reasonable ($>15$ arcsec) for an identification. Almost all of these
stellar candidates turned out to be stars and most were ruled as unlikely
identifications on the basis of their X-ray/optical ratios (eg Stocke
\etal 1991).

We took spectra of the likely candidates objects to R=22 mag and the
results are discussed in detail in Section 4.1, and are tabulated in
Table~\ref{tab:main}. When an identification is considered certain it
is flagged with a `*' in column [o] of Table~\ref{tab:main}. Likely, but less
certain, identifications are flagged as `(*)' and possible
identifications have no flag.

We have defined a number of source classes. Stars are labelled with
their approximate spectral type.  QSOs are defined as having broad
(FWFM $>$1000 km $\rm s^{-1}$) emission lines and $L_{X} > 10^{43}$ ergs
s$^{-1}$ ($H_{0}= 50$ km s$^{-1}$ Mpc$^{-1}$, $q_{0}=0.5$).
Galaxies are divided into absorption line systems (`galaxies') and
those with narrow (FWFM $<$1000 km $\rm s^{-1}$) emission lines
(NELGs). There are also a small number of groups or clusters of
galaxies.
``?'' means there are no useful spectra of objects in the
errorbox and ``blank'' means that there are no objects brighter
than R=23mag in the errorbox.\\

Of the 96 sources of flux $\geq2.0 \times 10^{-15}$ erg cm$^{-2}$
s$^{-1}$, 70 lie within our sub-area (or 78 if we include mask K).
However in order to make the data as useful as possible to other
researchers the photometric data for all sources, and spectroscopic
results for all sources for which data are available, including some
outside the complete area but within the original 15 arcmin radius
survey area, are listed in Table~\ref{tab:main}.  It is hoped that
other observers may be able to add to the identifications and, should
they do so, we would be grateful if they could contact us so that we
may update Table~\ref{tab:main} and eventually publish a completely
identified version.  Where there is anything of interest in the
optical or X-ray field over and above the simple identification
information given in Table~\ref{tab:main}, notes
are given below. The notes concentrate on sources within the complete
sample.

The sources are arranged in count rate order (which very closely
matches the flux order) so that the distribution of
identification class with flux can be seen clearly.  It will be noted
that not all source numbers are present. This is because our original
source list comprised all sources out to 20 arcmin radius but we have
ignored those beyond 15 arcminutes in our subsequent analysis. As the
source number is merely a label we choose to keep the original source
number to avoid any possibility of mixing sources up.

\begin{table}
\centering
\caption {\bf SUMMARY OF X-RAY AND OPTICAL OBSERVATIONS}
 
\begin{tabular}{ll}
[a] & Our source number.\\
{[b]} & Observed 0.5-2.0 keV counts per 10,000s.\\
{[c]} & 0.5-2.0 keV flux corrected for off-axis angle.\\
{[d]} & Hardness ratio, ie the ratio of counts in the\\
      & 0.5-2.0 keV band to those in the 0.1-0.5 keV band.\\
      & See section 5 for details. \\
{[e,f]} & RA and Dec (J2000) of the X-ray centroid.\\
{[g]} & Off-axis angle of the source (arcmin).\\
{[h]} & ``R'' indicates that the source is detected in the \\
    & preliminary 20cm radio map.\\
{[i,j]} & RA and Dec of optical object. In most cases \\
      &	this object is the most likely optical counterpart\\ 
      & but where the optical counterpart is uncertain the\\ 
      & object listed may be simply the brightest optical\\
      & object in the errorbox, or the object nearest the\\ 
      & X-ray centroid. In the least certain cases no \\ 
      & optical coordinates are given.  Where there is the\\
      & possibility of X-ray emission from more than one\\
      & optical object, the probable dominant emitter is \\
      & listed. See notes for more details.\\
{[k]} & ``-''  means not in complete sample.\\ 
    & ``K'' means from CFHT mask K.\\
{[a]} & Repeat of source number.\\
{[l]} & X-ray/optical offset (arcsec).\\
{[m]}  &R-band magnitude.\\
{[n]} & Identification class.\\
{[o]} & Confidence of the identification. ``*'' means certain, \\
    & ``(*)'' means likely, blank means possible.\\
{[p]} & Redshift.\\
{[q]} & ``Y'' means notes are given below.\\
\end{tabular}
\label{tab:main}
 
\vspace{2mm}
\noindent{\bf Note:} Table~\ref{tab:main} is a separate
postscript file (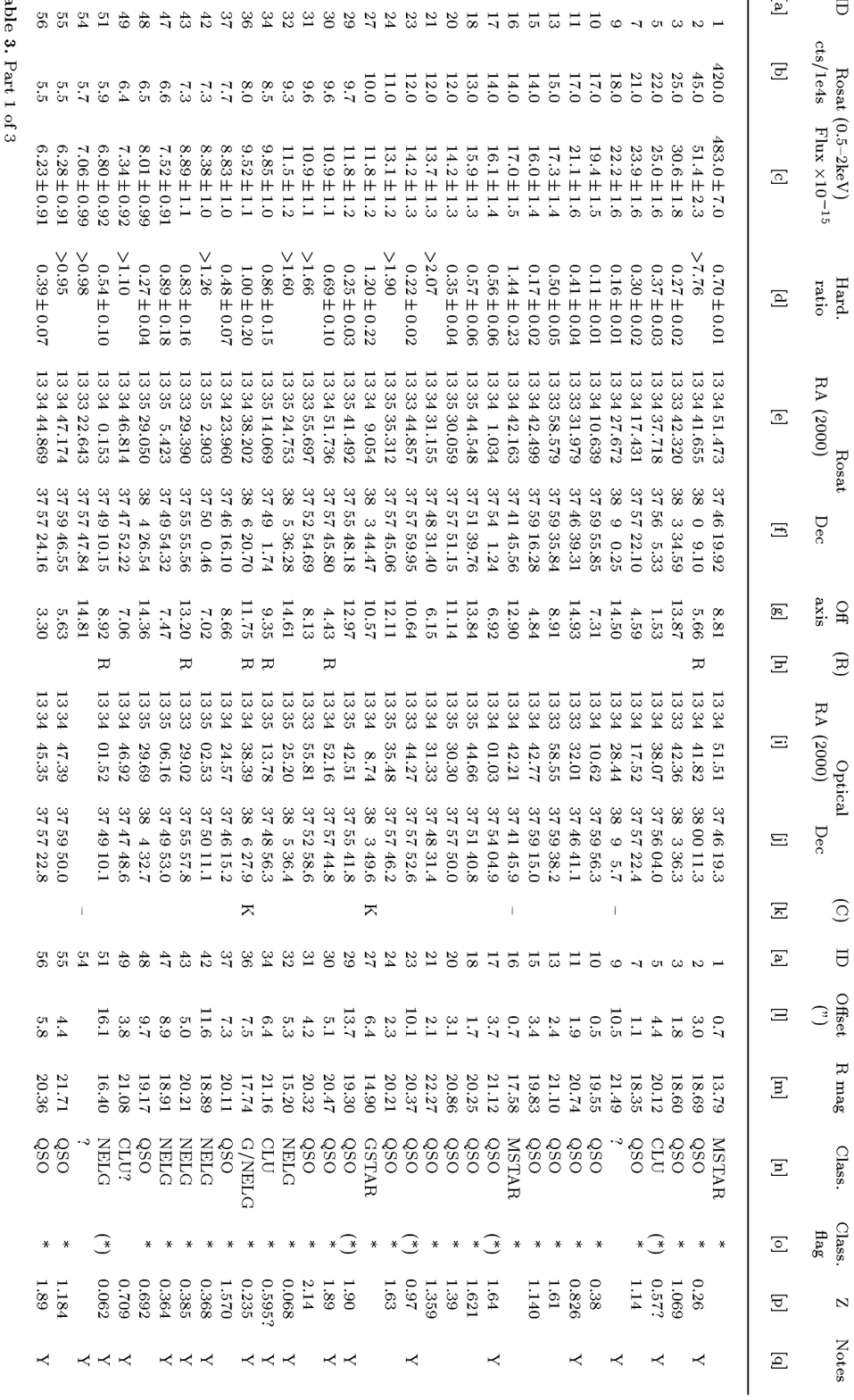).
\end{table}

\subsection{Notes On Individual Sources}

\hspace*{6mm}{\bf 2} Broad line radio galaxy. Extended radio emission.
Has an extremely high hardness ratio. The source certainly is very
hard, perhaps indicating some absorption above galactic, but the
hardness ratio may be overestimated due to the difficulty of detecting
the source in the soft (0.1-0.5 keV) image as it is very close to the
very soft source, 15.

{\bf 5} The X-ray source, appears pointlike, with a very weak
extension to the south-west. Fitting the image weakly supports the
visual impression.  The optical field within an arcminute of the X-ray
centroid contains 6 fairly bright spiral and elliptical galaxies (R
$\sim 18$) and a large number of fainter galaxies (R$<$20) which may
be a distant rich cluster. The brightest objects within the 10 arcsec
radius errorbox are two elliptical galaxies of R=20.1 and 21.5
respectively, the brighter of which looks like the brightest galaxy of
the distant cluster.  The fainter galaxy has a probable redshift of
0.57 with possible very weak [OII] emission.  There are no strong
features in the spectrum of the brighter galaxy and we are not able to
determine its redshift.  NOTE that the redshift given in
Table~\ref{tab:main} is the redshift of the fainter galaxy, not the
galaxy whose position is listed in Table~\ref{tab:main}, on the assumption that
both galaxies are members of the apparent distant cluster.  Two of the
brighter (foreground?)  galaxies have measured redshifts, both being
0.247, and one spiral shows $H_{\alpha}$ and [SII] emission. None lie
within the errorbox.  The most likely identification is therefore with
the distant cluster with the more nearby non-X-ray emitting group
superposed.  The weak extension overlays another group of 21mag
galaxies which may be part of the distant cluster, but no redshifts
are available.  Note however that the overall hardness ratio of the
whole source (H=0.37) is too low for all of the emission to be
explained by hot cluster gas and requires some contribution from a
steep power-law (ie AGN) - see Table~\ref{tab:hard}.

{\bf 9} This source is nearly at the edge of the 15 arcmin field, and
is also very soft, more consistent with AGN-like emission. Even so,
there is some evidence for extension with a weak component to the
west.  There are no candidates within 10'' of X-ray centroid but there
are some 21mag compact objects at 10-20'' with the nearest listed in
Table~\ref{tab:main}. The weak X-ray component overlays a relatively
bright (17 mag) stellar object $\sim 30''$ west of the X-ray centroid.
No useful spectra have been obtained.

{\bf 17} Visual inspection of the X-ray image appears to show a faint
extension to the south, overlaying a 20 mag elliptical galaxy.
However the X-ray flux of the extension is below the survey limit, and
fitting of the image provides no statistical evidence of extension.

{\bf 23} The X-ray image is definitely extended and provides our
strongest statistical case of confusion. The image is best fit by two
point sources separated by 20-30'' along a NE-SW axis. Approximately
2/3 of the counts come from the southern source which overlays the
20.4 mag QSO listed in Table~\ref{tab:main}.  The fainter source
overlays a 17.7 mag elliptical galaxy with weak [OIII] and
$H_{\alpha}$ emission.  The coordinates of the galaxy are RA 13 33 45.35,
Dec 37 58 07.7 and its redshift is 0.175. Although we have not included it
in our complete sample of NELGs, this galaxy does have a valid claim
to be included.  The overall source spectrum is rather soft,
consistent with the majority of the counts coming from the QSO.

{\bf 29} Although slightly further from the X-ray centroid than
expected (13.7''), there are no other likely identifications. The
X-ray position is pulled to the NW by a nearby source (154) which,
although real and identified with a redshift 2.278 QSO, is below the
flux cut-off for this catalogue.

{\bf 30} Note that the redshift of this QSO is identical to that of
the QSO identified with source 56 which is 2 arcmin away.

{\bf 32} The X-ray centroid is coincident with the brighter of 2
interacting spiral galaxies. Both have the same redshift and both show
narrow emission lines.

{\bf 34} The X-ray source is best fit by an extended source.  A
reasonably compact group of faint (21 mag) galaxies lies within the
errorbox, confirming the identification with a distant cluster of
galaxies.  We have spectra of two galaxies. Neither show strong
emission lines.  The spectra are not of sufficient quality to
determine a definite absorption line redshift but we make a tentative
estimate of 0.595 based on possible H+K and G-band.
The relatively hard spectrum (H=0.86) is in agreement with
the identification with a cluster.

{\bf 42} Interacting galaxy. Measured position of X-ray source is
probably displaced from its true position towards neighbouring source
47 (and away from the optical galaxy). Note source 47 is at the same
redshift. There are many galaxies in this field.

{\bf 43} Narrow line radio galaxy. Member of a rich cluster whose centre
is approximately 80 arcsec south of this galaxy. We have spectra of 4
more galaxies all with redshift 0.383. A source which shows up only in the
soft band (0.1-0.5 keV), and so is not listed here, is coincident with the
cluster centre.

{\bf 47} Measured position of X-ray source is probably displaced from
its true position towards neighbouring source 42 (and away from the
optical galaxy). Note source 42 is at the same redshift. There are
many galaxies in this field.

{\bf 49} We give the coordinates of the brightest galaxy within the
10'' radius errorbox. We can only detect absorption features. At such
a high redshift (0.709) the luminosity ($> 10^{43}$ ergs/s) implies emission
from a group or poor cluster rather than from an individual galaxy.
However, perhaps not unexpectedly, a rich cluster is not visible on
our CCD images.  We also note a bright (R$\sim$18) K star at
\del$\sim$20'' to the southwest. For a relatively bright source only 7
arcmin off axis, 20'' is much further from the errorbox centre than we would
expect for a real identification, however the star is quite bright so
cannot be entirely ruled out. 
The source is not detected in the soft band (0.1-0.5 keV), which
is quite consistent with the hard spectrum expected from a
cluster.

{\bf 51} The X-ray source is slightly extended ($\sim30''$) in a NW-SE
direction, consistent either with an elliptical source  or with the
existence of a fainter companion source to the SE of the X-ray
centroid.  Optically the field is crowded.  The most likely
identification is with a bright (16.4mag) narrow line elliptical radio
galaxy, redshift 0.061, which lies $\sim16$ arcsec east of the X-ray
centroid. However a group of six 18-19 mag galaxies lie 10-40'' from
the centroid along the extension of the X-ray source.  Two of these
have the same redshift (0.257 - absorption line) so weak emission from
the group may be responsible for the X-ray extension. There is also
another narrow emission line galaxy
at redshift 0.025 (not a misprint), 40 arcsec NW of the X-ray centroid,
lying just at the edge of the X-ray extension.

{\bf 54} No candidates brighter than 23 mag within 10 arcsec; nearest
object is stellar, R=19, at 20 arcsec. No spectra have been
obtained. Note large off-axis angle.

{\bf 56} Note that the redshift of this QSO is identical to that of
the QSO identified with source 30 which is 2 arcmin away.

{\bf 58} A second redshift is available in this cluster, 0.307.
The X-ray source is complex/extended. It is not detected in the
soft (0.1-0.5 keV) band, which is consistent with the hard
spectrum expected from a cluster.

{\bf 60} An object of R=22.7 which is either a galaxy of high central
surface brightness or possibly a QSO is dead centred (\del= 1.4
arcsec) in the errorbox and is a reasonable candidate identification
as the next nearest object is a galaxy on the edge of the errorbox
(\del=11 arcsec) of R=21.6.  The R=22.7 galaxy shows narrow emission
lines at 5900 and $6565\rm \AA$ which may be badly subtracted night sky
NaD and $H_{\alpha}$, but we note that the 5900 line is over twice as
strong as the 6565 line and that the very strong [OI] 5577 night sky
line subtracts out perfectly.  No broad emission features are
detected. Alternatively, the 5900 line may be real. If it is [OII]
3727, this may be a high redshift (z=0.58) NELG.
The brighter galaxy on the edge of the box
displays only one narrow line at $6600\rm \AA$. This line is seen
independently in spectra from CFHT and from the WHT. If it is [OII]
3727, the redshift is 0.76. The source is harder (H=$0.86\pm0.23$)
than expected from a QSO-like power-law spectrum but, on
the other hand, does have a high X-ray/optical ratio
(see figure~\ref{fig:lxlopt}), more consistent with a QSO than a
starburst galaxy.

{\bf 62} Large R=18.6 elliptical galaxy 9 arcsec from X-ray centroid.
No emission lines. Redshift 0.251. There are a lot of nearby fainter (22mag)
galaxies and a few other brighter (R$\sim19$) galaxies. The R=18.6
galaxy may be the dominant galaxy in a group.

{\bf 73} A 23mag galaxy is well centred (\del=4.7). The nearest object
noticeably brighter is a 20mag stellar object (\del =20'').

{\bf 74} Identical absorption line redshifts of 0.382 exist for both a galaxy
well within the errorbox and one just outside it.

{\bf 77} Group of galaxies. A second galaxy with a similar absorption
line spectrum has a close redshift (0.304).

{\bf 78} The nearest object to the X-ray source is listed in
Table~\ref{tab:main} (R=22.07, \del=15 arcsec). A galaxy of R$\sim 21$
lies further away (\del $\sim 25''$). However note large off-axis
angle. The field is just off the edge of our deepest CFHT images.  No
spectra were obtained.

{\bf 80} The errorbox is empty apart from a 23mag compact object at
\del = 2.7'' and the surrounding field is also very empty
so this object is the most likely identification.
The low signal/noise spectra are reasonably
clear of poor night sky subtraction features but there is one emission line 
at $\sim4940\rm \AA$ which is visible on spectra taken in 1994 and 1995.
If this is [OII]3727 the redshift would be 0.327.

{\bf 82} A reasonably compact object of R$\sim23.5$ is visible on the
deepest CFHT and WHT auxiliary port images very near the centre of the
errorbox.

{\bf 84} The nearest object to the X-ray centroid is a 20.1 mag compact
object at \del=13.7''. No spectra are available and the source is just
off the edge of our deepest CFHT image. Note large off-axis angle.

{\bf 85} The 20.7 mag galaxy listed is the brightest object in the
errorbox.  It is at \del=5.1''. The only other object brighter than
R=23 in the errorbox is an R=22.2 mag object, to the north of the 20.7
mag galaxy, at \del=7.8''. We cannot tell if the fainter object is
a galaxy or a star and no spectra are available.

{\bf 87} The coordinates and magnitudes given are of the nearest bright
(R$<$23) galaxy. Note large off axis angle.

{\bf 88} The identification is unclear.  There are two 21mag objects
within 3'' of the X-ray centroid. One is stellar, the other is
slightly fuzzy. The spectra are of poor S/N with no really strong
features.  A 21mag Mstar lies at \del=9'', but is too faint optically
to be a likely identification. In Table~\ref{tab:main} we give the
coordinates of a 17mag K star at \del=14'' which is just about bright
enough optically to be a reasonable identification but is rather
further from the errorbox centre than we would expect for a real
identification.

{\bf 90} There is only one possible candidate, a 21mag slightly fuzzy
stellar object at \del=2.2''. There is no spectrum but it is probably
a QSO.

{\bf 93} The galaxy listed is the only object brighter than R=23 in
the errorbox.

{\bf 94} $H_{\alpha}$ is slightly broadened (FWHM=1000 km/s). This is
a narrow line radio galaxy.

{\bf 96} The X-ray source is coincident with the centre of a
reasonably rich cluster.  The redshift given is that of the brightest
galaxy, a large elliptical, 11 arcsec from the X-ray centroid.
The elliptical shows no optical emission lines but does host an extended
radio source. Some contribution to the X-ray emission from an active
nucleus cannot therefore be ruled out.

{\bf 97} The brightest object within 20 arcsec of the X-ray centroid
is a 22.7mag compact galaxy, with faint extension to NW, 6 arcsec from
the X-ray centroid. This may be a low redshift QSO, however no optical
spectrum is available.

{\bf 98} 
An elliptical galaxy with a bright nucleus, which appears to be the
brightest in a cluster is near the centre of the errorbox
(\del=3.7''). Based on possible H and K the redshift is 0.255. There is a hint
of [OIII] emission but it is right on top of the 6300 night sky line
and so may not be real. No other features are visible.

{\bf 99} The brightest galaxy in the errorbox is $\sim23$ mag with a
suspicion of a bright nucleus. A possible extremely distant, but quite
rich, cluster with all galaxies of R=23 or fainter is centred about 20
arcsec south of the X-ray centroid.  

{\bf 100} There is nothing brighter than 23 mag within 10'' of the
X-ray centroid but there are two somewhat flattened 20mag galaxies on
opposite sides of the box, each of \del=20''. Note the large off axis
angle. There are no spectra.

{\bf 105} A compact group of 3 bright galaxies lies within the
errorbox, two very close together; a fourth lies just outside. The
coordinates of the brightest are given. No spectra were obtained.

{\bf 107} There is nothing brighter than R=23.5 within 10''. There are
a number of 21 mag galaxies slightly further away and the coordinates
of the nearest is given. We have low S/N spectra of three of them but
cannot distinguish any features. There is also a much brighter galaxy
($R \sim 16$) about 30'' north of the errorbox, but, although very
bright, it is too far away to be a reasonable candidate for a source
only 10 arcmin off axis.

{\bf 109} This is a complicated region. There are two other sources
(which we know as 170 and 169) which are detected only in the 0.1-0.5
keV band, lying 30 and 70 arcsec northeast respectively of source 109.
Likely identifications of sources 169 and 170 are an M star and a
z=0.256 NELG. Also to the NE, at \del=$\sim$20'' is a z=2.12 QSO,
which may be the identification of source 109.  The nearest object
brighter than R=23 to source 109 is a 21.9mag absorption line galaxy
of probable redshift 0.226 at \del=12'', to the NE but this is
unlikely to be the identification.

{\bf 112} There is nothing brighter than R=23 within 10''. The
coordinates given are of the nearest object, an R=21.7 galaxy but this
is too far (16'') from the X-ray centroid for a source almost on-axis
to be a reasonable identification.
The spectrum is of poor S/N.

{\bf 113} There is a group of 3 faint (R$\sim23$)
galaxies within 10''. The coordinates
of the brightest is given. There are 2 brighter (R$\sim 21.3)$ galaxies both
at \del=15'', one to the north, one to the southeast. There are no spectra.

{\bf 116} The coordinates given are of the brightest galaxy within 10''.
It appears to be a double/interacting galaxy. There is nothing brighter than
21mag within 40''. No spectra were obtained.

{\bf 119} Within 10'' there is a 17 mag K star at \del=8'' and a 20
mag absorption line galaxy at \del=4'' with a possible redshift of
0.368. Both are possible candidates.  There is no indication of a
group or cluster of galaxies.

{\bf 120} There are quite a few very faint (R$\sim23$) galaxies
nearby. The coordinates of the nearest bright (R$>$21) galaxy are
given. This is a possible cluster candidate. No spectra were obtained.

{\bf 122} There is one 21mag galaxy at \del=18'' south and another at
\del=$\sim22''$ to the northeast. Given the large off axis angle and
faint flux, these are feasible candidates.  There are no spectra.

{\bf 123} There are a couple of very faint (R$\sim23$) galaxies within
10'' but more likely candidates are two 20.5mag stellar objects both
at \del=20'', one to the north and the other to the south (listed in
Table~\ref{tab:main}). There are no spectra.

{\bf 124} There is a 21.7mag galaxy at \del=5'' (poor spectrum) which
is a possible identification, as is a 16.4mag stellar object at
\del=18'' (no spectrum) to the south and a 19.9mag M star at \del=19''
to the southwest. 

{\bf 126} A number of faint (R$>$22) galaxies around the errorbox.
The only object brighter than 22mag is what looks like an R=$\sim21$
disc galaxy with smaller companion  at \del$\sim20$'', to the south.  There
are no spectra.

{\bf 127} A 20mag galaxy with quite bright nucleus showing strong
H$_\alpha$ and [SII] but H$_\beta$ is not detectable and [OIII]5007 is
very weak, lies at \del=9'' to the southeast (coordinates given). It
is a weak VLA radio source. 
This is a likely identification but
some contribution from a 19.7mag M star, 13'' to the northwest cannot
be ruled out. Note the very similar redshift of this galaxy to the
nearby cluster identified with source 62. Two further galaxies with
identical redshifts, one of which is also a weak radio source but
neither of which is an X-ray source, also lie in the same general
area.

{\bf 128} A likely identification is with a 19mag absorption line disc
galaxy with a bright nucleus at \del=6''. The only
other possible candidate is a 20.7mag object which is either a very
compact galaxy or a QSO/star at \del=15'' (poor spectrum).

{\bf 129} Although the galaxy is not particularly bright, the spectrum
is of good quality and many emission lines are visible including
[OII], [OIII], $H_\alpha$, [SII] but not $H_\beta$.

{\bf 130} A very empty field. The nearest object brighter than R=23.5
to the X-ray centroid is probably a galaxy of R$\sim22$ at \del=20'' to
the southwest. 

{\bf 131} Strong narrow [OII] and [OIII] emission lines are detected
from the listed galaxy, which is diffuse, with a nearby fainter
companion. It is a reasonable candidate but a more compact object,
probably also a galaxy, of similar total magnitude and at a similar
distance east of the X-ray centroid cannot be ruled out. (We do not
have a good spectrum of the latter object.)

{\bf 132} X-ray source is centred (\del=5'') on the brighter of two
interacting galaxies.  Both galaxies have the same redshift.  The
brighter galaxy shows moderately strong $H_\alpha$ and [SII] emission
but weak or absent [OIII]. The spectrum of the fainter galaxy
(R$\sim21$) is similar, but of poorer quality; $H_\alpha$ is right at
the end of the spectrum and is weaker, if present at all.

{\bf 133} The most likely identification is with a 20.6mag stellar
object 9'' south of the X-ray centroid.  There is nothing else
brighter than 22mag within 20''. In terms of optical/X-ray fluxes, 
a QSO is more likely but an M star is possible. No spectra were obtained. 

{\bf 134} As almost the faintest source in the list, the offset of the
NELG from the X-ray centroid (22 arcsec) is just about consistent with
the NELG being the identification. The offset is larger than expected
for a typical source at this flux but there is nothing brighter than
23.5 mag any closer to the X-ray centroid.

{\bf 135} The listed 20.6mag galaxy at \del=13'' has a strong narrow
emission line spectrum ([OII], $H_\beta$, [OIII]4959 - 5007 is in
night sky A band, $H_\alpha$) and is the most likely
identification. Another fainter galaxy lies very close to it.
However a 20.3mag Mstar, also at \del=13'' cannot be ruled out.
The surrounding field is somewhat richer than average in galaxies.

\subsection{X-ray/Optical offsets for QSOs and NELGs}

\begin{figure}
\begin{center}
  \leavevmode
  \epsfxsize 1.0\hsize
  \epsffile{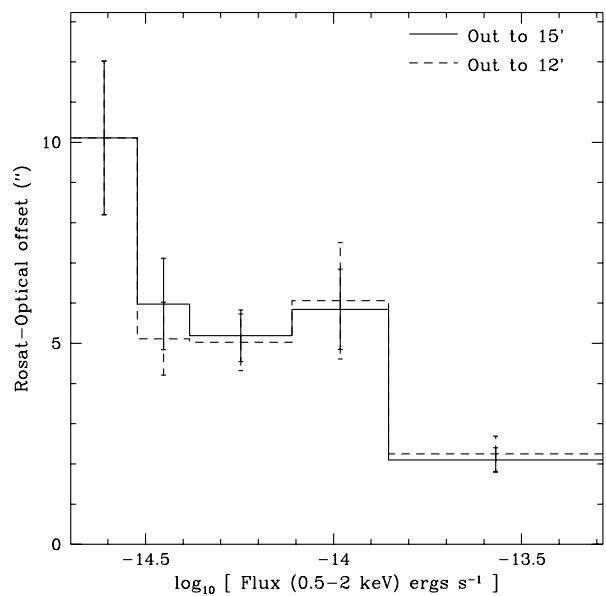}
\end{center}
\caption{Average offset between the X-ray centroid and the
optical identification, as a function of X-ray flux. The solid line
represents the complete sample out to 15 arcmin radius and the dashed line
is the restricted sample out to 12 arcmin. In the lowest flux bin the dashed
and solid lines lie at exactly the same offset as all the sources in that
bin lie within the 12 arcmin radius.}
\label{fig:offsets}
\end{figure}

The accuracy of our X-ray positions is defined fundamentally by the
simulations shown in figure~\ref{fig:positions}. However a consistency
check is provided by the average X-ray/optical offsets of
identifications as, besides position, secondary criteria such as
spectral type and morphology are taken into account before optical
candidates are classed as reasonable identifications.  We present
these offsets, as a function of flux, in figure~\ref{fig:offsets} for
the firmest (ie `*') identifications.  For this purpose we ignore
clusters and groups as their X-ray centroid is not always well
defined, and we ignore the `(*)' identifications as, in some cases the
reason that the identification class is `(*)' rather than `*' is
because of a suggestion of a weak companion source which might, of
course, affect the position of the X-ray centroid. However including
the `(*)' sources has little effect on the distribution.  Reassuringly,
figure~\ref{fig:offsets} agrees quite closely with our simulations
(figure~\ref{fig:positions}).

We have separately computed the offsets for sources out to 12
arcmin and those out to 15 arcmin. We see little difference in the
distributions, implying that flux, rather than offset from the field
centre, is the main factor affecting positional uncertainty. Note,
however, that in the very faintest flux bin we have no sources and
hence no identifications beyond 12 arcmin. 

Interestingly we note that, if we consider only the middle of our flux
range which is equally populated by QSOs and NELGs, the mean distance
of the optical identification from the errorbox centre is $4.8 \pm
0.7$ arcsec for the QSOs and $7.4 \pm 0.8$ arcsec for the NELGs. 
Although both offsets are well within the offsets expected for real
identifications, the difference is significant at greater than 95\%
confidence using the Student t-test. One possible explanation of 
the difference is that the small number of expected incorrect
identifications of NELGs has increased the average NELG offset
whereas there will be almost no incorrect identifications with
QSOs (for QSO number counts see, eg, Hall \etal 1996).
Alternatively we note that previous ROSAT PSPC
observations of bright nearby Seyfert 2 galaxies (Turner \etal 1993)
have shown a factor of 8 enhancement of faint X-ray sources (typically
5-50\% of the Seyfert 2 flux) within 100 kpc. The nature of these
companion sources is not yet known but, if similar sources surround
our NELGs, their fluxes would be below our detection limit but their
effect would be to shift the centroid of the NELG X-ray emission by a
few arcsec.  We note that many of the NELGs in the present survey are
in interacting systems (see notes on individual
sources above). In a small number of cases we see emission line
spectra from the companion galaxy as well as from the larger (NELG)
galaxy. Thus it is quite likely that X-rays are also emitted from the
companion galaxy, thereby accounting for the larger X-ray/optical
offset of the NELGs as the optical coordinates given are those
of the nucleus of the larger galaxy.

\subsection{ Identification Content as a function of flux}

\begin{figure*}
\begin{center}
  \leavevmode
  \epsfxsize 1.0\hsize
  \epsffile{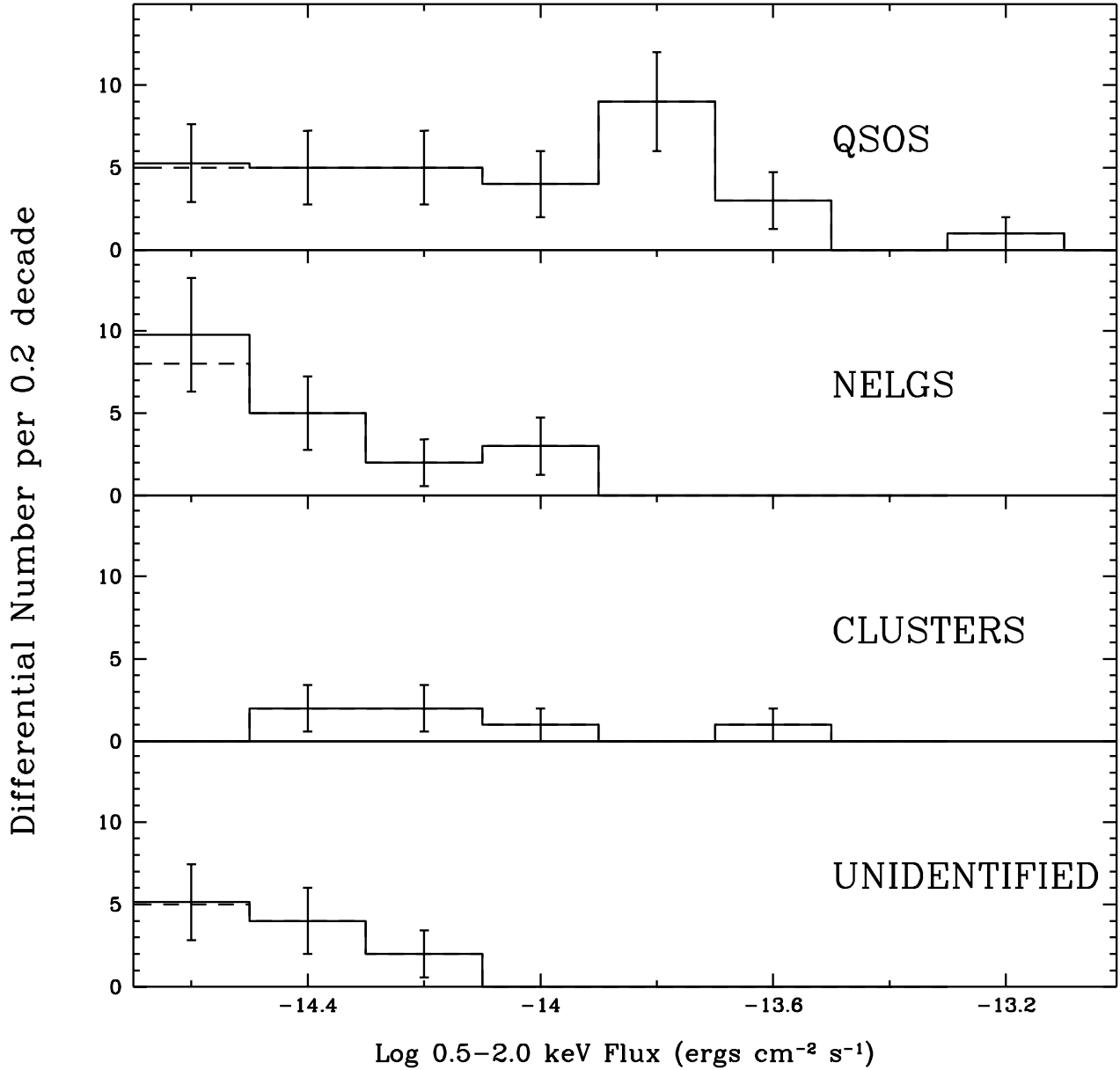}
\end{center}
\caption  {Differential number of QSOs, NELGs, clusters and
unidentified sources in the complete area as a function of source
flux in 0.2 decade flux bins.
The dashed line is the actual number of sources counted; the
solid line is the number corrected for the reduction in effective
area at low fluxes. No correction is applied here for sources lost
due to confusion.
}
\label{fig:diff_ids}
\end{figure*}

In figure~\ref{fig:diff_ids} we show the number of sources in our
complete area for the major identification classes, as a function of
flux.  The dashed line gives the number actually detected and the
solid line gives the number corrected for effective area as described
above.  No correction is applied here for sources lost due to
confusion.  However such a correction is applied to the calculation of
the source count slopes given in Table~\ref{tab:slopes}.  The only
objects not plotted in figure~\ref{fig:diff_ids} are the 3 definite
identifications with stars.  Only those sources for which the
identification is reasonably firm, ie confidence classes `*' or `(*)', are
plotted in specific astrophysical classes (ie QSOS, NELGS, CLUSTERS);
the 5 objects of possible but uncertain identification listed in
Table~\ref{tab:main} are plotted as `UNIDENTIFIED' together with the
blank fields and the two sources whose identification class is given
as `?' in Table~\ref{tab:main}.

\begin{figure*}
\begin{center}
  \leavevmode
  \epsfxsize 1.0\hsize
  \epsffile{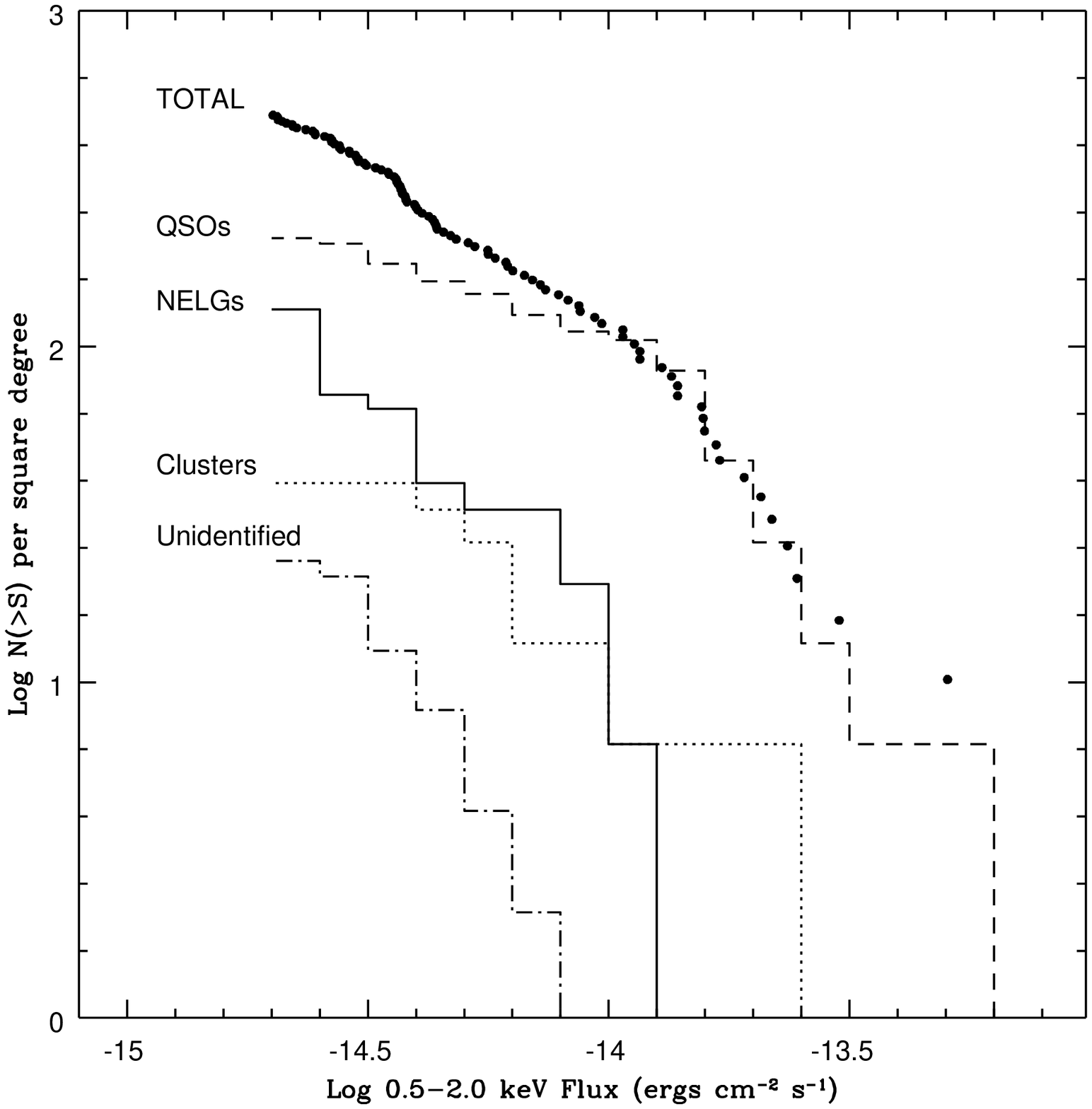}
\end{center}
\caption {Integral surface density of QSOs (dashed line), NELGs (solid
line), clusters (dotted line) and unidentified sources (dot-dash line)
from the complete sample, together with the total surface density of
sources from the full 15 arcmin radius area (filled dots), as a
function of flux. The only identified sources not plotted are the
the 3 stars. The numbers are corrected for incompletness of
the area at faint fluxes and the unidentified sources are shifted down
by 0.5 decade in surface density to avoid overlapping the NELG plot.
No correction is applied here for sources lost
due to confusion.
}
\label{fig:int_ids}
\end{figure*}

\subsubsection{QSOs}

There are 32 QSOs in our sample and they dominate the bright end of
the source counts, but not the faint end.  Above $10^{-14}$ erg
cm$^{-2}$ s$^{-1}$ 16 of the 19 sources (ie 84\%) in the complete
sample are QSOs (the others being 1 NELG, 1 cluster and 1 star). At
bright fluxes our results are therefore similar to those of earlier
less deep ROSAT observations, eg the 30ksec observation of Shanks
\etal (1991). In the latter observation 24 of the 39 X-ray sources
were identified with QSOs, 5 with stars but with no other class of
object being a signficant contributor.  However in our survey below
$10^{-14.5}$ erg cm$^{-2}$ s$^{-1}$, only 5 (26\%) of the 19 sources
are QSOs.

The distribution of our X-ray selected QSOs rises with increasing
magnitude to R=21 (3 of R$<19$, 5 of R=19-20, 12 of R=20-21). Comparison
of surface densities indicates that, at these brighter magnitudes
(R$<21$), we have probably detected almost all of the QSOs that would
have been picked out by optical surveys. 
In our survey we find a QSO surface density of
129$\pm$28 deg$^{-2}$ at R$<21$, which corresponds to B$\lesssim$21.5,
assuming that the the mean optical QSO colour of B-R$\approx$0.5 (e.g.
Boyle, Jones \& Shanks 1991) at the redshift of the majority of the QSOs
(z$<$2.9) applies here. The surface densities found in B band optical
surveys at the fainter limit of B$<$22 are slightly less than this value
(77$\pm$10 deg$^{-2}$ and 115$\pm$16 deg$^{-2}$ were found by Boyle,
Jones \& Shanks (1991) and Zitelli et al. (1992) respectively),
suggesting that at R$<21$ we are complete. At fainter magnitudes the
distribution falls off (10 of R=21-22, 2 of R$>$22) and so we might ask
whether we have missed any faint QSOs.

\begin{figure}
\begin{center}
  \leavevmode
  \epsfxsize 1.0\hsize
  \epsffile{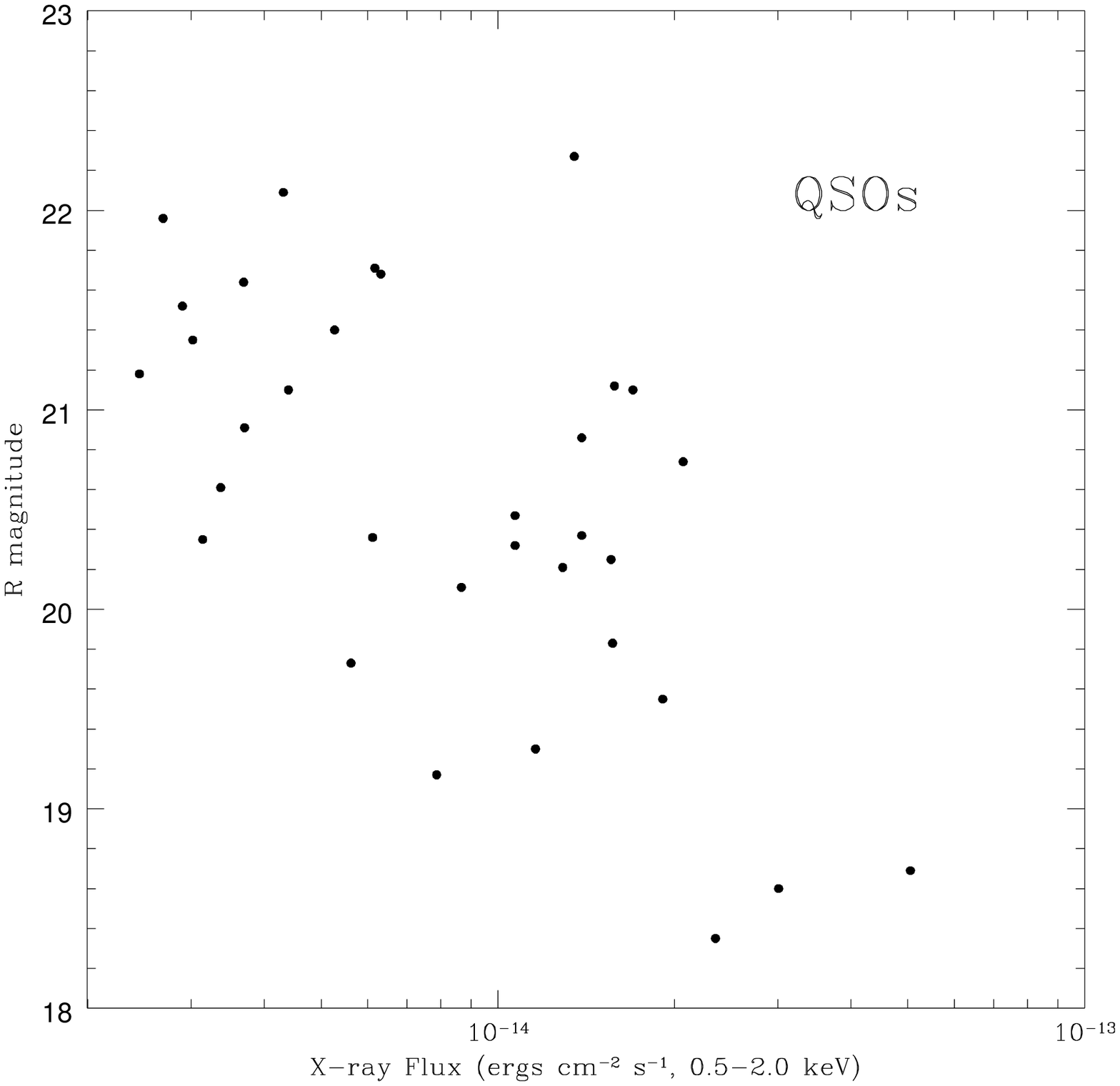}
\end{center}
\caption{ Distribution of X-ray fluxes and R-band magnitudes for the QSOs
identified in the present survey. Errorbars are not shown to avoid cluttering
the diagram but the X-ray flux errors are given in Table~\ref{tab:main}
and typical optical errors are 0.1 magnitude, or better.
}
\label{fig:qsorx}
\end{figure}

The obvious answer as to why there are so few QSOs of R$>22$ is that
the X-ray fluxes of any such QSOs would lie below the survey limit.
We can crudely address this possibility by reference to the
distribution of X-ray fluxes and R-band magnitudes which is given in
figure~\ref{fig:qsorx}. For any given X-ray flux away from the survey
limit the range of observed R-band magnitudes is typically 2
magnitudes. However at the survey flux limit the range is somewhat
truncated to about 1 magnitude. The optically brightest QSO near the
X-ray survey limit has R$\sim21$ but the faintest QSO only has
R$\sim22$. Thus if the X-ray/optical ratio, and dispersion therein, is
the same at all fluxes, it is likely that we have missed a few fainter
QSOs. 

At present the faintest optical survey sensitive to QSOs at all
redshifts (as is our X-ray survey) is that of Schade \etal (1996) who
found 6 QSOs to I$_{AB}<$22.5 (or B$<$23) serendipitously in the CFRS
survey. The corresponding QSO surface density is 200$^{+120}_{-80}$
deg$^{-2}$, similar to our value of 210$\pm$40 deg$^{-2}$ at R$<$22,
suggesting that we are reasonably complete to at least R=22, in
agreement with the distribution shown in figure~\ref{fig:qsorx}.
Although our spectroscopic limit for galaxies is $R\sim22$, we
generally made efforts to obtain spectra of even fainter stellar
objects and so we suspect that any missing QSOs have R$\sim23$, or
fainter. Until we have complete identification of our survey we 
cannot say how many of such QSOs there may be although
we note that 3 of the sources classed as
`unidentified' have objects of R=23 near the centres of the
errorboxes, and there are also 4 blank fields.  
In addition up to 5 of the NELGs may be misidentifications
(see next subsection). 
It is unphysical to suppose that all the unidentified sources
will turn out to be QSOs and that there will be no further
identifications with optically fainter NELGs or clusters of galaxies
(see NELG and QSO source counts in figure~\ref{fig:diff_ids})
however it is quite reasonable that up to half of the uncertain
identifications will later turn out to be associated with
faint QSOs.

\subsubsection{NELGs}

NELGs make almost no contribution to our survey above $10^{-14}$ erg
cm$^{-2}$ s$^{-1}$, but they are the major contributor 
below $10^{-14.5}$ erg cm$^{-2}$ s$^{-1}$ (where
there are 8 NELGs (42\%), 5 QSOs, 1 star and 5 unidentified sources).
From their rising source counts (see below for details) compared to
the flat QSO source counts, we can see that NELGs represent a potentially
large contribution to the soft XRB.

Galaxies with narrow emission lines have, of course, been detected as
X-ray sources since the earliest days of X-ray astronomy.  Most of the
early detections of narrow emission line galaxies were of relatively
high excitation Seyfert 2 galaxies whereas our NELGs have a mixture of
excitations, generally somewhat lower than that of Seyfert 2's.  NELGs
with similar optical spectra to ours have recently been detected in
less deep X-ray surveys, eg EMSS (Stocke \etal 1991), Boyle \etal
(1995), Carballo \etal (1995).
At brighter fluxes ($>2 \times 10^{-14}$ erg
cm$^{-2}$ s$^{-1}$) where the QSO source counts are
still rising, Boyle \etal also find a rising NELG source count.
However here we show that, at faint fluxes, whilst the QSO
source counts fall, the NELG source counts are still rising.

To determine how many of the NELG identifications are chance
associations we assume an average errorbox radius of 10 arcseconds
which, as we have shown above, is a somewhat pessimistic figure.  We
expect 0.25 unrelated galaxies of \underline{any} spectral type
brighter than R=22.0 in each error box (Metcalfe \etal 1991) or 0.10
unrelated galaxies brighter than R=21.0.  Ignoring the fields in which
we have an unambiguous identification with a QSO, cluster or star, we
therefore expect 1 galaxy of R$\leq20.0$, 3 galaxies of R$\leq21.0$ or
7.5 with R$\leq22.0$ in the remaining 30 errorboxes.  (As not all
field galaxies show emission lines like our NELGs (eg Tresse \etal
1996) we could reduce these chance coincidences by about 20\% but we
do not consider that small deviation here.)  There are 10
identifications with NELGs having R$\leq20.0$, 6 with NELGs having
$20 \leq R <21$ and 2 with NELGs having $R\geq21$.  We therefore conclude
that 13 of the 16 NELGs of $R\leq21$ are real associations.  However
although the 2 NELGs with $R\geq21$ are, as far as we can tell, the
most likely identifications of their respective sources, the number of
NELGs with $R\geq21$ does not exceed the number of expected chance coincidences.
Thus we conclude that 5 of the 18 NELGs are likely to be chance
coincidences.  We refer readers to the notes on sources for discussion
of individual identifications.  Note that, as the X-ray/optical ratio
of the NELGs varies considerably (see figure~\ref{fig:lxlopt}), we
cannot easily correct the NELG source counts
(figures~\ref{fig:diff_ids} and figures~\ref{fig:int_ids}) to take
account of possible incompleteness as a function of optical magnitude
and so these figures remain as observed distributions.  Similarly we
do not correct the derived source count slopes
(Table~\ref{tab:slopes}) or the contribution of NELGs to the X-ray
background (figure~\ref{fig:contrib}) although we do make some
broad comments in the appropriate sections.

\subsubsection{CLUSTERs}

There are 6 clusters or groups in our survey within the complete area.
These clusters were classified initially on the basis of a visual
estimate of an overdensity of close companions relative to the field
and so may not all be real physical associations.  Subsequently we
have obtained more than one similar redshift of galaxies in the
putative clusters in sources 58 and 77 (and 74 in mask K). Also two of
the clusters (sources 34 and 58) are associated with extended X-ray
emission. Thus at least 3 of the 6 possible clusters are 
real. However, although we always
took spectra of the brightest galaxies in the errorboxes, the
possibility of AGN-type activity from a fainter nearby galaxy, rather
than true diffuse cluster emission, does exist.

Note that there are no isolated absorption line `galaxies' left as
firm identifications in the complete area; the only `galaxy' left is
of uncertain identification and so classed here as `UNIDENTIFIED'. All
other galaxies showing only absorption line spectra lie in what
we classify as `CLUSTERS' (which here includes groups). 
With one exception the clusters all lie in the mid-flux range of the
survey, below $10^{-14}$ erg cm$^{-2}$ s$^{-1}$ and above $10^{-14.5}$
erg cm$^{-2}$ s$^{-1}$. Cluster luminosities lie in the range
$10^{42}$ to few $\times 10^{43}$, characteristic of groups or poor
clusters locally rather than either rich clusters or individual
(non-AGN) elliptical galaxies. We have not yet performed a proper
analysis of the richness of the clusters but eye-ball estimates do
generally indicate poorer rather than richer clusters.  

The 6 clusters correspond to a surface density of 40 deg$^{-2}$ above
a flux limit of $2 \times 10^{-15}$ \ecs.  The lower limit on the
number of real clusters is 3, as stated above, corresponding to 20
deg$^{-2}$.  The upper limit is harder to determine until we have
achieved satisfactory identifications for the unidentified sources but
our best guess, based on visual examination of the optical fields, is
about 11, corresponding to 73 deg$^{-2}$. (Note that no correction is
made here, or in figure~\ref{fig:int_ids}, to account for missing
objects at fluxes $>$2x10$^{-14}$ \ecs which are too rare to be
sampled in this pencil beam survey. For clusters the additive
correction is +6 clusters deg$^{-2}$.)  The survey flux limit for
clusters is only approximate because in general they are extended
sources and we have only included the flux within the psf; however
only two of the clusters had a measurable extent and thus the
approximation should be reasonably accurate.  The number of clusters
expected has been obtained by integrating the local cluster X-ray
luminosity function of Ebeling et al (1997) over redshift (0$<$z$<$2)
and luminosities $>$10$^{42}$ erg s$^{-1}$.

The number predicted assuming no evolution of the cluster luminosity
function and q$_{\circ}$=0.5 is 22 deg$^{-2}$ (or 32 deg$^{-2}$ for
q$_{\circ}$=0).  Thus the number observed is consistent with a
no-evolution model, or is possibly slightly higher (ie weak positive
evolution).  Since 5 of the 6 clusters have measured redshifts of
z$>$0.3, this result implies that the low luminosity clusters observed
here do not evolve at the same (negative) rate as that claimed for the
high luminosity EMSS ($\sim$10$^{45}$ erg s$^{-1}$) clusters by Henry
\etal (1992) in approximately the same redshift range. However our
results may be consistent with the results of Henry \etal as the
latter provide no strong evidence for evolution of the lowest
luminosity clusters ($\sim$10$^{44}$ erg s$^{-1}$) in the EMSS
sample. Strong negative evolution had earlier been claimed for more
nearby ($z\leq0.2$) high luminosity clusters by Edge \etal (1990) but
Ebeling \etal (1997), from ROSAT all sky survey observations, strongly
dispute that result.  From observations of extended sources found
serendipidously in ROSAT pointed observations, Castander \etal (1995)
found few clusters, again supporting negative evolution, however
Collins \etal (1997) and Scharf \etal (1997), searching the same
database, dispute the Castander \etal result and find little evidence
for any evolution. The Collins \etal sample covers the same redshift
range as our clusters (z$>$0.3) and the clusters are of generally
intermediate luminosity ($\leq$10$^{44}$ erg s$^{-1}$).

Thus our observations support the growing concensus that there is
little evidence for evolution in low luminosity clusters even
at quite high redshifts ($z\sim0.5$); so far only the
Henry \etal observations sample high luminosities at high
redshifts. The detailed implications of our results for models of the
growth of structure in the Universe will be investigated in a future
paper.

We note that some current cluster surveys (eg Collins \etal) rely on
detecting extended sources in many separate ROSAT observations. Thus
differences in source detection algorithm may affect the number of
clusters found, particularly for marginally resolved clusters. However
in the present survey, which strives to classify all sources in a
selected area, source extension is not the only characteristic used in
classification. Supporting optical information is used too.  Thus low
luminosity, very distant clusters, which may not appear resolved in
the X-ray data, are not selected against.  With regard to the optical
information we note that our clusters are all relatively distant
($z>0.3$) and even the brightest cluster members are not that bright
(typically $m \geq 19$).  Some of the clusters which are quite obvious
on the 8K$\times$8K images were not at all obvious on our earlier CCD
images.  It is possible that other surveys with less deep optical
imaging may miss distant clusters and it is not impossible that we may
find even more, very distant, clusters by the time we have identified
all the sources in our complete area.

\subsubsection{STARs}

There are only 3 definite identifications with stars in our complete
sample. One of these, a 13.79 mag M star, happens to be the brightest
source in the 15 arcmin field by a factor of $\sim 10$ over the second source.
The two other stars are also late type stars; source 64 is a 13.41 mag
G star and source 115 is a 20.23 mag M star. The implied X-ray/optical
ratios are typical of coronal emission and similar to the ratios
found by Stocke \etal 1991 for stars of the same class.

\subsubsection{UNIDENTIFIED SOURCES AND BLANK FIELDS}

Eleven sources in the complete sample as classified as `unidentified'.
These sources do not occur in the upper third of Table~\ref{tab:main},
but otherwise are distributed more or less evenly throughout the lower
flux range.  Of these, seven errorboxes contain objects such as faint
galaxies or stellar objects which could be the identification. In some
cases (49, 60, 80) there is only one reasonable candidate, each one
lying close to the errorbox centre, but the optical spectra are
inconclusive.  Given the accuracy of our X-ray positions these
candidates must be considered fairly likely identifications.  In the
other 4 cases there is more than one optical candidate and the spectra
are again either inconclusive or not available.  For 5 of the 7 we
make a tentative best guess at the identification class, finding 1
QSO, 1 NELG, 1 cluster, 1 star and 1 absorption line galaxy. We refer
to the notes on individual objects for more information.

In 4 cases we find no objects brighter than $\sim23$mag in the errorboxes.
Some of these `blank' fields contain even fainter objects in the errorbox,
but nothing more than one would expect by chance at such magnitudes.
Given the accuracy of our X-ray positions (figure~\ref{fig:positions})
we are confident that the positions of these
blank fields are good and so we must ask what the identifications
could be. Possibilities include very distant clusters or highly
reddened AGN but we can draw no further conclusions on the basis of
the present data. The results of deep infrared imaging
are reported elsewhere (Newsam \etal 1997).

\begin{table}
\centering
\caption {\bf IDENTIFICATION SUMMARY}
 
\begin{tabular}{lccc}
         & BRIGHT & FAINT & ALL \\
         & $>10^{-14}$ & $\leq  10^{-14.5}$ &  \\
                         &  & erg cm$^{-2}$ s$^{-1}$ & \\
 & & & \\
QSOs                     & 16 & 5 & 32 \\
NELGs                    & 1  & 8 & 18 \\
Clusters                 & 1  & 0 &  6 \\
Stars                    & 1  & 1 &  3 \\
Unidentified             & 0  & 5 &  11 \\
\end{tabular}
\label{tab:ids}
\end{table}

\subsection{Source Count Slopes}

\begin{table*}
\centering
\caption {\bf QSO AND NELG DIFFERENTIAL SOURCE COUNT SLOPES AT FAINT FLUXES}
\begin{tabular}{llll}
              & All Sources              & QSOs      &   NELGS  \\
              &                          &           &          \\
Uncorrected   &  $-1.42^{+0.19}_{-0.26}$ & $-1.26^{+0.50}_{-0.55}$&
$-2.06^{+0.50}_{-0.58}$ \\
Corrected for Confusion&$-1.53^{+0.19}_{-0.28}$& $-1.40^{+0.48}_{-0.61}$ &
$-2.23^{+0.53}_{-0.63}$ \\
Corrected for Missing Area and Confusion &$-1.46^{+0.19}_{-0.28}$ &$-1.33^{+0.53}_{-0.55}$ &
$-2.15^{+0.50}_{-0.65}$   \\
&&&\\
Including 12 CRSS NELGs to flux & & & \\
$1 \times 10^{-13}$ erg cm$^{-2}$ s$^{-1}$& & & $-2.46^{+0.14}_{-0.17}$ \\
\end{tabular}
\label{tab:slopes}
\end{table*}

We can see from figure~\ref{fig:diff_ids}
that the number of QSOs per logarithmic flux bin is
constant below $10^{-14}$ erg cm$^{-2}$ s$^{-1}$ whereas the number of
NELGs per logarithmic flux bin is increasing. In order to quantify
this effect we have performed a maximum likelihood fit to find the
best single power-law slope to the distribution of sources.  We
perform the fit over the flux range $2 \times 10^{-15}$ erg cm$^{-2}$
s$^{-1}$ to $1 \times 10^{-14}$ erg cm$^{-2}$ s$^{-1}$.  For the NELGs
the choice of upper flux limit does not affect significantly the
derived slope but the error increases as the flux range is reduced.
For example taking an upper flux cut off of $10^{-13.6}$ erg cm$^{-2}$
s$^{-1}$ gives a slope of $-2.37^{+0.32}_{-0.58}$ (uncorrected) or
$-2.44^{+0.33}_{-0.56}$ (corrected).  However for the QSOs the choice
of upper flux limit does affect the derived value as the QSO source
counts steepen above $1 \times 10^{-14}$ erg cm$^{-2}$ s$^{-1}$.  As
we are interested in the most realistic extrapolation of the source
counts to fluxes below our survey limit, we restrict our fitting to
the sources of flux below $10^{-14}$ erg cm$^{-2}$ s$^{-1}$.  The best
fit slopes to the differential source counts are given in
Table~\ref{tab:slopes}.

As already stated in section 4.3.2, we have not attempted to correct
the NELG source count slopes for chance associations as we do not
know which these are. However of the 8 NELGs with R$\geq20$, of
which 4 are expected to be chance coincidences, 5 have X-ray
fluxes $<2.25 \times 10^{-15}$ erg cm$^{-2}$
s$^{-1}$, ie just above our survey limit. Thus removal of
chance coincidences will probably flatten the NELG source counts.
However, as we have 11 sources classed as `unidentified',
whose source counts rise at a similar rate to that of the NELGs,
it is quite possible that new NELGs may be discovered to
compensate for the chance associations. Given these uncertainties
we do not alter the observed distributions.

Although not plotted in figure~\ref{fig:int_ids}, the integral source
count slope derived from the 12 NELGs in the Boyle \etal (1995;
CRSS) survey at fluxes $>10^{-14}$ erg cm$^{-2}$ s$^{-1}$ does join on
quite smoothly to the present NELG source count slope at fainter
fluxes. It is not certain that the CRSS NELGs are exactly
the same sort of animal as our present NELGs (eg the CRSS NELGs are
slightly more luminous, $10^{42} < L_{X} < 10^{43}$ ergs s$^{-1}$),
but their optical spectra are rather similar.  We have therefore
calculated the maxiumum likelihood source count slope for the
combination of the 18 NELGs from the present deep survey and the 12 CRSS
NELGs. We perform the fit over the flux range 
$2 \times 10^{-15}$ erg cm$^{-2}$ s$^{-1}$ to
$1 \times 10^{-13}$ erg cm$^{-2}$ s$^{-1}$ and the fit is given
at the bottom of Table~\ref{tab:slopes}. The combined fit is quite consistent
with the fit based solely on the present deep survey NELGs, and is consistent
with a Euclidean slope. The error on the combined slope is, of course, less
than on the slope of the deep survey NELGs alone. 

On the basis of these data, the NELG source counts
are quite consistent with a Euclidean distribution but the QSOs source
counts definitely are not and imply that we are seeing beyond a
cut-off in the redshift distribution of X-ray QSOs.  We noted, in
section 2.1.2, that the faint end source count slope for all sources
in the complete area are slightly flatter (although only at the $1
\sigma$ level) than those in the full 12 arcmin or 15 arcmin radius
X-ray survey area.  We repeat the complete area slopes in
Table~\ref{tab:slopes} as those are the ones we should compare the QSO
and NELG counts with. In Section 6 we use these slopes to extrapolate
the contribution of various classes of objects to the soft XRB to flux
limits below that of the present survey.

\section{X-RAY SPECTRA}

Ideally any major contributor to the XRB should have a similarly hard
spectrum.  The 0.1-2.0 keV spectrum of that part of the X-ray
background which remains unresolved in the deepest ROSAT observations
is best fit (GBR), assuming an obscuring column at the galactic value,
by a combination of a very
cool (kT=0.1 keV) thermal bremsstrahlung component and a power-law of
energy index 0.7 which dominates above 0.4 keV.  Recent ASCA observations
(Gendreau \etal 1995) indicate that the XRB spectrum may be even
harder ($\alpha \sim 0.4$ in the 1-10 keV range).  

The work of GBR and others (eg Romero-Colmenero \etal 1996) confirms
that the X-ray spectra of QSOs in the ROSAT band, $\alpha \sim 1$, are
steeper than the spectrum of the diffuse XRB within the ROSAT band. If
the QSO spectra extrapolate to higher energies then they cannot
contribute all of the high energy XRB.  In Table~\ref{tab:main} we
present the hardness ratios of the brighter ROSAT sources. The
hardness ratio is defined as the ratio of counts in the 0.5-2.0 keV
band to those in the 0.1-0.5 keV band.  For those sources which were
undetected in the soft (0.1-0.5 keV) band we assume a soft count rate
equal to that of the faintest source detected in the soft band.
The counts in the soft and hard bands were derived by psf fitting to images
made in those bands rather than by simply counting photons within
a certain radius of the X-ray centroid. Thus they should be reasonably
robust to the presence of nearby sources. However when a particularly
soft source (eg 15) is very close to a particularly hard one (source 2),
the hard source may then not be detected at all in the soft band and so
the hardness ratio of the hard source may be overestimated.
Thus, although the hardness ratios do not contain as much spectral
information as a proper spectral fit, they do provide a reasonable
description of the X-ray spectra.  The hardness
ratios cease to be useful below about source 80 where the lower limits
on the source hardness can only be used to eliminate sources with
extremely soft spectral indices.

In Table~\ref{tab:hard} we give the hardness ratios expected
from various spectral models. We can see that the hardness ratios
actually provide a remarkably good diagnostic of spectral type.
For example, power law models (assuming a galactic absorbing
column), unless noticeably flatter than $\alpha=1$, are readily
distinguished from Raymond-Smith thermal models. For the
Raymond-Smith thermal models we also note that the gas temperature
does not have a great effect on the hardness ratio. However the 
metallicity is very important
because of the dominant iron-L emission at $\sim$1 keV.
Thus high metallicity thermal sources
(eg stars) can, in general, be distinguished from low metallicity
thermal sources such as clusters of galaxies.

Examination of Table~\ref{tab:main} shows that the NELGs are, on
average, significantly harder than the QSOs.  For example, taking an
arbitrary hardness ratio of 0.8, we can see that all NELGs brighter
than source 60 have hardness ratio above 0.8. (Below source 60 the
lower limit on the hardness ratio for sources undetected in the soft
band is generally below 0.8).  However only a few of the QSOs are
harder than 0.8 and these include sources 2 and 24 which are not
detected in the soft band because of the presence of a very soft
nearby confusing sources (15 and 20 respectively). Unfortunately there
are insufficient photons to simultaneously provide useful constraints
on both the power-law index and the absorbing column but a fit to the
summed spectra of the 5 brightest NELGs shows no evidence of strong
absorption, ie the column is $2.2^{+2.6}_{-1.8} \times 10^{20}$
cm$^{-2}$ and the energy index is $1.06^{+0.7}_{-0.6}$.  Examination
of the spectra of individual bright NELGs indicates possible
differences in absorbing column between individual NELGs but the
sources are too faint for definite statements to be made.  The above
hardness ratio results were noted by \mch (1995).  Carballo \etal
(1995) also calculated hardness ratios for the sources in their sample
and concluded that galaxies without broad emission lines had very hard
spectra.  A more detailed X-ray spectral analysis of the galaxies in
the present sample was then presented by Romero-Colmenero \etal
(1996).  Romero-Colmenero \etal found that the average NELG spectral
index, $<\alpha>=0.45 \pm 0.09$ (assuming a galactic absorbing column)
whilst that of the QSOs is $0.96 \pm 0.03$.  Thus the average spectrum
of the NELGs is in good agreement with the background spectrum and is
certainly a better fit to it than is the average QSO spectrum.  A
similar result was subsequently found by Almaini \etal (1996) from
examination of a somewhat X-ray brighter sample of NELGs.  Almaini
\etal also conclude that the average absorbing column of the NELGs is
close to the galactic value but do find some sources with higher than
galactic absorbing columns.

\begin{table}
\centering
\caption {\bf HARDNESS RATIOS FOR SPECTRAL MODELS}
 
\begin{tabular}{lll}
   HR    &$<E>$(keV)  &      Model              \\
         &           &                         \\
 0.13    &    0.3    &   PL with $\alpha=2$  \\
 0.5     &    0.5    &   PL with $\alpha=1$  \\
 0.9     &    0.6    &   1 keV RS model (Z=0.3)\\
 0.94    &    0.65   &   5 keV RS model (Z=0.3)\\
 1.7     &    0.7    &   1 keV RS model (Z=1)  \\
\end{tabular}
\label{tab:hard}
\end{table}

An X-ray spectrum without observable absorption can, of course, be
produced by scattering even though all direct low-energy emission
might be completely absorbed. A column density of $\sim 5 \times
10^{23}$ atoms cm$^{-2}$ would be sufficient to prevent direct
radiation from the brightest known (2-10 keV) AGN being detected in
our ROSAT observations. However if we were seeing scattered, rather
than direct, X-radiation from our NELGs, then their implied
luminosities at energies $>$10 keV would be very large $>10^{45}$ ergs
cm$^{-2}$ s$^{-1}$ (eg see ASCA and OSSE observations of NGC4945 by
Done \etal 1996) and the integrated emission from all NELGs would
greatly exceed the observed hard X-ray background.

Taking the X-ray spectral index of the NELGs to be 0.45, and taking
the differential source count slope to be -2.2 we can calculate the
contribution of NELGs to the X-ray background at any arbitrary energy.
The only variable is then the lower flux limit in the ROSAT band at
which one cuts off the contribution of NELGs.
For the X-ray background spectrum at energy, $E$, in the energy range
3 to 60 keV, we use the expression of D E Gruber, given in
Fabian and Barcons (1992, equation 2) as:

\[ I(E) = 7.877 E^{-0.29}exp(-E/41.13)  keV s^{-1} sr^{-1} cm^{-2} keV^{-1}\]

The resultant percentage contributions at energies of 5, 10 and 20 keV
are given in Table~\ref{tab:nelg_hard}. The percentages in brackets
are those applicable to a differential source count slope of -2.5.

\begin{table}
\centering
\caption {\bf NELG PERCENTAGE CONTRIBUTION TO THE HARD X-RAY BACKGROUND}
 
\begin{tabular}{llll}
Lower flux Limit                    &5 keV       &  10 keV    & 20 keV     \\
(ergs cm$^{-2}$ s$^{-1}$, 0.5-2 keV) &            &            &            \\
                                    &            &            &            \\
$2\times 10^{-15}$                  & 10.0 (7.8) & 10.2 (7.9) & 11.6 (9.0) \\
$5\times 10^{-16}$                  & 20.6 (23.7)& 20.8 (23.9)& 23.8 (27.3)\\
\end{tabular}
\label{tab:nelg_hard}
\end{table}

If we ignore all other possible contributors to the hard X-ray background
then Table~\ref{tab:nelg_hard} directly gives us the lower limit to
the fraction of the NELG X-ray emission which we can allow to be
scattered if we are not to exceed the total X-ray background flux.
Thus if the scattering model is to succeed we require a scattering
percentage of at least 10\%, and probably greater than 20\%,
which is a much higher percentage than the ball park figure of $\sim1\%$
which is usually assumed. Including contributions to the hard X-ray
background from other sources such as clusters and Seyfert galaxies
which we already know to exist only raises the lower limit to the
allowed scattering fraction. 

Moreover high fluxes in the hard X-ray band are likely to be
associated with high optical fluxes, ie bright galaxies (eg see Elvis
\etal 1978), and we note that NGC4945 is a 9th magnitude galaxy
whereas our NELGs are 18-21 mag.  Scattering might also be expected to
produce weak broad bases to the permitted lines. However, with the
exception of object 51 where no such broad bases are seen, the present
optical spectra are not good enough to put useful limits on the
strength of any possible broad lines.  We conclude that it is much
more likely that we are looking at direct, rather than scattered,
radiation from the NELGs and so the majority of them probably do have
low intrinsic absorbing columns.

A number of other points related to the identification content of the
survey can be made from examination of the hardness ratios. It is
quite clear that there is a wide range in the X-ray spectra of the
QSOs. Some are extremely soft with implied power-law energy indices
steeper than 2.  Only the QSOs have such soft spectra.  We may
therefore conclude that the unidentified source 9, which also has a
very soft spectrum, is probably also a QSO and, conversely, the
unidentified hard source 54 is more likely to be a NELG than
a QSO, although we do note that there are 
some quite hard QSOs. On the basis of the hardness ratio alone we
cannot rule out the possibility that a high hardness ratio in a QSO is
a result of absorption although, with few exceptions,
X-ray selected QSOs are rarely highly absorbed.

With the exception of source 5, which is slightly confused, all the
proposed cluster identifications have relatively high hardness ratios,
consistent with what we expect from a low metallicity hot gas. Thus we
strengthen our belief in the majority of cluster identifications and
confirm our suspicion, based on examination of the X-ray and optical
fields, that source 5 may contain emission from more than one discrete
location.

With the exception of the rather unusual emission line M star, source
1, the other stellar identifications, 16 (M star) and 27 (G star),
have high hardness ratios (1.44 and 1.20 respectively) entirely
consistent with high metallicity thermal plasma emission.

Thus a rather consistent pattern emerges between the identifications
and their X-ray spectra, further strengthening our confidence in the
identifications and providing some grounds for guessing at what at
least the brighter of the presently unidentified sources might be.

\section{CONTRIBUTION OF THE IDENTIFIED SOURCES
TO THE SOFT X-RAY BACKGROUND}

\begin{figure}
\begin{center}
  \leavevmode
  \epsfxsize 1.0\hsize
  \epsffile{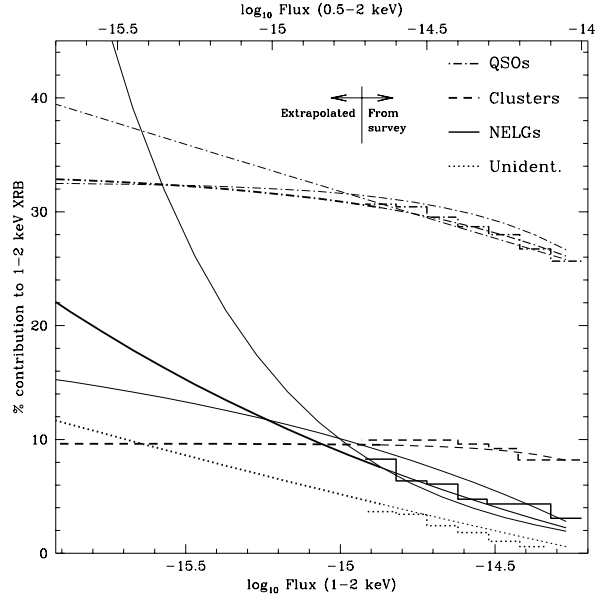}
\end{center}
\caption{ Contribution to the 1-2 keV X-ray background from sources of
different types, above a given flux. Below our survey limit the
extrapolations are based on maximum likelihood fits to the corrected
differential source counts given in figure~\ref{fig:diff_ids}.
The flux ranges over which the source count slopes were calculated
are given in the text. For the NELGs and for the QSOs we present
$1 \sigma$ confidence limits in the form of thinner lines of the same
type (ie solid for the NELGs, dot-dash for the QSOs).
}
\label{fig:contrib}
\end{figure}

In figure~\ref{fig:contrib} and Table~\ref{tab:xrb} we present an
estimate of the contribution of various classes of object to the
diffuse 1-2 keV XRB, above a particular flux, based on their
differential source counts, corrected for missing area and confusion
losses.  The source counts have been extrapolated down to $2\times
10^{-16}$ ergs cm$^{-2}$ s$^{-1}$ (0.5-2 keV) at which point almost
all of the soft XRB would be accounted for.  For the QSOs and NELGs we
plot the best fit prediction as thick lines (solid for the NELGs,
dot-dash for the QSOs) and also plot 1$\sigma$ error bounds as thin
lines. The upper uncertainty bound for the NELGs is particularly large
and identification of X-ray sources only a little fainter than the
limits of our present X-ray survey would greatly reduce the
uncertainty.  Similarly, inclusion of NELGs with brighter fluxes would
also reduce the uncertainty. We have earlier calculated the NELG
source count slope from a combination of our present NELGs and the 12
CRSS NELGs. To avoid confusion we do not add further lines to
figure~\ref{fig:contrib} but we note that the steeper combined NELG
source counts predict a contribution to the soft XRB somewhat in
excess of the best fit shown in figure~\ref{fig:contrib} (ie around
30\% at $2\times 10^{-16}$ ergs cm$^{-2}$ s$^{-1}$). However the
reduced error on the combined slope means that the upper bound is
consequently reduced (to around 40\% at the same flux level).  

As stated in section 6, removal of NELG chance coincidences will
probably flatten the NELG source counts and so reduce the contribution
of NELGs to the soft XRB at faint fluxes. However the uncertainties
associated with NELG chance associations, and with any NELGs which may
not yet be identified, prevent us from making any quantitative
correction.  Of course a smooth extrapolation to $2\times 10^{-16}$
ergs cm$^{-2}$ s$^{-1}$ followed by a sharp cut-off is unrealistic and
so it is likely, for example, that the NELG source counts will turn
down at some flux limit below our survey limit. However as yet we have
no evidence as to what that flux limit might be.  We remind reader
that the unidentified sources provide a major uncertainty in
determining the origin of the XRB. We have already argued (section
4.3.1) that some unidentified sources will be faint QSOs and so we
underestimate the contribution of QSOs to the soft XRB. However as it
is impossible to estimate precise numbers, we do not attempt any
correction to the QSO contribution.  It is therefore unrealistic to
extrapolate the unidentified sources as if they were a uniform class
of objects.  Despite these provisos, figure~\ref{fig:contrib} does
provide a first order prediction which future deeper surveys can
address.

In Table~\ref{tab:xrb} we also list the contribution to the 1-2 keV XRB
from resolved sources in various flux ranges higher than those covered
by our own survey.  These higher flux contributions are also included
in figure~\ref{fig:contrib}.  This source-count based approach clarifies
the contribution of various classes to the XRB in particular flux
ranges and allows simple extrapolation to fainter fluxes (although, of
course, such extrapolations should not be pushed too far).  Also it is
simple to take into account possibilities such as all of the
unidentified sources being NELGs.

To avoid contamination of the extragalactic XRB by diffuse soft
emission from our own galaxy we follow previous workers (Hasinger
\etal 1993) and restrict discussion to energies above 1 keV.  We
assume a 1-2 keV X-ray background intensity of $1.46 \times 10^{-8}$
erg cm$^{-2}$ s$^{-1}$ sr$^{-1}$ (Chen, Fabian \& Gendreau 1996),
which is consistent with the value of $1.4 \times 10^{-8}$ erg
cm$^{-2}$ s$^{-1}$ sr$^{-1}$ found by GBR. Fluxes have been converted
to the 1-2 keV band assuming an energy index of 1
for the QSOs, 0.45 for the NELGs and clusters and 0.7 for the 
unidentified sources, 
and a column density
N$_{H} = 6.5 \times 10^{19}$ cm$^{-2}$.

\begin{table*}
\centering
\caption{\bf 1-2 KEV X-RAY BACKGROUND PERCENTAGES
AS A FUNCTION OF FLUX}
 
\begin{tabular}{llll|l|l}
  &  &  &  &  &  \\
Flux erg cm$^{-2}$ s$^{-1}$ & 1x10$^{-11}$  & 1-5 x10$^{-14}$ &
   2-10 x10$^{-15}$   & Total to & 0.5-2x10$^{-15}$ \\
(0.5-2 keV)  & -5x10$^{-14}$  &  &  & 2x10$^{-15}$  & Extrapolation  \\
  &  &  &  &  &  \\
All extragalactic&11$^{\dag}$ &23$^{\ddag}$ & 18$^{\sharp}$ & 52 & $\sim18$ \\
QSOs            & 7$^{\flat}$ &18$^{\ddag}$ & 6$^{\sharp}$  & 31 & 2  \\
NELGs           &$<$0.7$^{\S}$&$\sim$1$^{\oplus}$ & 6$^{\sharp}$ &$\sim8$ & 9 \\
Clusters        &1.4$^{\diamondsuit}$&   5.5$^{\diamondsuit}$ & 3$^{\sharp}$ & 10 & 0.5 \\
Unidentified    &             &             & 3$^{\sharp}$       & 3  & 6 \\
  &  &  &  &  &  \\     

\end{tabular}
\label{tab:xrb}

\raggedright
\noindent{\small

$^{\dag}$ From the model fit to the extragalactic Einstein Medium
Sensitivity Survey (EMSS) of Gioia \etal (1990a).\\ 
$^{\ddag}$ From Shanks \etal (1991), Table~\ref{tab:main}. \\ 
$^{\sharp}$ Present work. \\
$^{\flat}$ From EMSS AGN data (Maccacaro \etal 1991).\\ 
$^{\S}$ From EMSS data (Stocke \etal 1991). This is an upper limit
since  it includes all EMSS galaxies which may have narrow lines.\\
$^{\oplus}$ Estimate from the 12 NELGs in 5 ROSAT fields found by 
Boyle \etal (1995).\\
$^{\diamondsuit}$ From Jones \etal, in preparation.\\
}

\end{table*}

As is already clear from figures~\ref{fig:diff_ids} and
\ref{fig:int_ids}, we see again in Table~\ref{tab:xrb} and
figure~\ref{fig:contrib} that QSOs are the dominant contributor to the
1-2 keV XRB at bright fluxes whereas NELGs dominate at low fluxes. By
extrapolating the differential source counts to fluxes below our
survey limit we see, in Table~\ref{tab:xrb}, that QSOs contribute very
little more to the soft XRB whereas the contribution from NELGs
continues to rise. Indeed, if the NELG source counts continue to rise
then the overall source counts should start to show an upturn at
fluxes not far below our survey limit.  Although not required by the
ROSAT deep field fluctuation analyses (Hasinger \etal 1993; Barcons
\etal 1994) the predicted upturn is consistent with these analyses. A
useful constraint on NELG evolution would therefore be provided by a
survey which extended the soft X-ray overall source counts, without
the need for identifications, to a flux limit a factor of only a few
below the present survey limit.

In Table~\ref{tab:xrb} we separately list the contribution to
the soft XRB of the unidentified sources. 
The unidentified sources have a rising source count slope similar
to that of the NELGs and some unidentified sources are very probably
NELGs, but obscured QSOs or very distant clusters cannot be ruled
out at this stage. The unidentified sources provide the major
uncertainty in the contribution of the various identified classes
of sources to the XRB and so completion of the identifications
of our sample remains a high priority.

An estimate of the contribution of various classes of object to the
XRB can also be made by modelling the evolution of their luminosity
functions.  The evolution of the QSOs in the present sample, together
with QSOs from the EMSS sample (Stocke \etal 1991), has been
considered by Jones \etal (1997) and, considering also RIXOS and other
samples, by Page \etal (1997a). The evolution of the NELGs in the
present sample has been considered by Page etal (1997b) and Pearson
\etal (1997).  Jones \etal deduce that, depending on the evolutionary
model chosen, between 31\% and 51\% of the soft XRB can be contributed
by QSOs and Page \etal (1997b) agree, giving an upper limit of 45\%.
The present QSO results are in agreement with the previous work
although towards the lower end of the Jones \etal range.
Page \etal (1997b) conclude that NELGs contribute between 15\% and
35\% of the soft XRB, in approximate agreement with extrapolations from
previous surveys eg Boyle \etal (1995). Our NELG results are also
in agreement with previous work.

We have resolved a total of 52\% of the 1-2 keV X-ray background into
point sources at the flux limit of 2x10$^{-15}$ erg cm$^{-2}$
s$^{-1}$.  Most of the resolved flux (31\%) arises from QSOs. 
Approximately 8\% arises from NELGs, and the rest comes from clusters
and groups of galaxies ($\sim10\%$), with some sources as yet unidentified. 
However extrapolating the differential source counts to a flux a
factor 4 below the survey limit ($5 \times 10^{-16}$ ergs cm$^{-2}$
s$^{-1}$), we expect that $\sim70\%$
of the 1-2 keV XRB would be resolved.  The NELG surface density at $5
\times 10^{-16}$ ergs cm$^{-2}$ s$^{-1}$ would be between 1000 and 2500
deg$^{-2}$ sufficient, unlike the surface density of QSOs, to explain
the isotropy of the XRB (see Fabian and Barcons (1992) for a review of
the isotropy of the XRB).

\subsection{Implications for the Hard X-ray Background}

The NELGs have hard X-ray spectra, similar to that of the XRB, and
harder than those of Seyfert galaxies in the 2-10 keV range ($<\alpha>
\sim0.8$). If their X-ray spectra extrapolate smoothly to higher
energies (see section 7)
they will be significant contributors to the XRB at energies
above those considered in the present survey ($0.5-2$ keV).  As their
spectra are similar to the XRB then, to first order, they will
contribute similar fractions of the XRB in higher energy bands as they
do in the ROSAT band (Table~\ref{tab:nelg_hard}).  Thus they will be
significant contributors but, assuming that we can ignore the
scattering model for their X-ray spectra discussed in section 5, they
will not explain all of the presently unresolved hard XRB and, as in
every new band, new populations of sources will be needed.  However,
as in the present study, NELGs will only be detectable as significant
contributors at faint fluxes, well below the level of all sky surveys
such as that of Ariel V (\mch \etal 1981). NELGs should be detectable
in small numbers in deep ASCA observations but in large numbers in XMM
and AXAF observations.

\section{ NELGs - AGN or STARBURST GALAXIES?}

To determine what the NELGs might be, and how they might relate to
other established galaxy types, we consider next their photometric and
spectroscopic properties.

We have observed the majority of the survey field in the V-band as well
as R-band, and some I-band observations have also been made.
Thus we have observed V-R colours for the NELGs. These colours,
which correspond to approximately rest frame
B-V colours, or even bluer in most cases, lie in the range 0.1-1.0,
with a mean around 0.5, more indicative of spiral than elliptical
galaxies.  

Morphologically, we have obvious visual examples of spiral host
galaxies in the more nearby NELGs (eg source 32). We also note that
some galaxies appear flattened (eg ROSAT 93 and 103) whilst others are
more circular (ROSAT 121). Thus some NELGs are certainly spiral
galaxies, but the possibility that some are elliptical galaxies
cannot be ruled out at this stage.  We also note some rather distorted
morphologies and evidence of interactions with neighbouring galaxies
in many cases. The results of visual examination are given in the
notes on individual sources.  However a proper morphological
classification of the majority of NELGs on the basis of our present
CCD data requires model fitting. A proper morphological and colour
analysis of the NELGs will be presented in a future paper (Newsam
\etal, in preparation).

The observed NELGs lie in the redshift range 0.061 to 0.590, with the
majority lying in the range 0.2-0.4.  Within the present small sample
there is no evidence for evolution in the NELG population or for a
redshift cut-off, principally because of the uncertainty associated
with the nature of the unidentified sources.
However when the present sample are combined with NELGs from the RIXOS
sample, there is some evidence for evolution (Page \etal 1997b), but
at a slower rate than found for broad line AGN.  From analysis of a
small sample of NELGs with brighter fluxes, Boyle \etal (1995) find a
broadly similar result to that of Page \etal.  Griffiths \etal (1996)
claim a higher rate of NELG evolution but their sample consists of
NELGs from both ROSAT and Einstein observations and so may be subject
to additional uncertainties.
The details of NELG evolution are therefore far from clear
and even deeper X-ray surveys are required to resolve the isssue.

The absolute magnitudes of the NELGs lie in the range $\rm M_{R}$ -20
to -23, typical of large spirals or ellipticals. However their X-ray
luminosities lie in the range $3 \times 10^{41}$ to $5 \times 10^{42}$
ergs s$^{-1}$, factors of 10-100 more than normal large spiral
galaxies and at the upper bound for large non-cluster ellipticals (see
Fabbiano 1989 for a review of X-ray emission from normal galaxies).
AGN can easily produce such X-ray luminosities, which are towards the
lower end of the distribution of AGN X-ray luminosities. However
typical AGN X-ray spectra are steeper (energy index $\sim 0.8$) than
the observed average NELG X-ray spectral index ($\sim 0.45$).
Blazar-like AGN (eg 3C273, Turner \etal 1990; Leach \etal 1995) have
similarly flat X-ray spectra but have much higher observed
luminosities ($\geq 10^{45}$ ergs s$^{-1}$).  Recently there has been
some interest in advection dominated accretion onto black holes (eg
Narayan and Yi 1995). Such accretion produces a very hard X-ray
spectrum and so advection dominated accretion onto a very massive
black hole could account for both the observed luminosities and X-ray
spectra of the NELGs.

X-ray data on starbursts galaxies is not as good as
for AGN because of the generally lower X-ray fluxes of starburst
galaxies. Many `starburst' galaxies of apparent high X-ray luminosity
turn out, on closer inspection, to be AGN (eg Moran \etal 1994).  The
maximum luminosity so far detected (0.5-2 keV) is $\sim2 \times
10^{41}$ ergs s$^{-1}$; Griffiths and Padovani 1990; Read and Ponman
1997).  However with these provisos we note that Rephaeli
\etal (1995) measure the average X-ray spectral index of 51 candidate
starburst galaxies to be $0.47 \pm 0.26$ (1-100 keV), very similar to
the average spectral index of the NELGs. Rephaeli \etal also consider
various possible X-ray emission mechanisms.  They conclude, as does
Fabbiano (1989), that X-ray binary sources, which have hard X-ray
spectra extending out to at least 20 keV, can contribute up to
$10^{41}$ ergs s$^{-1}$.  Such binary systems, which would be
distributed throughout the bulge and disc, would be resolvable in many
of our NELGs by AXAF. Other possibilites include supernovae remnants
(which have a variety of spectral indices), Compton scattering off
relativistic electrons from supernovae (hard spectrum) and thermal
emission from a supernova-driven galactic wind or from an extended
gaseous halo (soft spectrum). These possibilities require high
supernovae rates. 

A simple scenario is that starburst/X-ray binary activity
dominates the X-ray emission from the lower luminosity NELGs but
that AGN activity dominates the higher luminosity ones. 
More detailed observations, particularly X-ray spectral observations
of individual NELGs such as will be possible with XMM,
are required to clarify the issue.

\begin{figure*}

\begin{center}
  \leavevmode
  \epsfxsize 0.48\hsize
\epsffile{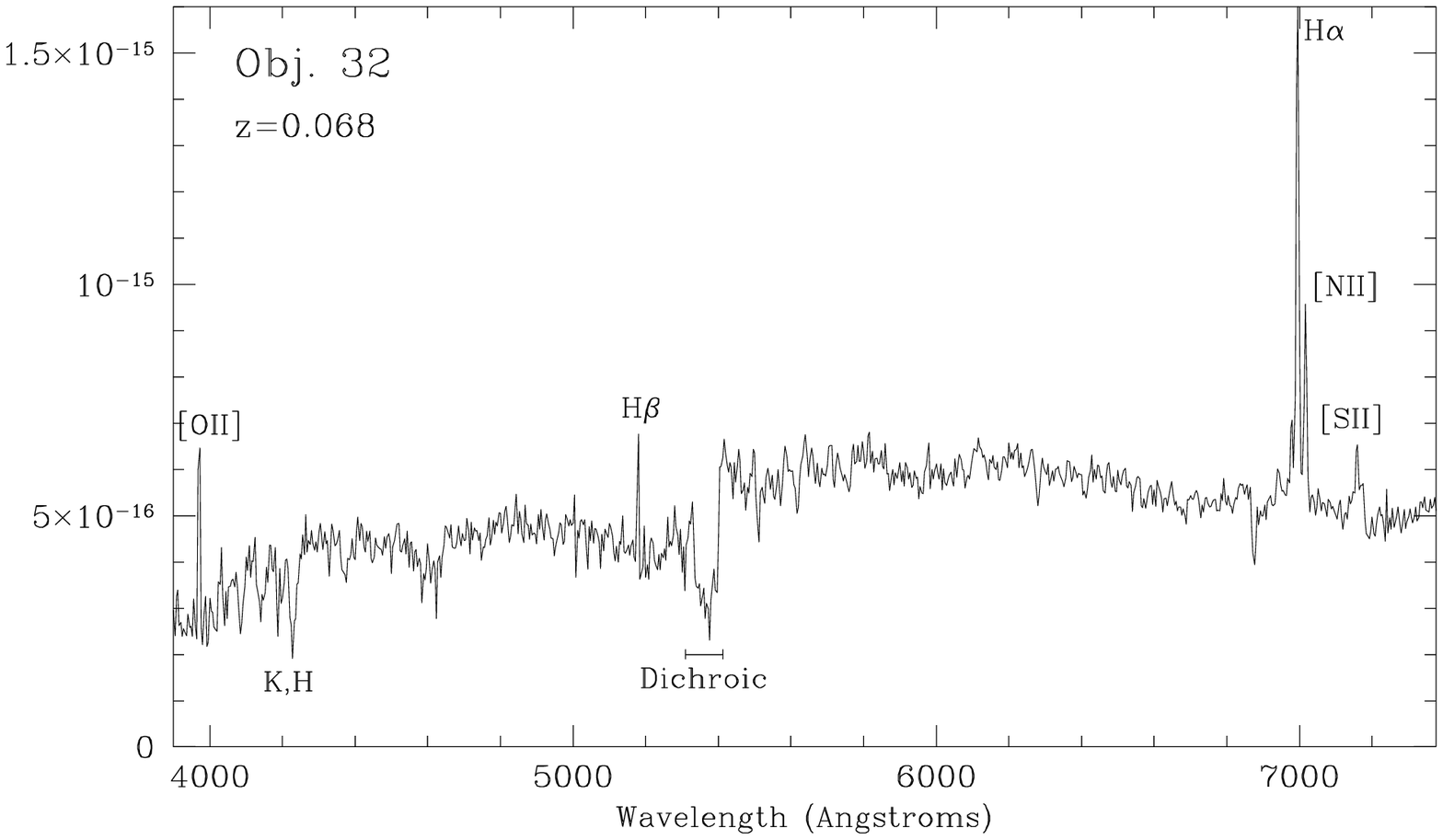}
  \epsfxsize 0.48\hsize
\epsffile{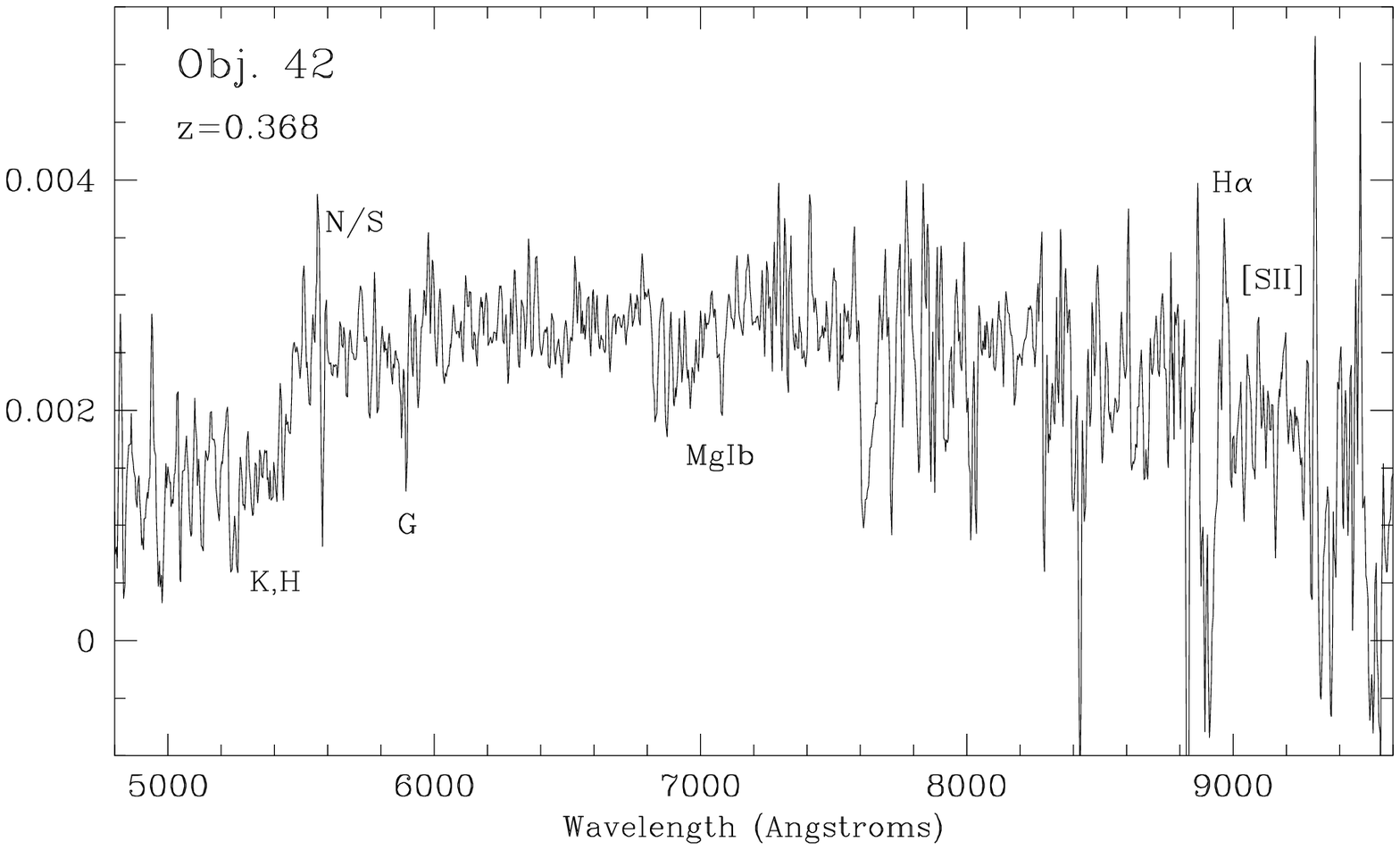}
\end{center}

\begin{center}
  \leavevmode
  \epsfxsize 0.48\hsize
\epsffile{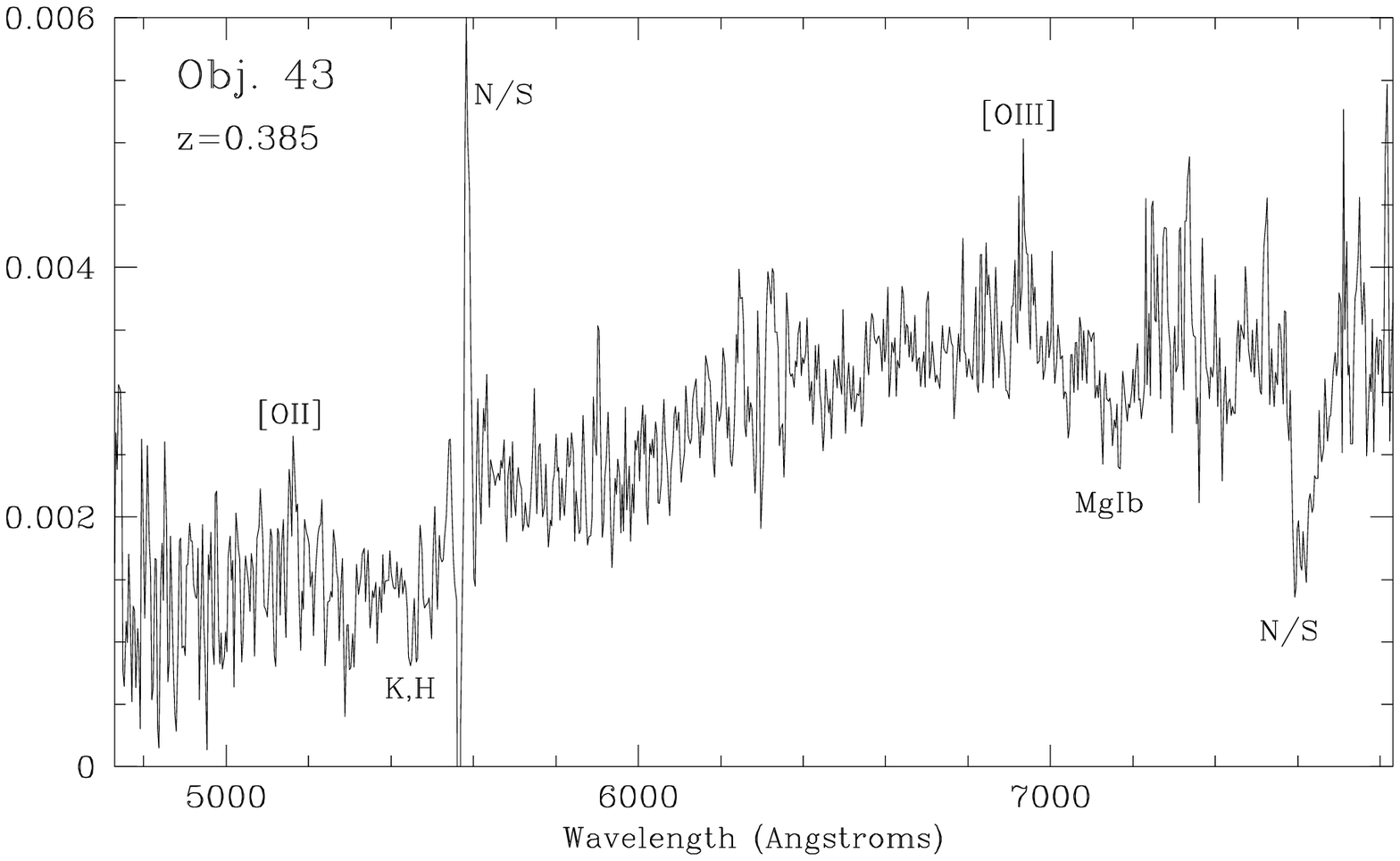}
  \epsfxsize 0.48\hsize
\epsffile{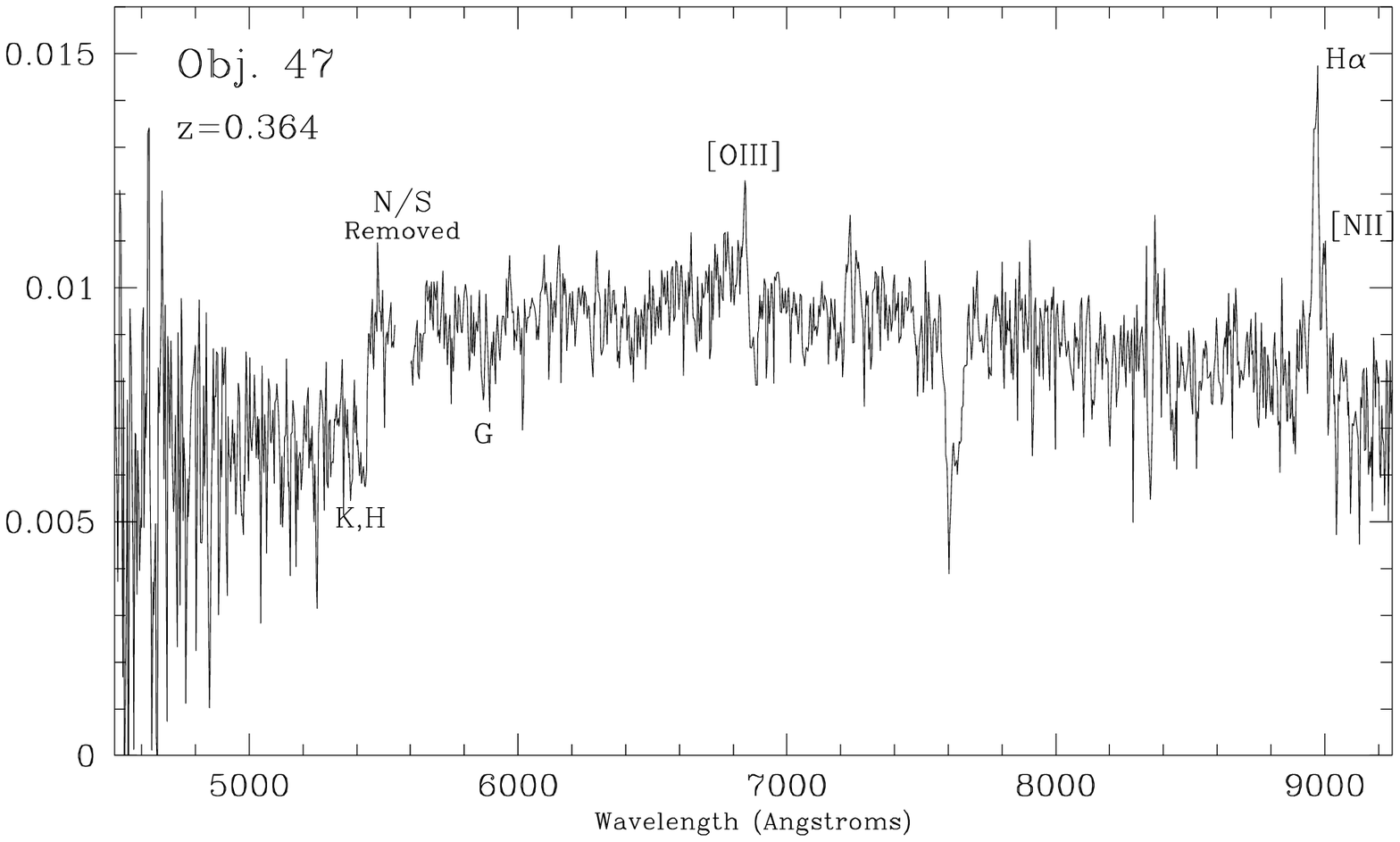}
\end{center}

\begin{center}
  \leavevmode
  \epsfxsize 0.48\hsize
\epsffile{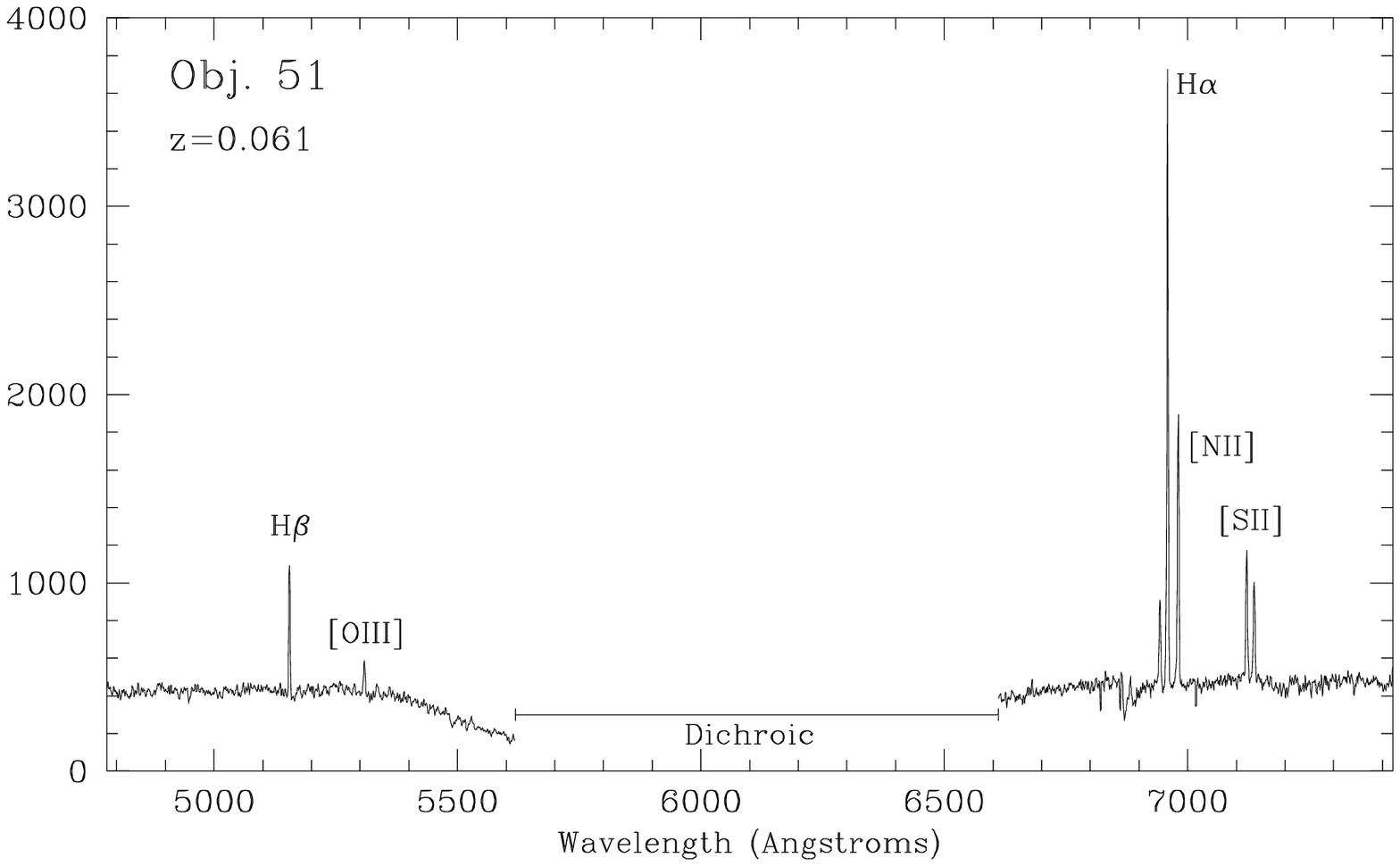}
  \epsfxsize 0.48\hsize
\epsffile{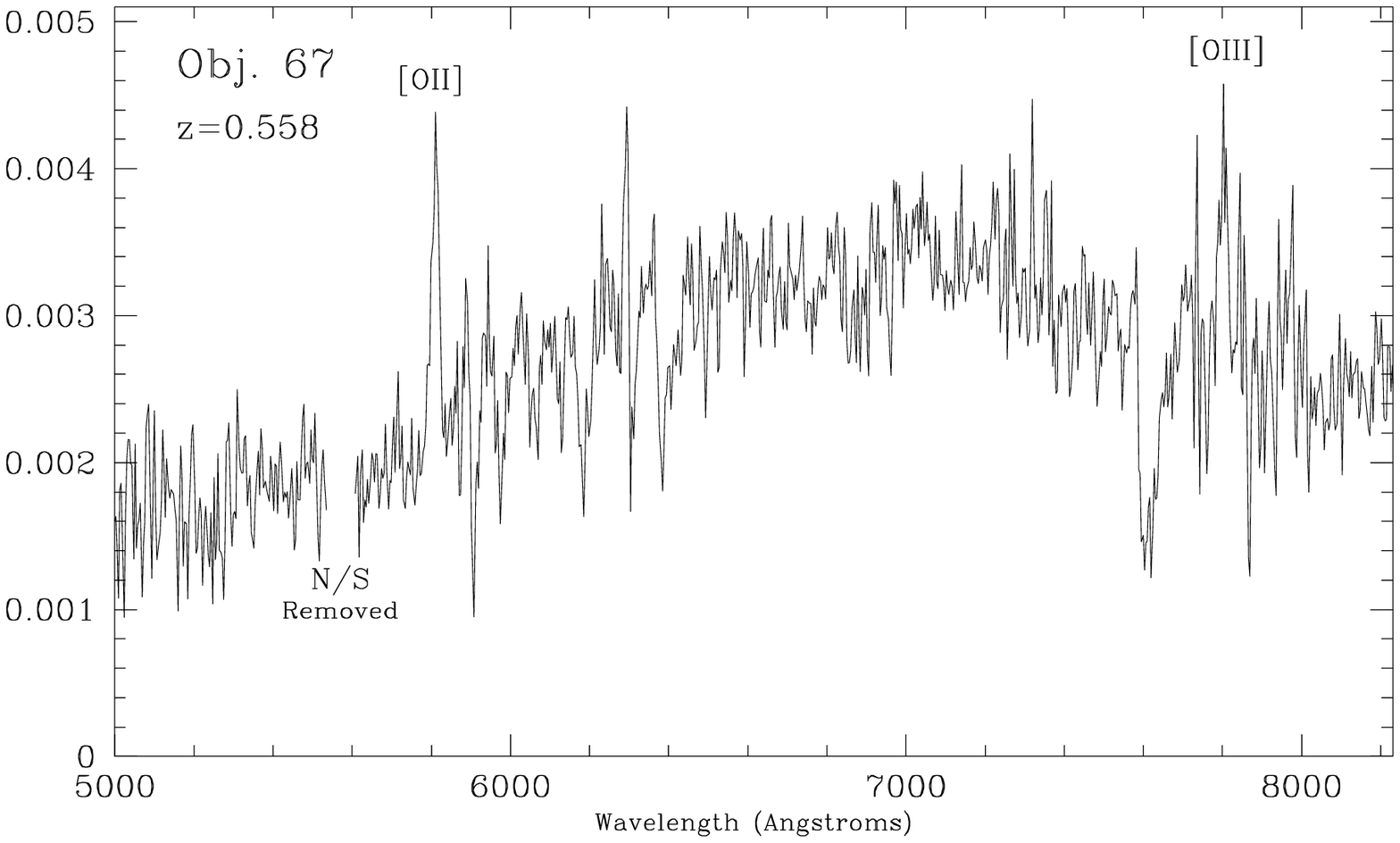}
\end{center}

\begin{center}
  \leavevmode
  \epsfxsize 0.48\hsize
\epsffile{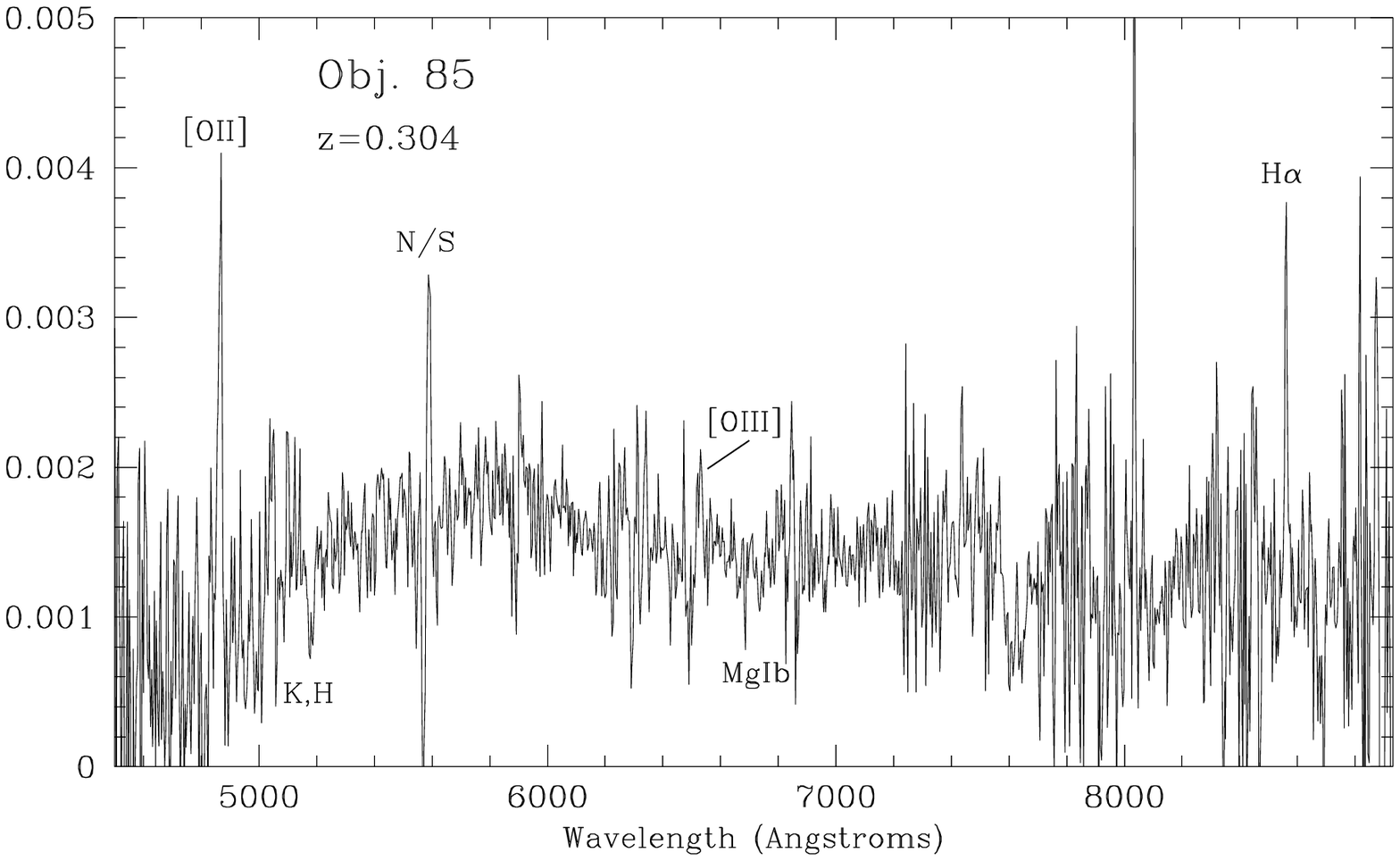}
  \epsfxsize 0.48\hsize
\epsffile{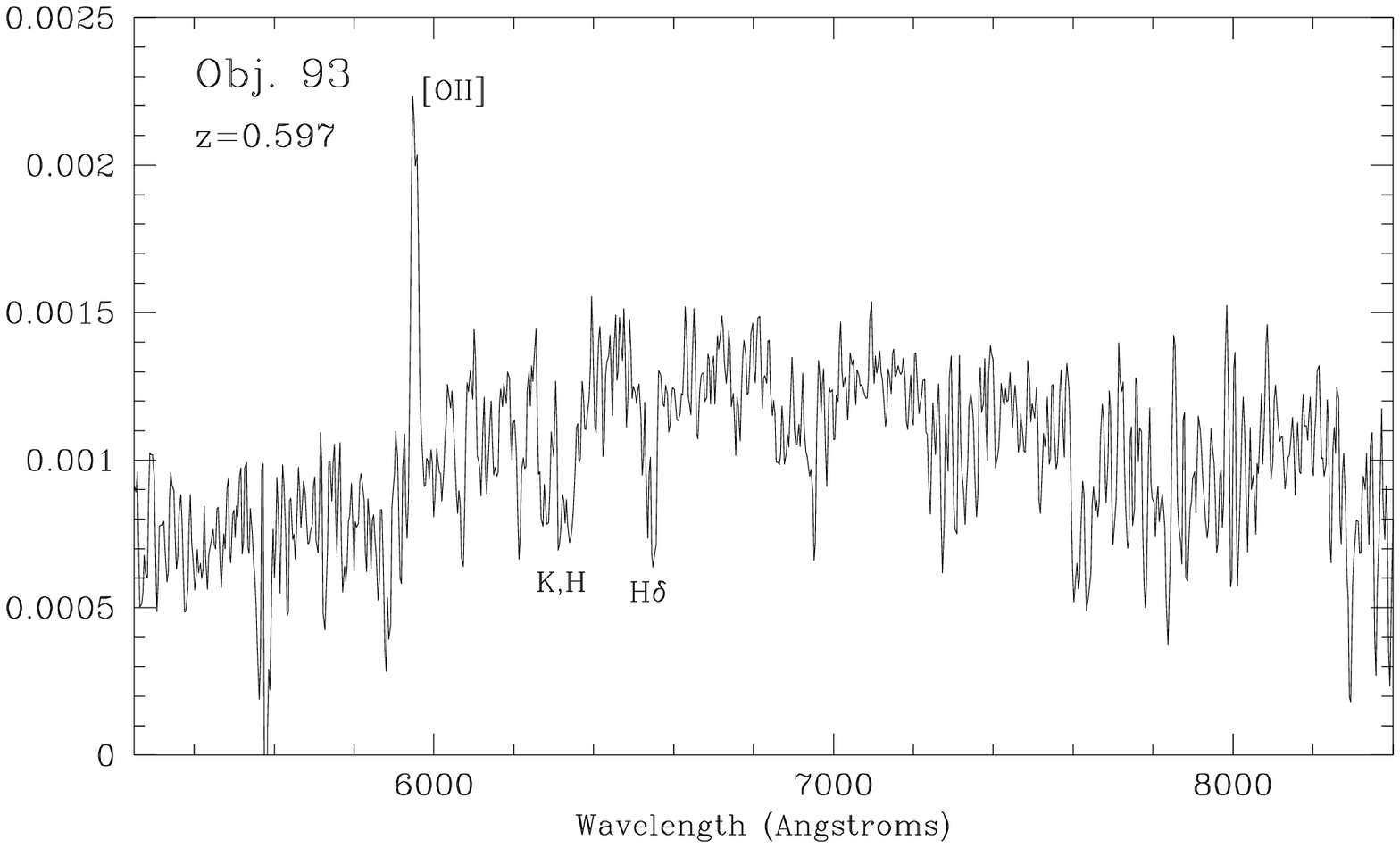}
\end{center}

\caption{ NELG optical spectra. 
With the exception of object 51
the spectra have been flux calibrated
but the flux scales are arbitrary.
The high resolution spectrum of object 51 was taken on a different
observing run to all the other (low resolution) NELG spectra.
}
\label{fig:spectra1}
\end{figure*}

\begin{figure*}

\begin{center}
  \leavevmode
  \epsfxsize 0.48\hsize
\epsffile{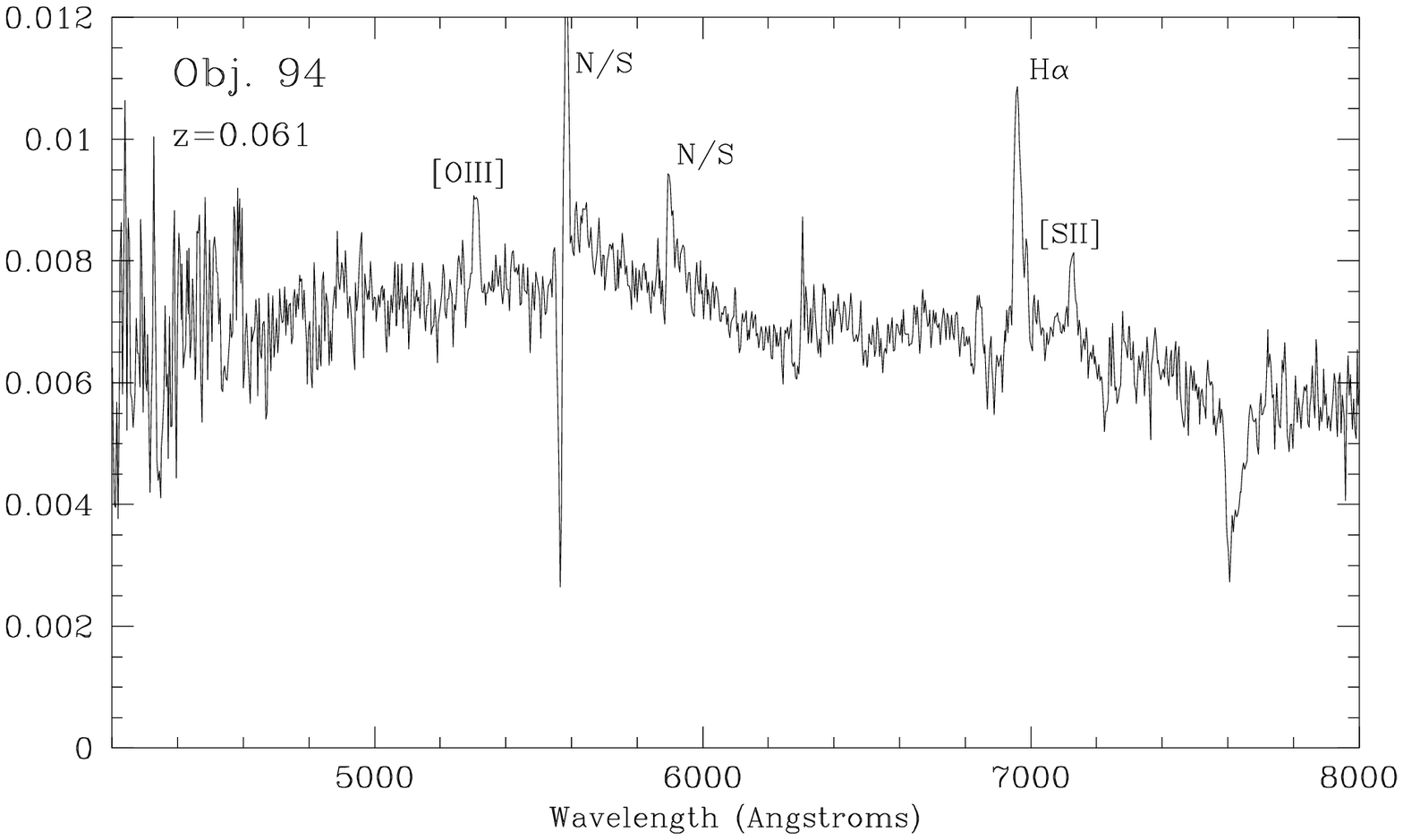}
  \epsfxsize 0.48\hsize
\epsffile{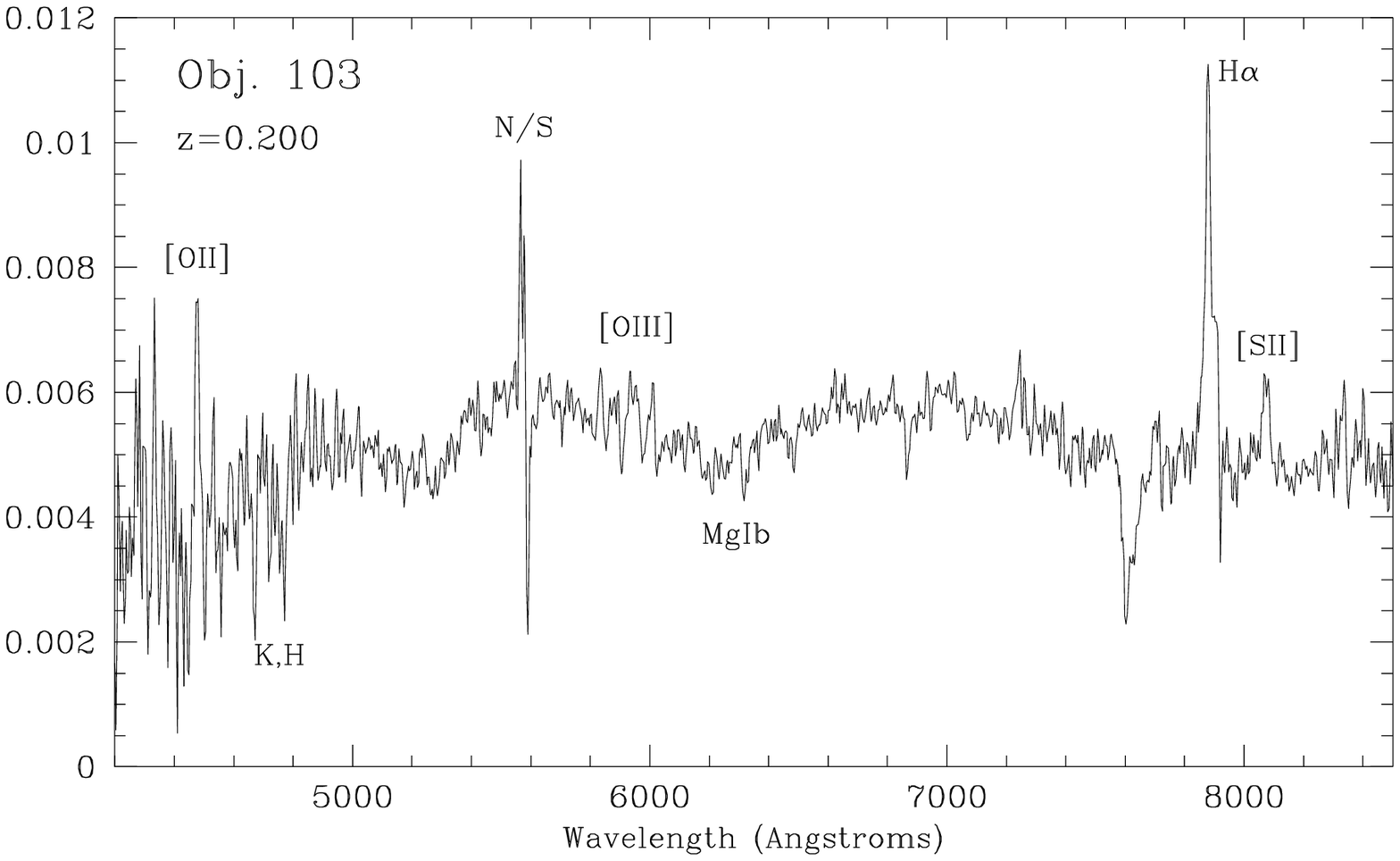}
\end{center}

\begin{center}
  \leavevmode
  \epsfxsize 0.48\hsize
\epsffile{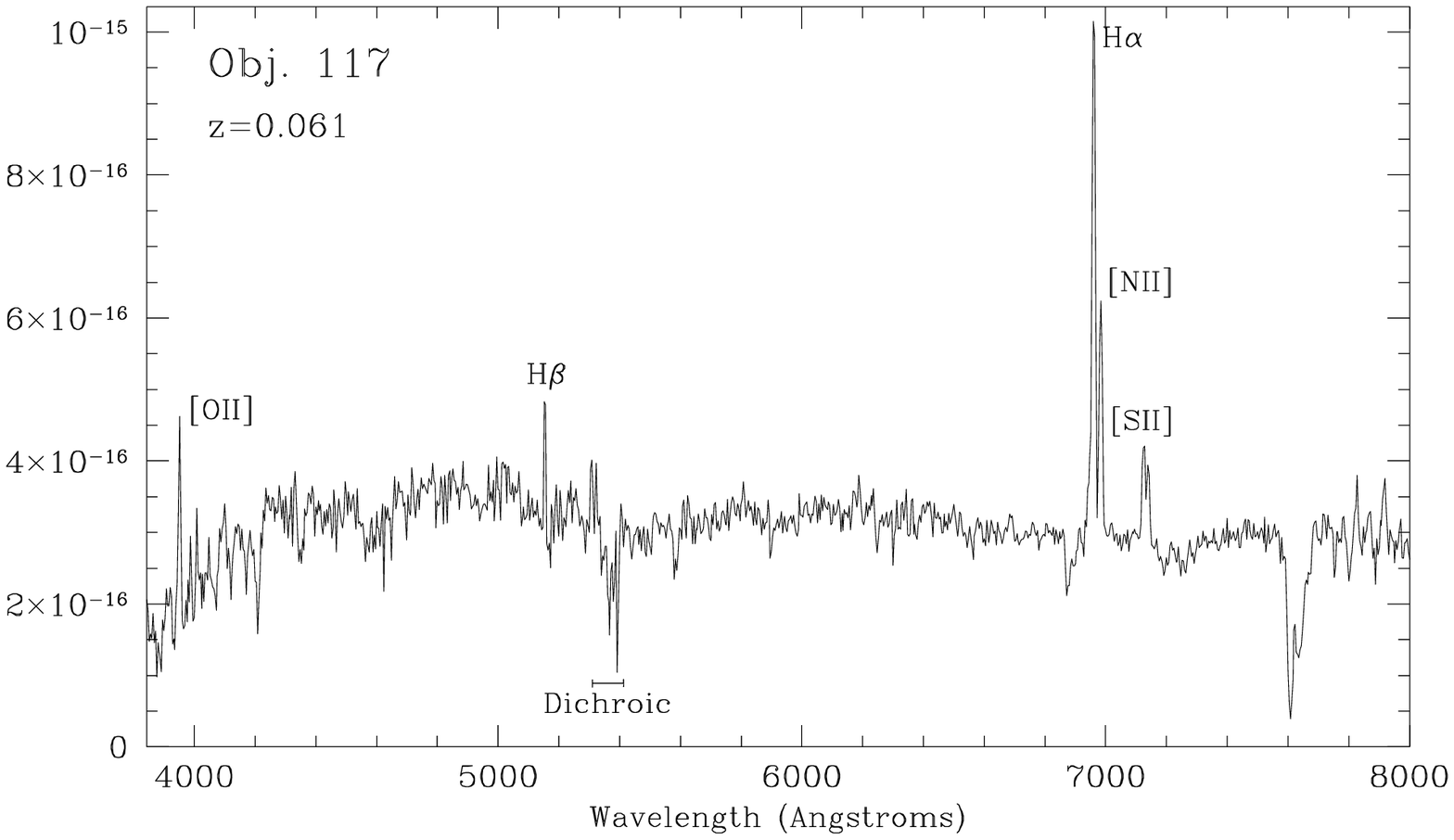}
  \epsfxsize 0.48\hsize
\epsffile{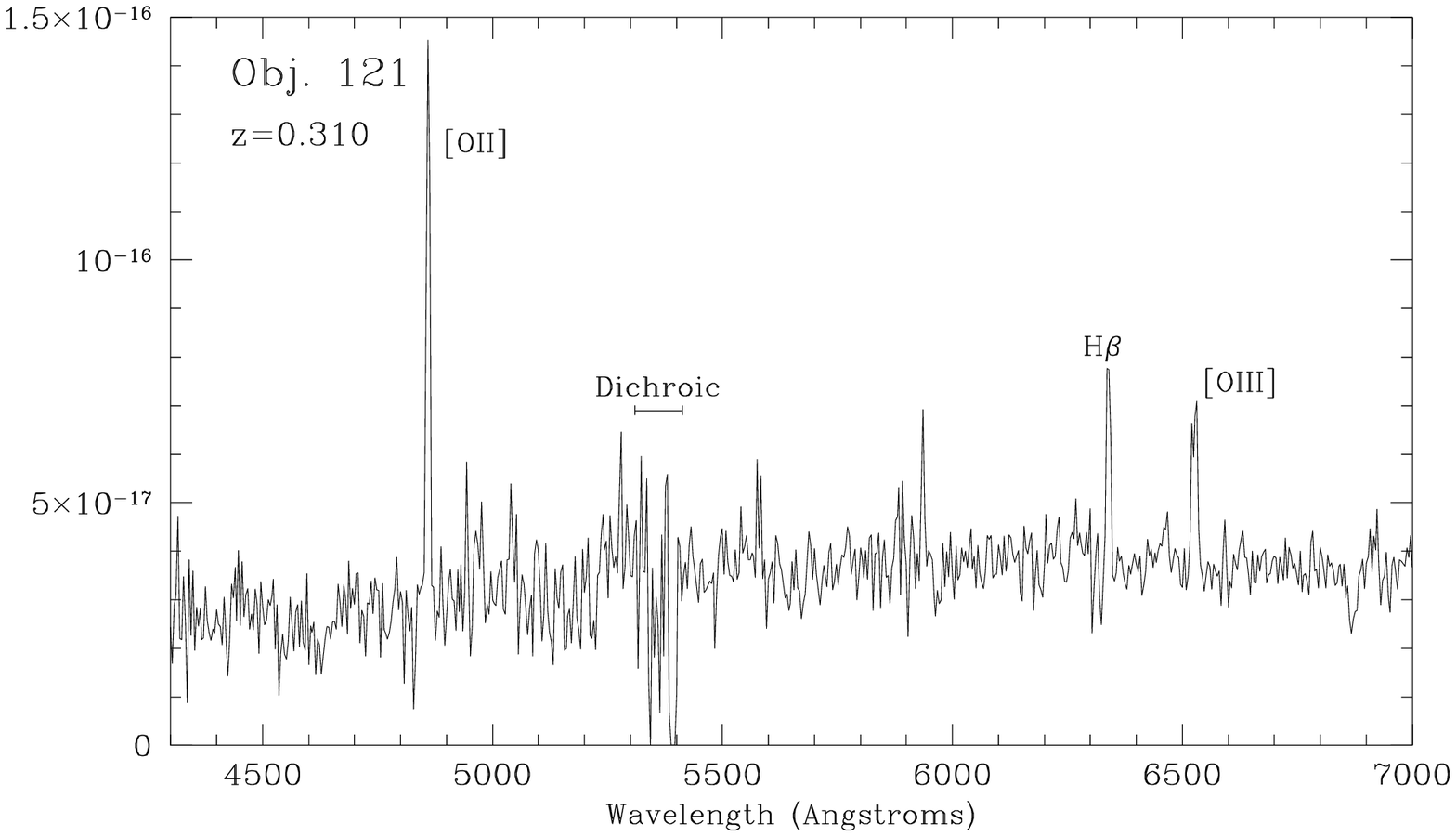}
\end{center}

\begin{center}
  \leavevmode
  \epsfxsize 0.48\hsize
\epsffile{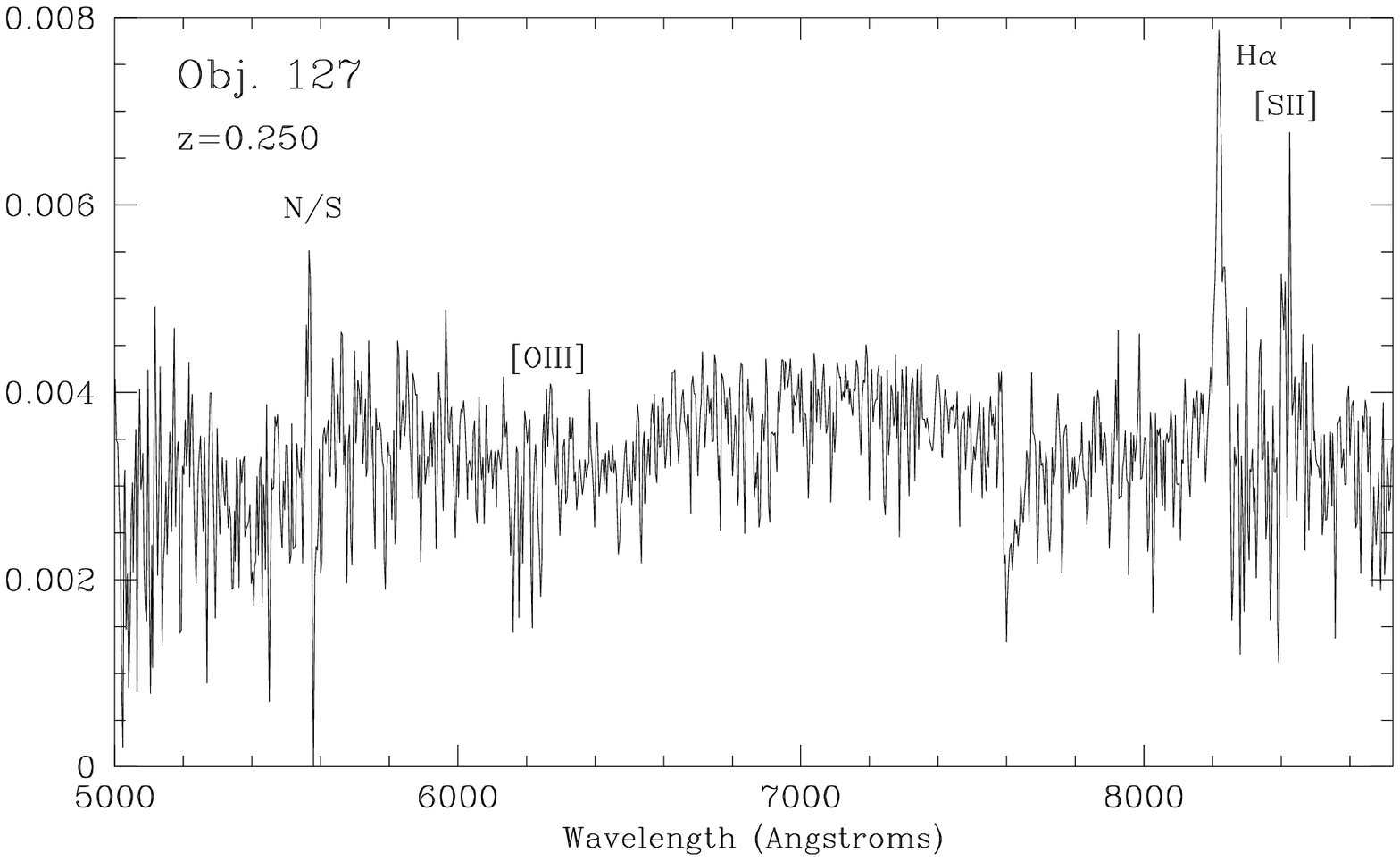}
  \epsfxsize 0.48\hsize
\epsffile{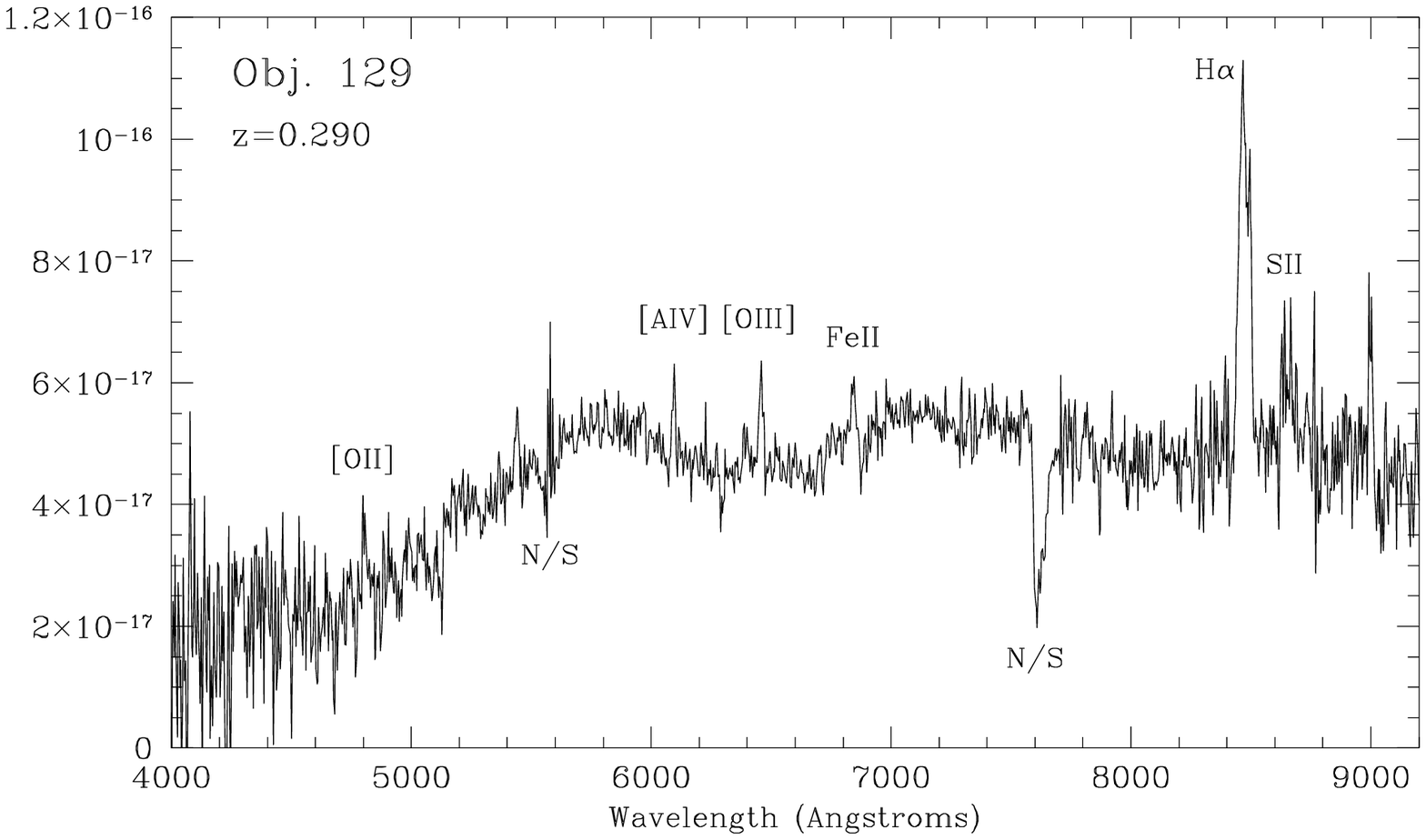}
\end{center}

\begin{center}
  \leavevmode
  \epsfxsize 0.48\hsize
\epsffile{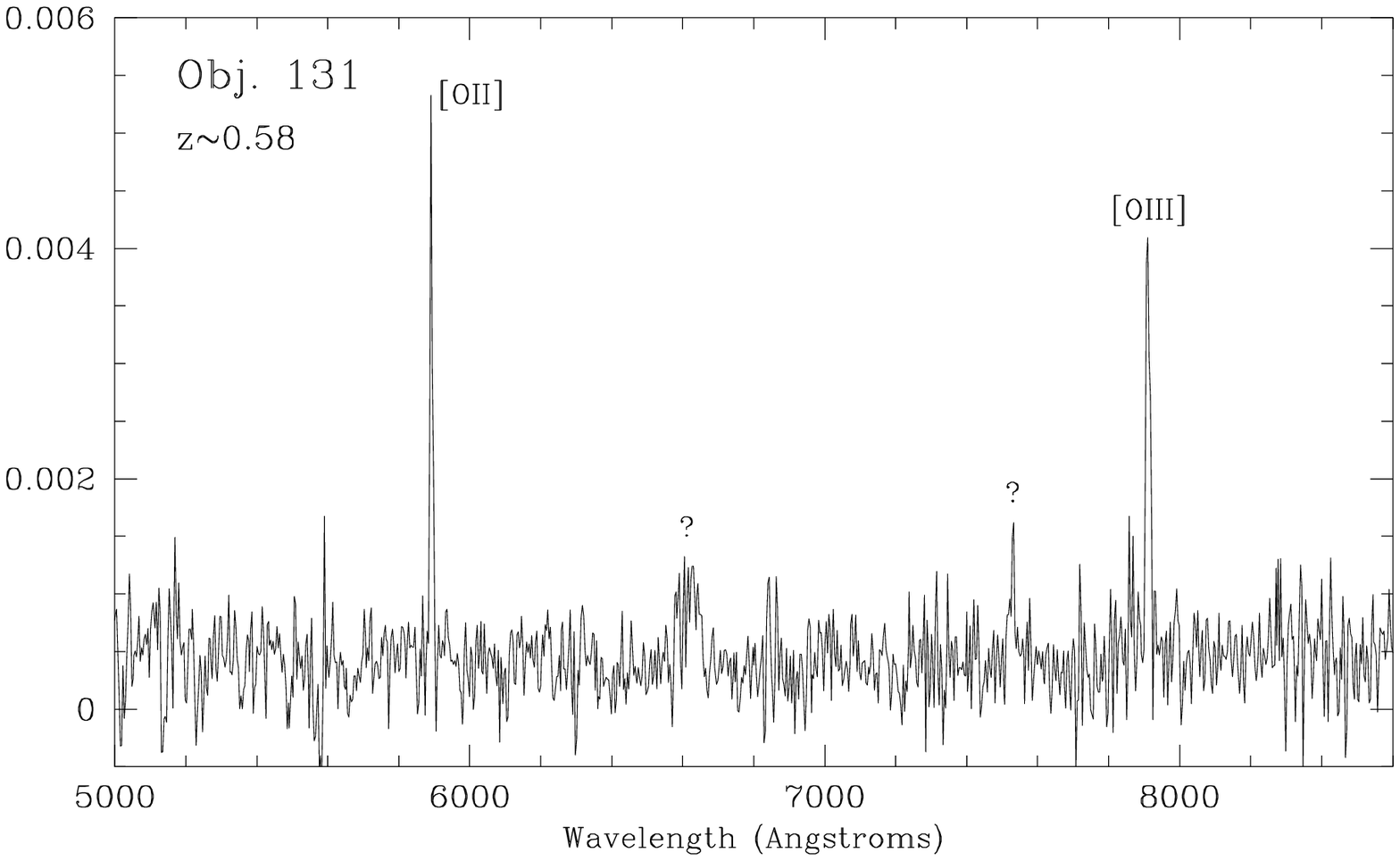}
  \epsfxsize 0.48\hsize
\epsffile{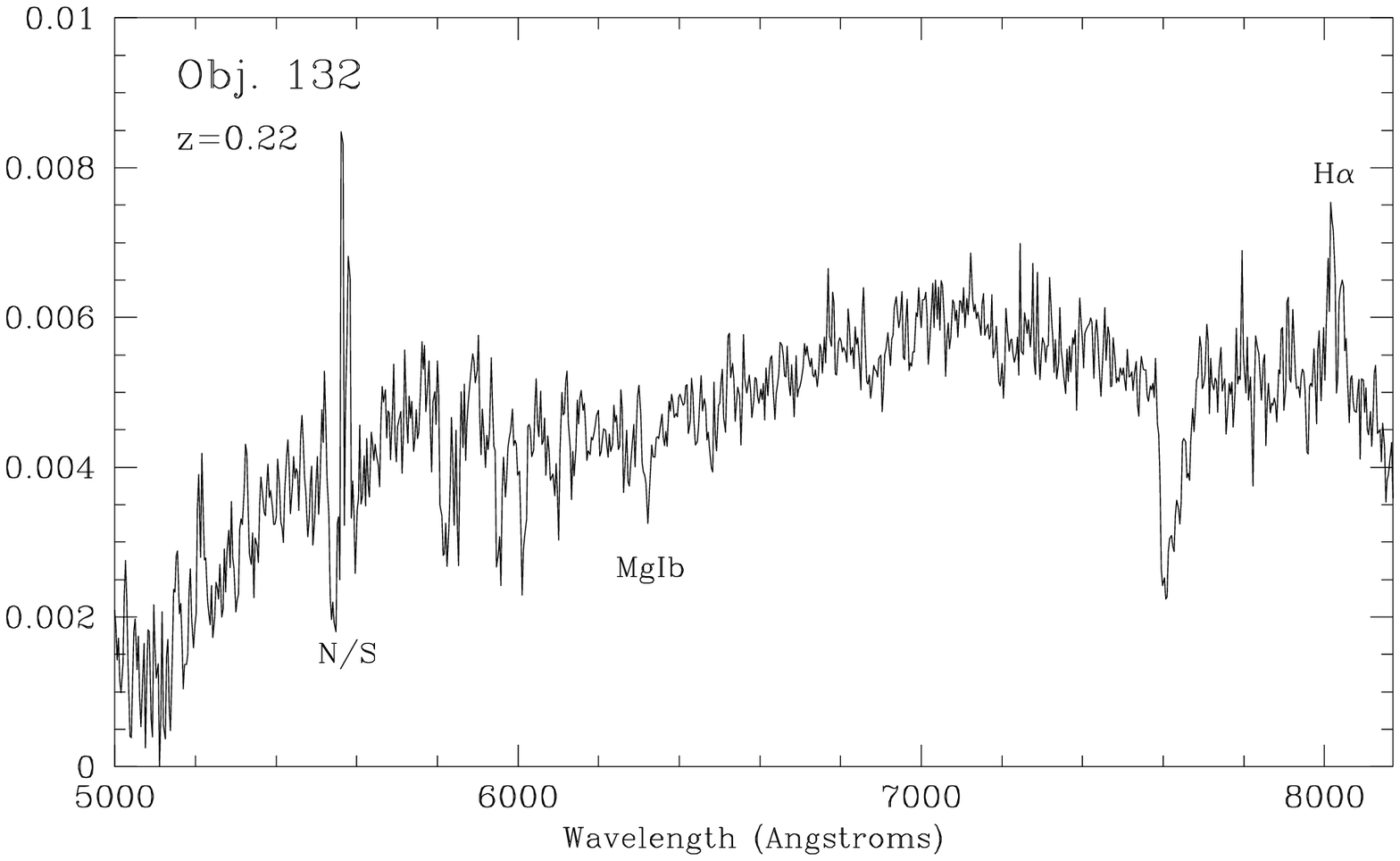}
\end{center}

\caption{ NELG optical spectra.}
\label{fig:spectra2}
\end{figure*}

\begin{figure*}

\begin{center}
  \leavevmode
  \epsfxsize 0.48\hsize
\epsffile{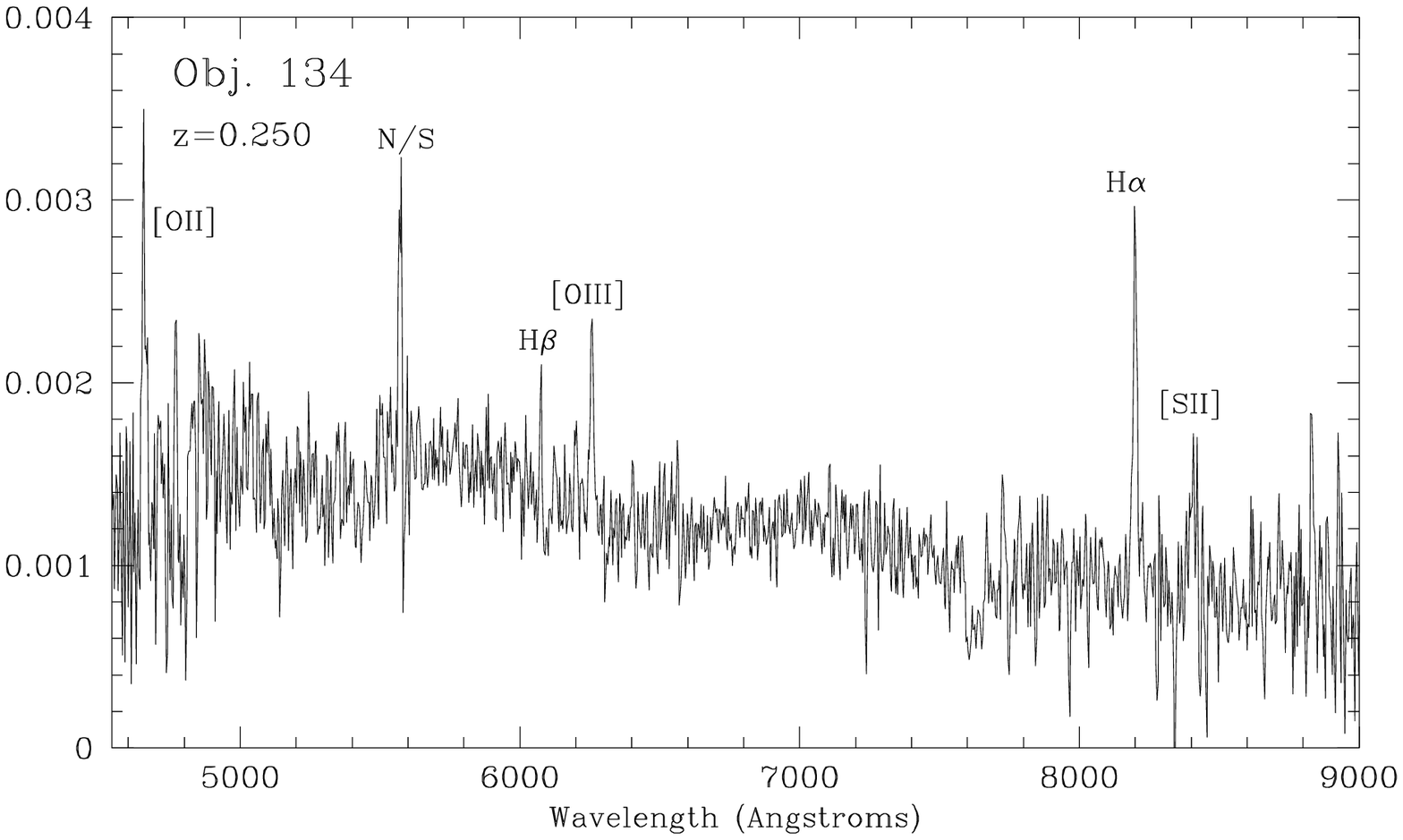}
  \epsfxsize 0.48\hsize
\epsffile{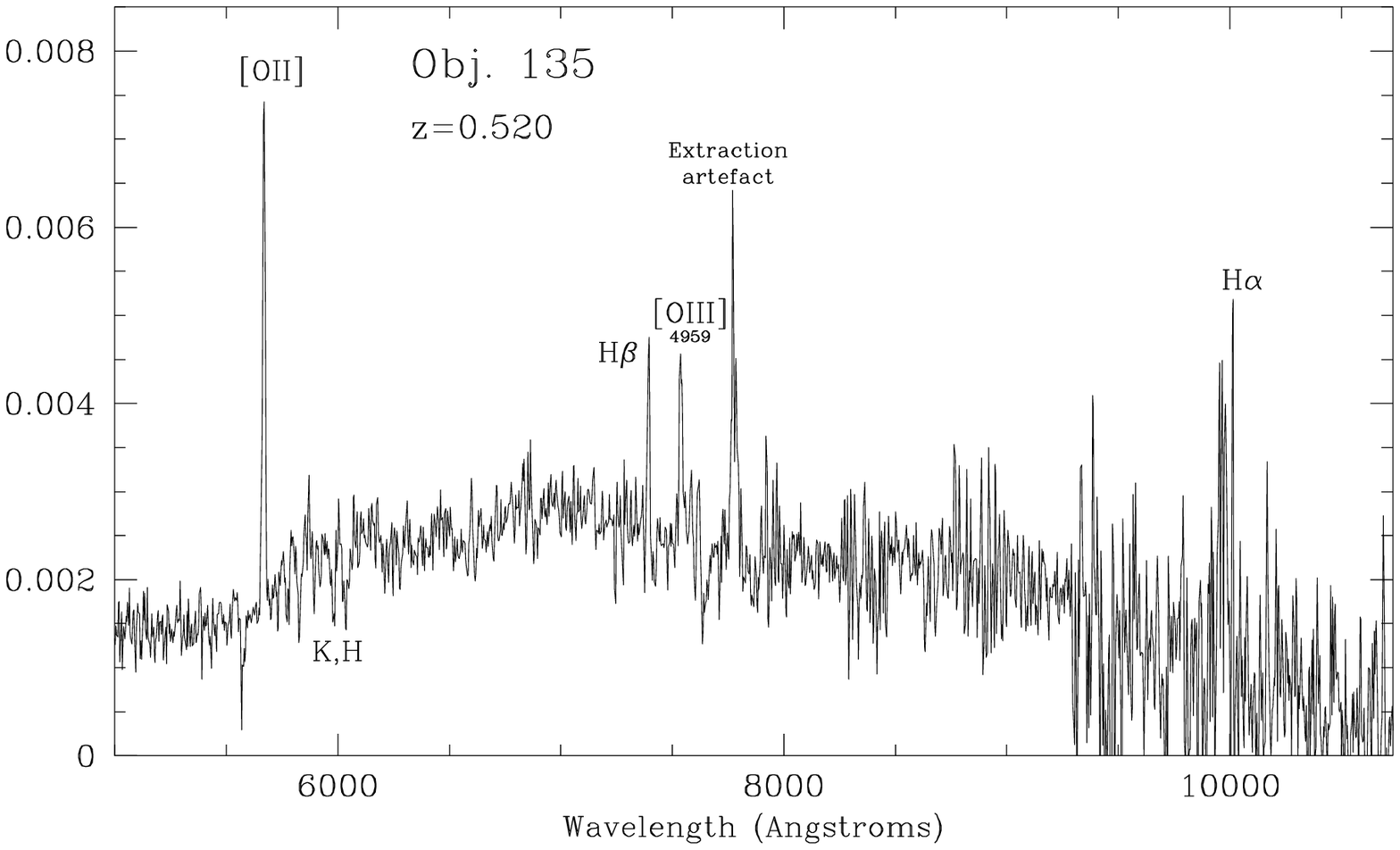}
\end{center}

\caption{ NELG OPTICAL SPECTRA. Note that in object 135 the [OIII]
line is 4959, not 5007 which is obscured by a night sky feature.
Also note the nearby extraction artefact. This artefact is not
visible on another spectrum of this object which confirms 
[OII]3727, $H_{\beta}$ and [OIII]4959 but does not cover $H_{\alpha}$.
The artefact arises because the object lies right on the edge of
the slit.}
\label{fig:spectra3}
\end{figure*}

Our optical spectra
(figures~\ref{fig:spectra1},~\ref{fig:spectra2} and~\ref{fig:spectra3}) show
a variety of narrow (FWHM $<$ 1000 km/s) emission lines.  $H_{\alpha}$
is often redshifted out of the detectable spectral band but [OII] is
usually within range, and present. The rest frame equivalent width of
[OII] is typically 20-60$\rm \AA$, larger than that found in nearby
galaxies except for irregular galaxies and Markarian galaxies
(Kennicutt 1992).  $H_{\beta}$ and [OIII] are usually in range and
sometimes present. Although the spectra are not all of the same signal
to noise, it is clear that the spectra are not all identical.  Some
show strong lines, others weak, and the line ratios vary between the
spectra.

\begin{figure}
\begin{center}
  \leavevmode
  \epsfxsize 1.0\hsize
\epsffile{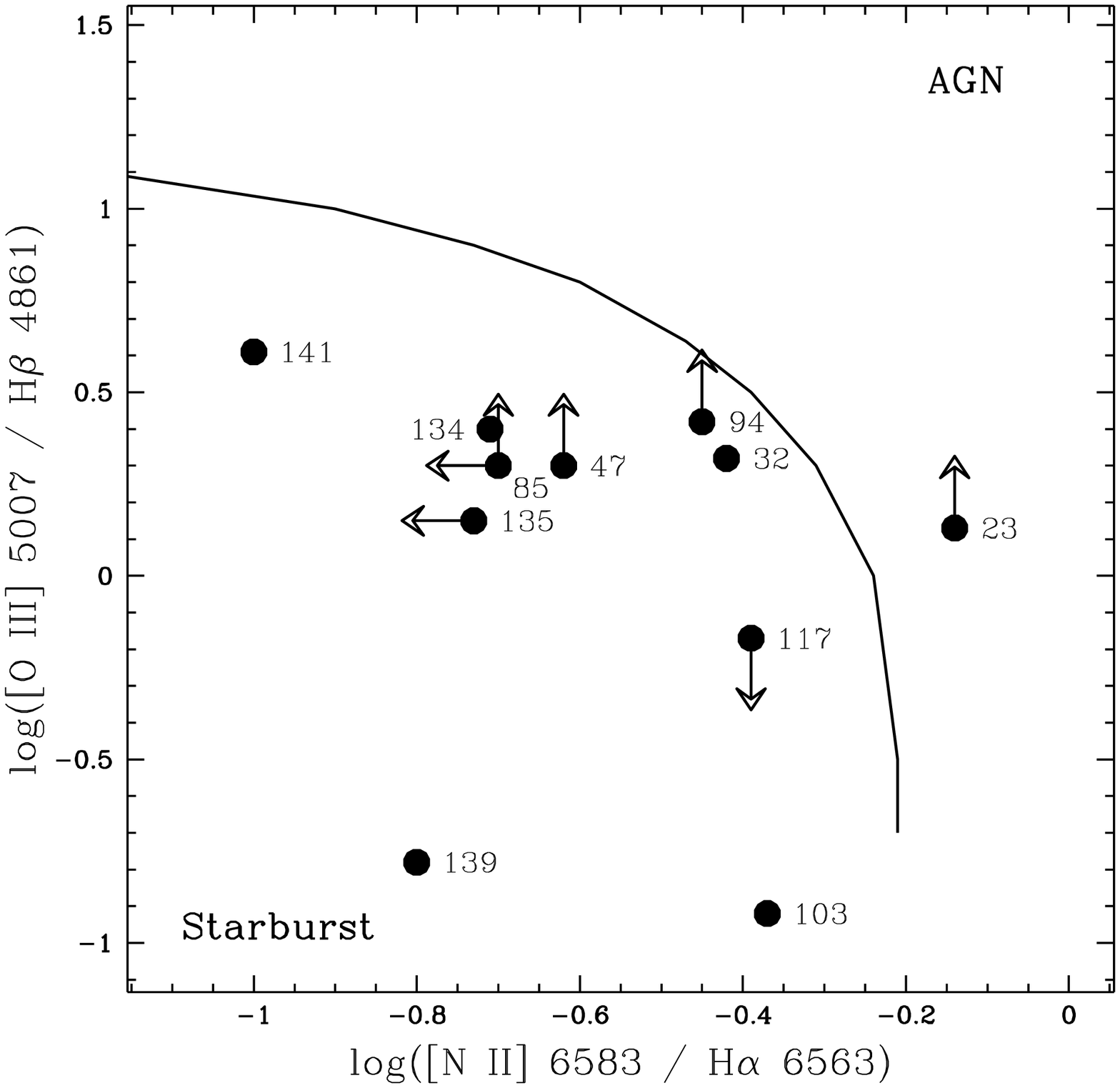}
\end{center}
\caption {$[OIII]5007/H_{\beta}$ vs $[NII6583]/H_{\alpha}$ line ratio
plot for all NELGs for which we have optical spectra within our 15 arcmin
full survey area. Thus we include sources 139 and 141 which are below
the flux limit of our complete survey. We also include the NELG which
contributes part of the X-ray emission in source 23.  Not all NELGs
for which we have optical spectra are plotted as, in some cases, some
of the spectral lines are not covered by our spectra.  The solid line
is the dividing line between AGN and HII region galaxies from Veilleux
and Osterbrock (1987)}
\label{fig:NII}
\end{figure}

\begin{figure}
\begin{center}
  \leavevmode
  \epsfxsize 1.0\hsize
\epsffile{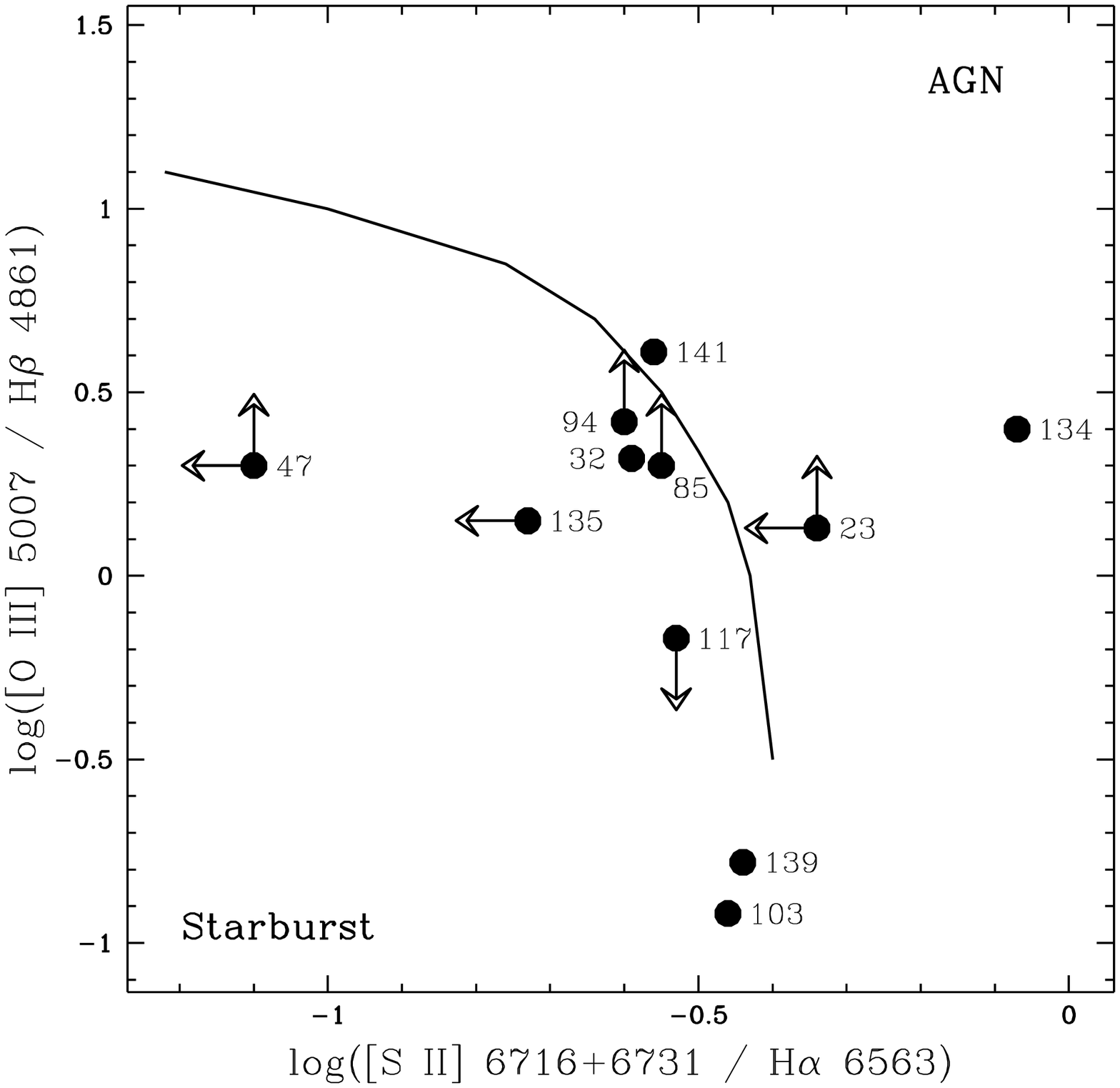}
\end{center}
\caption { $[OIII]5007/H_{\beta}$ vs $[SII6716]/H_{\alpha}$ line ratio
plot for the same sample as in figure~\ref{fig:NII}.}
\label{fig:SII}
\end{figure}

In order to quantify the spectra we have plotted the standard
diagnostic line ratios (cf Veilleux and Osterbrock 1987) of
[OIII]5007/$\rm H_{\beta}$ {\it vs.} [NII]6583/$\rm H_{\alpha}$
(figure~\ref{fig:NII}), and [OIII]5007/$\rm H_{\beta}$ {\it vs.}
[SII]6716+6731/$\rm H_{\alpha}$ (figure~\ref{fig:SII}).  Veilleux and
Osterbrock show that these line ratios provide a reasonable
discriminant between starburst galaxies and true AGN.  We find that
the NELGs are distributed between the higher ionisation bound of HII
region (ie starburst) galaxies and the lower ionisation area of true
AGN. Although not plotted here, a further diagnostic diagram is
provided by the ratio of [OIII]5007 to [OII]3727 (eg plotted as
[OIII]5007/$\rm H_{\beta}$ {\it vs.} [OII]3727/$\rm H_{\beta}$ by
Tresse \etal (1996) in their figure 3). For our sample [OIII]5007/$\rm
H_{\beta}$ is usually slightly greater than unity and [OII]3727/$\rm
H_{\beta}$ is a factor of a few greater than unity, placing our NELGs
in the region of parameter space occupied by LINERs (low ionisation
nuclear emission region galaxies, eg Ferland and Netzer 1983).  Thus,
the NELGs appear to be a mix of galaxy types, generally of low
ionisation, including  LINERs and starbursts but also some Seyfert II
galaxies.

\begin{figure}
\begin{center}
  \leavevmode
  \epsfxsize 1.0\hsize
  \epsffile{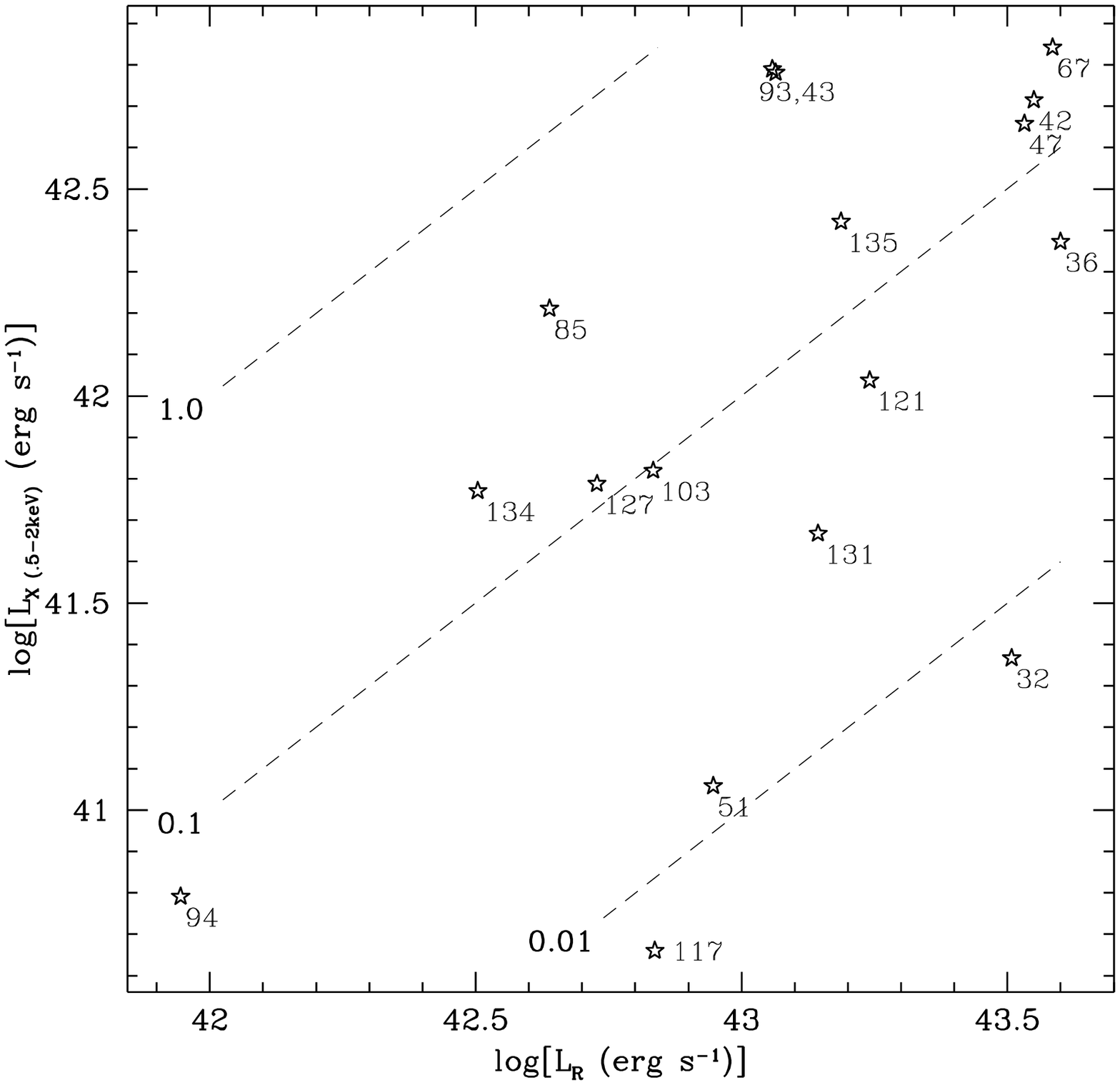}
\end{center}
\caption 
{The rest frame X-ray and optical luminosities for the NELGs in our
complete sample. $L_{X}$ is the integrated flux from 0.5-2keV and
$L_{R}$ is the R band luminosity integrated over a 1000A band. $H_0 =
50, q_0 = 0.5$, $\alpha_{opt} = 1$ and $\alpha_{X}$ = 0.7. The dashed
lines are lines of constant X-ray/optical ratio.  On average we expect
AGN to have ratios near 1 and starburst galaxies to have ratios around
0.001. The majority of our NELGs lie somewhere inbetween indicating
a possible mixture of X-ray emission mechanisms.}

\label{fig:lxlopt}
\end{figure}

As a further diagnostic of the NELG emission mechanism we plot, in
figure~\ref{fig:lxlopt}, the NELG X-ray/optical ratios. Typically
accretion powered objects, ie AGN, have X-ray/optical ratios close
to unity, whereas starburst galaxies have ratios nearer $10^{-3}$
(eg Moorwood 1996).
Our NELGs have a wide range of ratios and
are distributed in between these two values. Taken together with
the line ratio plots, the X-ray/optical ratios indicate that there
is probably not one single emission mechanism which is responsible
for the X-ray emission in the NELGs. It is more likely that there
is a mixture of emission mechanisms not only within the sample as
a whole but within individual objects. It is likely that both
starburst and true AGN emission exists together in many of the NELGs.

In addition to the X-ray and optical surveys, we have carried out deep
radio observations of the whole 30 arcmin diameter X-ray survey area
with the Very Large Array (VLA) reaching source detection limits of
$\sim 0.3$ mJy at both 20 and 6 cm.  A full analysis of these data
will be presented elsewhere and here we simply note that, of the 10
X-ray brightest NELGs, 4 are detected in our 20cm VLA survey with
fluxes $>$0.5 mJy.  These are noted in Table~\ref{tab:main}. The
luminosities are high ($\sim 10^{23}$ W Hz$^{-1}$ at 20 cm) by normal
spiral galaxy standards (Condon 1992) but, with one exception, well
below those of classical radio galaxies. Condon shows that
radio emission in non-AGN galaxies is related to the recent star
formation rate; similarly Benn \etal (1993) find that the optical
spectra of sub-mJy radio sources are similar to those of the faint
starburst galaxies detected in observations with the Infrared
Astronomical Satellite Observatory (IRAS).  Our radio observations are
therefore consistent with the suggestion above that some of the NELG
X-ray emission arises in starburst activity.

Some galaxies with similar optical spectra to our NELGs have been
detected in previous X-ray surveys with brighter flux limits (eg
Boller \etal 1993; Boyle \etal 1995; Carballo \etal 1995). However in
these surveys they represented a small fraction of the identified
sources. It was therefore not possible to determine reliably whether
they were likely to be a major contributor to the XRB at fainter
fluxes or not. Our observations unambiguously show that they are. The
statistical association of galaxies with faint X-ray sources (in the
form of X-ray fluctuations) has also been reported previously.  Lahav
\etal (1993) find an association between GINGA X-ray fluctuations and
UGC galaxies (of unknown spectral type) and, of particular relevance
to the present work, Roche \etal (1995) find a $5\sigma$ correlation
between galaxies of R$<$23 and fluctuations in a 50ksec ROSAT PSPC
observation.  However optical spectra are generally not available for
the galaxies in the Lahav \etal and Roche \etal correlations and so
their nature was unknown.  Our observations indicate that those
galaxies probably included a mixture of NELGs and the absorption line
galaxies such as we find to be associated with groups and clusters.

\section{COMPARISON OF THE NELGS WITH OTHER FAINT GALAXY POPULATIONS}

Above we have discussed the likely nature of the NELGs. We see that
they are generally of low ionisation but that some true AGN
contribution, as well as a starburst contribution to the X-ray and
optical emission, is likely in at least some NELGs. We have
preliminary indications that the NELGs may be related to the faint
radio galaxy population, and they are certainly related to starburst
galaxies which are already well known to be moderately luminous X-ray
sources (eg Griffiths and Padovani 1990; Rephaeli \etal 1991;
Fruscione and Griffiths 1991; Moran, Halpern and Helfand 1994).
However we note that the X-ray luminosities of the NELGs ($3 \times
10^{41}$ to $5 \times 10^{42}$ ergs s$^{-1}$) are all well above the
X-ray luminosities of typical starburst galaxies (maximum detected
luminosity $\sim2 \times 10^{41}$ ergs s$^{-1}$; Griffiths and
Padovani 1990; Read and Ponman 1997).  Thus either the
NELGs are extreme examples of starburst galaxies or they contain
another X-ray emission source, eg an AGN.  If they are starbursts,
then their implied far infrared luminosities should be detectable in
deep ISO observations.  There has been much speculation as to the
likely contribution of starburst galaxies to the XRB but uncertainty
in their evolution and in extrapolating to faint flux levels has
prevented any reliable conclusions from being reached. We note that
the NELGs are predominantly blue, and faint, and so we may speculate
about their relationship to the faint blue galaxy population (eg
Colless \etal 1990).  However we note that, based on their angular
correlation function, the faint blue galaxies have likely redshifts of
0.2-0.3 (Landy \etal 1996), a little lower than that of the NELGs. For
that assumed redshift range, the faint blue galaxies have absolute
magnitudes $\sim 3$ magnitudes fainter than the NELGs.

The obvious question therefore arises as to whether the various faint
galaxy populations are really different in any way or merely different
manifestations of broadly the same basic phenomenon.  It is therefore
interesting to compare our NELGs with the field galaxy population in a
similar redshift range. Tresse et al (1996) have examined the spectra
of 138 galaxies selected from an I-band magnitude limited sample
($17.5 \leq I \leq 22.5$) out to redshift 0.3. As their sample selects
galaxies on the basis of their old stellar population it avoids bias
towards starburst galaxies which dominate B-band or IRAS selected
samples. However they still find that 85\% of their sample display at
least narrow line $H_{\alpha}$ in emission and 53\% display, in
addition, several forbidden emission lines; 17\% of their galaxies
display spectra consistent with Seyfert 2 galaxies.  Only 15\% display
pure absorption line spectra.  The Tresse \etal sample shows a very
similar distribution of emission line spectra to our NELGs and their
emission line galaxies have a very similar distribution to our NELGs
in the standard line ratio plots (Veilleux and Osterbrock 1987),
straddling the boundary between AGN (Seyfert II) and HII region
spectra. The distribution of [OII]3727 equivalent widths is very
similar in our NELGs and in the Tresse \etal sample.

The average absolute magnitude of Tresse et al's emission line
galaxies, $M_{B} = -18.4$, is less luminous than that of our NELGs
($\rm M_{R}$ -20 to -23) by around 3 magnitudes, but is similar
to that estimated for the faint blue galaxies by Landy \etal
Our NELGs may
therefore just represent the more luminous examples of the emission
line field galaxies.  If the optically less luminous, but more numerous,
emission line field galaxies produce similarly lower luminosity X-ray
emission, then we may reasonably expect that the contribution of NELGs
to the XRB will continue to rise at lower X-ray fluxes, supporting
the extrapolations made in Section 6.

The Tresse \etal sample is selected up to redshift 0.3, whereas our
NELGs reach redshift 0.6 thus the possibility of evolutionary
differences between the two samples exists. Similarly, in the very
local universe Huchra and Burgh (1992) find very few active galaxies
indicating possible evolution in the activity of field galaxies
between redshift 0 and 0.3. However, as discussed by Tresse \etal, the
fraction of field galaxies which are active which is given by Huchra
and Burgh (2\%) should be treated cautiously and is probably an
underestimate of the overall number of emission line galaxies
(including galaxies other than Seyferts). Evolutionary uncertainties
therefore prevent detailed extrapolation of the contribution of NELGs
to the XRB at very faint fluxes, however we do note that Page \etal
(1997b) find no strong evidence for evolution in the NELG populations
found in the present survey and in the RIXOS survey and so
the simple extrapolations given in Section 6 may not be too far wrong.

Thus we conclude that a substantial
fraction of the XRB is contributed by field galaxies. 
The IRAS galaxies, and sub-mJy radio sources, are most
likely simply the starburst-dominated fraction of these same emission
line field galaxies.

\section{\bf CONCLUSION}

In the deepest optically identified X-ray survey so far made, we have
resolved approximately half of the X-ray background at 1 keV.  The
identifications are spectroscopically 85\% complete to $2 \times
10^{-15}$ erg cm$^{-2}$ s$^{-1}$ (0.5 -2.0 keV). At brighter fluxes
($\geq 10^{-14}$ erg cm$^{-2}$ s$^{-1}$) we confirm the results of
previous less deep X-ray surveys with 84\% of our sources being
QSOs. However at the faint flux limit the survey is dominated by a
population of galaxies, mainly with narrow emission lines
(NELGs). Whereas the QSO differential source count slope below $
10^{-14}$ erg cm$^{-2}$ s$^{-1}$ is $\sim$-1.4, severely
sub-Euclidean, the differential NELG slope is close to Euclidean
($\sim-2.4$).  To the survey limit QSOs still contribute the largest
identified fraction of the XRB, $>$31\% as opposed to 8\% for NELGs
and 10\% for clusters.  We note that a small number of the fainter
NELGs may be misidentifications and that some of the unidentified
sources are almost certainly faint QSOs, although some are also likely
to be either NELGs or clusters.  However as we are unable to correct
our observed contributions to the XRB until these sources are properly
identified, we simply note that all of our contributions are
observed values.  However if the observed differential source counts
can be reliably extrapolated to fainter fluxes, clusters will
contribute almost nothing more to the XRB and QSOs will contribute
only a small amount more. On the other hand the NELG contribution to
the XRB will double by a flux limit a factor 4 below the present
survey.

The NELGs observed so far lie in the redshift range 0.1-0.6, 
as one might expect given an identification limit of R=23.
It is not yet clear whether there is any astrophysical
significance to the maximum observed redshift.
The NELGs have
generally blue colours, and have optical spectra similar to that of
the active field galaxy population at a similar redshift.  Some NELGs
are definitely large spirals but the possibility that some are
ellipticals cannot be ruled out on the basis of the present
analysis. Many of the NELGs are in disturbed or interacting systems.
The NELGs, both as a sample and individually, appear to be a mixture
of starburst galaxies and true AGN. By comparison with the field
galaxy surveys of Tresse \etal, the simplest interpretation of our
results is that the NELGs which we detect are simply the more luminous
members of the normal field galaxy population.

The average NELG X-ray spectrum is harder than that of the QSOs,
and similar to that of the remaining unresolved cosmic soft X-ray
background (XRB). NELGs should therefore be a major contributor to
the XRB at higher energies although without a detailed understanding
of the NELG X-ray emission mechanism it is not possible to say
up to which energies one might reasonably expect them to contribute.
However their fluxes at higher energies
will still be low ($<10^{-14}$ erg cm$^{-2}$ s$^{-1}$, 2-10 keV)
and so they may not show up in large numbers in ASCA surveys but
should be prominent in surveys with XMM and AXAF.

We also find that a small number of groups
or clusters of galaxies are identified as X-ray sources. So far we
have not found any X-ray identifications with entirely isolated
absorption line galaxies. All such galaxies
appear to have at least a few faint companions. However
the groups/clusters are generally not very rich. The number of groups
found is in approximate agreement with a zero-evolution scenario
for low luminosity clusters,
unlike the situation for rich clusters in the same redshift range
($Z>0.3$).

{}

\noindent
{\bf ACKNOWLEDGEMENTS}\\

\noindent
We thank the following observatories and their staff for support of
this project: the Canada-France-Hawaii Telescope, the UK William
Herschel Telescope, the Nordic Optical Telescope, the University of
Hawaii 88 inch Telescope the Mitchigan-Dartmouth-MIT Telescope and the
Very Large Array Radio Telescope.  
We thank Craig Collins for careful analysis of the extended X-ray sources.
This work was supported by grants
to a number of authors from the UK Science and Engineering Research
Council and Particle Physics and Astronomy Research Council. KOM
acknowledges support from the Royal Society.
FJC acknowledges partial financial support from the DGES under
project PB95-0122.

\end{document}